%% file: main.tex
\documentclass[SIG,biber]{nowfnt} 

\usepackage[utf8]{inputenc}
\usepackage{graphicx}
\graphicspath{{.}{./Figures/}}
\usepackage{csquotes}
\usepackage[cmex10]{amsmath}
\usepackage{longtable}
\usepackage{txfonts}
\usepackage{float}
\usepackage{multirow}
\usepackage{multicol}
\usepackage{amssymb}
\usepackage{tikz}
\usepackage[nolist,withpage]{acronym}
\usepackage{etoolbox}
\makeatletter
\newif\if@in@acrolist
\AtBeginEnvironment{acronym}{\@in@acrolisttrue}
\newrobustcmd{\LU}[2]{\if@in@acrolist#1\else#2\fi}

\newcommand{\ACF}[1]{{\@in@acrolisttrue\acf{#1}}}
\makeatother

\usetikzlibrary{arrows, shapes}

\input{math.tex}

\title{Wireless for Machine Learning: \\a Survey}

\maintitleauthorlist{
Henrik~Hellstr\"om \\
KTH - Royal Institute of Technology
\and
Jos\'e Mairton~B. da Silva Jr. \\
KTH - Royal Institute of Technology
\and
Mohammad~Mohammadi~Amiri \\
MIT - Massachusetts Institute of Technology
\and
Mingzhe~Chen \\
Princeton University
\and
Viktoria~Fodor \\
KTH - Royal Institute of Technology
\and
H.~Vincent~Poor \\
Princeton University
\and
Carlo~Fischione \\
KTH - Royal Institute of Technology
}

\issuesetup
{%
 copyrightowner={H.~Hellstr\"om},
 volume        = xx,
 issue         = xx,
 pubyear       = 2022,
 isbn          = xxx-x-xxxxx-xxx-x,
 eisbn         = xxx-x-xxxxx-xxx-x,
 doi           = yy.zzzz/XXXXXXXXX,
 firstpage     = 1, 
 lastpage      = x 
 }

\usepackage{mwe}

\author[1]{Hellstr\"om,Henrik}
\author[1]{Barros da Silva Jr.,Jos\'e Mairton}
\author[2]{Amiri,Mohammad~Mohammadi}
\author[3]{Chen,Mingzhe}
\author[1]{Fodor,Viktoria}
\author[3]{Poor,H.~Vincent}
\author[1]{Fischione,Carlo}

\affil[1]{KTH - Royal Institute of Technology, School of Electrical Engineering and Computer Science}
\affil[2]{MIT - Massachusetts Institute of Technology, MIT Media Laboratory}
\affil[3]{Princeton University, Department of Electrical and Computer Engineering}

\articledatabox{\nowfntstandardcitation}

\begin{document}

\input{acronyms.tex}

\makeabstracttitle

\begin{abstract}
As data generation increasingly takes place on devices without a wired connection, machine learning (ML) related traffic will be ubiquitous in wireless networks. Many studies have shown that traditional wireless protocols are highly inefficient or unsustainable to support ML, which creates the need for new wireless communication methods. In this survey, we give an exhaustive review of the state-of-the-art wireless methods that are specifically designed to support ML services over distributed datasets. Currently, there are two clear themes within the literature, analog over-the-air computation and digital radio resource management optimized for ML. This survey gives a comprehensive introduction to these methods, reviews the most important works, highlights open problems, and discusses application scenarios.
\end{abstract}


\chapter{Introduction}
\label{sec:introduction} 

With the increasing popularity of mobile devices and the continuous growth of \ac{IoT}, we are having increasing access to vast amounts of distributed data. According to a recent report from Ericsson, the global number of connected \ac{IoT} devices will rise to 4.1 billion by 2024~\cite{ericssonmobility}, which is four times the 1 billion observed in 2019. Simultaneously, breakthroughs in \ac{ML} are allowing us to analyze the data of edge devices so as to solve a wide range of complex problems, such as image recognition~\cite{he2016deep}, language processing~\cite{collobert2011natural}, and predictive modeling~\cite{branco2016survey}. However, since \ac{ML} was originally conceived in centralized settings where all data must be transmitted to a central sever, the application of \ac{ML} on distributed datasets over wireless networks is generating new challenges for the wireless networks, namely:
\begin{itemize}
    \item \textbf{Privacy:} Many \ac{ML} applications require the use of privacy-sensitive data. In these cases, it is either desirable or necessary that the training dataset cannot be inferred by listening to the \ac{ML} updates being transferred wirelessly \cite{so2021codedprivateml};
    
    \item \textbf{Security:} When an \ac{ML} model is trained distributively, a bad actor can corrupt the final model by transmitting malicious model updates \cite{vempaty2013distributed}. Wireless protocol design should inhibit an attackers ability to do so;
    
    \item \textbf{Communication and Energy Efficiency:} \ac{DML} requires the communication of high-dimensional model updates for hundreds or thousands of iterations before the model has converged. This communication of updates generally forms the performance bottleneck of the training process, imposes the risk of excessively draining the batteries of training devices and overwhelming the capacity of the wireless network \cite{shi2020communication}.
\end{itemize}
To address these challenges, a new approach toward communication protocol design has emerged \cite{zhu2020toward}. This new approach considers the design of completely novel wireless methods for carrying data needed for the \ac{ML} tasks. Unlike traditional wireless protocol design, the objective of Wireless for \ac{ML} is not to deliver bits as efficiently as possible, but to distill the intelligence carried within the data. The traditional communication protocols that are designed to maximize data rate and minimize bit errors have been shown to be greatly inefficient for carrying \ac{ML} related data~\cite{amiri2020machine,zhu2019broadband, nishio2019client, chen2020joint, liu2020data}. Instead, Wireless for \ac{ML} offers new methods that are better aligned with the \ac{ML} objective and invites us to rethink how wireless communication protocols are designed. Among the novel methods that have been proposed, two major themes arise, namely analog \ac{AirComp} and \ac{RRM} optimized for \ac{ML}. In \ac{AirComp}, the long-standing doctrine of interference avoidance is questioned and novel interference-promoting protocols are proposed. While in \ac{RRM} for \ac{ML}, the  new objectives lead to solutions that are fundamentally different from what is used today.

The idea of wireless protocols customized for ML, although not yet available in the current cellular wireless standards, is compatible with the current standard specifications. The new cellular standard 5G has introduced the concept of network slicing to improve flexibility and scalability~\cite{rost2017network}. Network slicing allows independent sets of network protocols to run on common physical infrastructure, to support services with conflicting requirements. As an example, video streaming requires high data rates and accepts high latency, while critical \ac{IoT} usually requires low latency and high reliability while accepting low data rates. As of today, these services cannot be supported using the same protocols, but with network slicing, they can be implemented on the same physical infrastructure~\cite{bennis2018ultrareliable}. Going beyond 5G, the demand for \ac{ML} services is projected to grow significantly and discussions have begun on a dedicated network slice for \ac{ML} in future-generation cellular networks such as beyond-5G and 6G~\cite{saad2019vision,zhang20196g,strinati20196g,guan2021customized}. Given this possibility, the investigation of Wireless for \ac{ML} becomes relevant not only for local-area networks but also for large-scale cellular networks.


\section{Related work}
Although the general intersection of \ac{ML} and wireless communications is currently a prolific field of research that has already generated multiple surveys, there is little review work on Wireless for \ac{ML}. The current surveys can roughly be classified into three categories: \textit{\ac{ML} for Wireless Communications}, \textit{Wireless for ML}, and \textit{Communication-Efficient DML}. We list a set of representative surveys in Table \ref{tab:relatedwork}. A brief description of the three areas follows.
\begin{enumerate}
    \item \textbf{Wireless for \ac{ML}} uses wireless communication protocols as a method to enable or significantly improve \ac{ML} training over wireless networks. Unlike in traditional wireless communication, the communication system is not oblivious to the meaning that the bits convey. Instead, Wireless for \ac{ML} is a task-oriented communication philosophy, where the goal of the communication system is to distill the intelligence carried within the data.
    \item \textbf{Communication-efficient \ac{DML}} has the same goal as Wireless for \ac{ML} but uses different methods. Instead of customizing the wireless protocols, advancements are made by modifying or redesigning the \ac{ML} algorithm. The results of these works are agnostic to the communication protocol so that they can be applied regardless of the specific technologies used to transmit data.
    \item \textbf{\ac{ML} for wireless} uses \ac{ML} as a method to design wireless communication protocols for general communication services. Therefore, its goal is the same as in traditional wireless communications, i.e., efficient and reliable transfer of arbitrary data. The communication system should support a wide variety of services and is therefore deliberately oblivious to the semantics of the transmitted bits.
\end{enumerate}

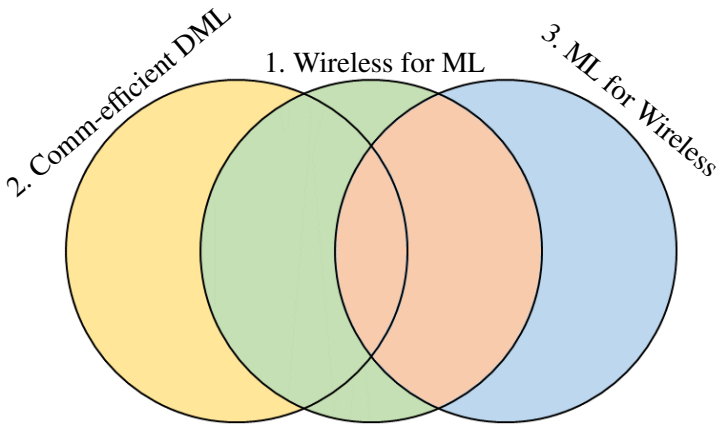
\begin{figure}[ht]
\centering
    \begin{tikzpicture}
        \node[inner sep=0pt] (russell) at (0,0)
            {\includegraphics[width=0.7\linewidth]{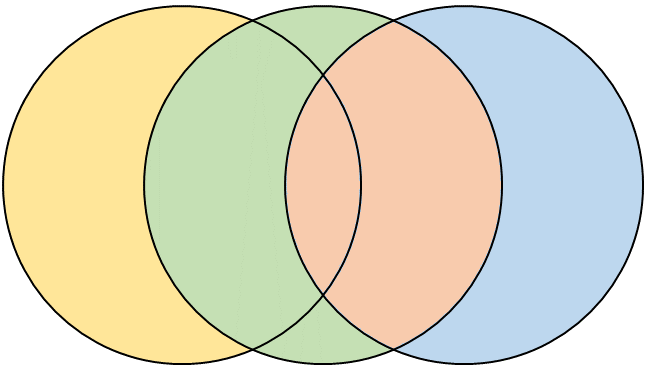}};
        \node[rotate=40] at (-3.4,2.0) {2. Comm-efficient DML};
        \node[rotate=-40] at (3.4,2.0) {3. ML for Wireless};
        \node[] at (0,2.5) {1. Wireless for ML};
    \end{tikzpicture}
    
    \caption{Illustration of the relationship between Wireless for ML and related fields. The first circle correponds to Communication-efficient DML, the second to Wireless for ML, and the third to ML for Wireless. The blue area corresponds to pure ML for Wireless, which is a very prolific field of research that has already generated a large number of review articles. Likewise, the yellow area corresponds to pure Communication-efficient DML which is also a well-covered area. In this survey, we focus on the green moon, i.e., pure Wireless for ML and its intersection with Communication-efficient DML. As far as we are aware, there are no published works in the red area.}
    
    \label{fig:relationship}
\end{figure}

In addition to the three categories above, their intersections can be considered as areas of their own, illustrated in Figure~\ref{fig:relationship}. The intersection of Wireless for \ac{ML} and communication-efficient \ac{DML} considers the co-design of the \ac{ML} algorithm and the wireless protocol. With such an approach, researchers attempt to reach some global optimality, which is lost when the two problems are treated in isolation. Additionally, one can consider the intersection between Wireless for \ac{ML} and \ac{ML} for Wireless, where \ac{ML} would be used as a tool to design a wireless protocol with the goal of supporting distributed \ac{ML} services. However, as far as we are aware, no works have been published in this direction. In this survey, we consider all works within Wireless for \ac{ML}, including its intersections, symbolized by the green moon in Figure \ref{fig:relationship}.

Some of the papers in Table \ref{tab:relatedwork} discuss Wireless for ML, but is not extensive since it is not the main purpose of the paper. The closest match to our survey is \cite{gafni2021federated}. However, despite describing some works within Wireless for ML, the paper is not a comprehensive survey of the field, instead its purpose is to introduce a new framework to describe Federated Learning. We believe that due to this gap, there is currently no one-stop survey that offers an overview of the Wireless for \ac{ML} literature, which motivates us to write this survey with the following contributions:
\begin{itemize}
    \item We provide an introduction to important concepts necessary to understand the field as a whole, such as \ac{DML}, over-the-air computation, and the distinction between generic wireless communication protocols and Wireless for \ac{ML};
    \item We describe the most important works of the field in a concise way to offer a thorough overview of the state-of-the-art. Both for analog over-the-air computation and digital communications;
    \item We discuss several important open problems and future research directions within Wireless for \ac{ML};
    \item We describe a number of application areas where Wireless for \ac{ML} can provide a benefit to society, such as vehicular communications and virtual reality, and describe the challenges associated with those applications.
\end{itemize}

\setlength{\extrarowheight}{1pt}
\begin{table*}[htbp]
    \caption{Surveys written within the intersection of ML and communications. The topics of ML for Communications and Communication-efficient DML have been covered in many surveys, unlike Wireless for ML. At most, Wireless for ML has been covered briefly in conjunction with Communication-efficient DML.}
    \resizebox{\textwidth}{!}{%
    \begin{tabular}{|p{0.07\linewidth}|p{0.35\linewidth}|p{0.1\linewidth}|p{0.3\linewidth}|}
    \hline
    \textbf{Year} & \textbf{Journal} & \textbf{Ref.}& \textbf{Research Area from Figure \ref{fig:relationship}} \\
    \hline
    2017 & IEEE Communication Surveys and Tutorials &~\cite{mao2018deep} & 3 \\
    \hline
    2018 & Proceedings of the IEEE &~\cite{park2019wireless} & 2 \\
    \hline
    2019 & Proceedings of the IEEE &~\cite{zhou2019edge} & 2 \\
    \hline
    2020 & IEEE Communication Surveys and Tutorials &~\cite{hussain2020machine} & 3 \\
    \hline
    2020 & IEEE Communication Surveys and Tutorials &~\cite{wang2020thirty} & 3 \\
    \hline
    2020 & IEEE Internet of Things Journal &~\cite{deng2020edge} & Mostly 2 with some 1 \\
    \hline
    2020 & IEEE Communication Surveys and Tutorials &~\cite{wang2020convergence} & 2 \\
    \hline
    2020 & IEEE Internet of Things Journal &~\cite{abdulrahman2020survey} & 2 \\
    \hline
    2020 & IEEE Communication Surveys and Tutorials &~\cite{Yang2020} & Mostly 2 with some 1  \\
    \hline
    2021 & IEEE Internet of Things Journal &~\cite{imteaj2021survey} & 2\\
    \hline
    2021 & Elsevier High-Confidence Computing &~\cite{xia2021survey} & 2 \\
    \hline
    2021 & arXiv &~\cite{gafni2021federated} & Mostly 1 with some 2 \\
    \hline
    \multicolumn{3}{|c|}{This survey} & 1 \\
    \hline
    \end{tabular}
    }
    \caption{Surveys written within the intersection of ML and communications. The topics of ML for Communications and Communication-efficient DML have been covered in many surveys, unlike Wireless for ML. At most, Wireless for ML has been covered briefly in conjunction with Communication-efficient DML.}
    \label{tab:relatedwork}
\end{table*}


\section{Notation and organization}
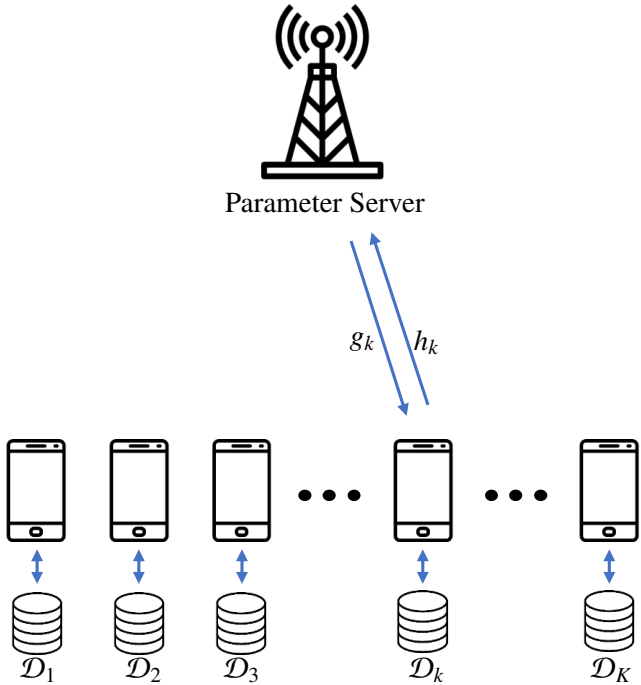
\begin{figure}[ht]
\centering
    \begin{tikzpicture}
        \node[inner sep=0pt] (russell) at (0,0)
            {\includegraphics[width=8.5cm]{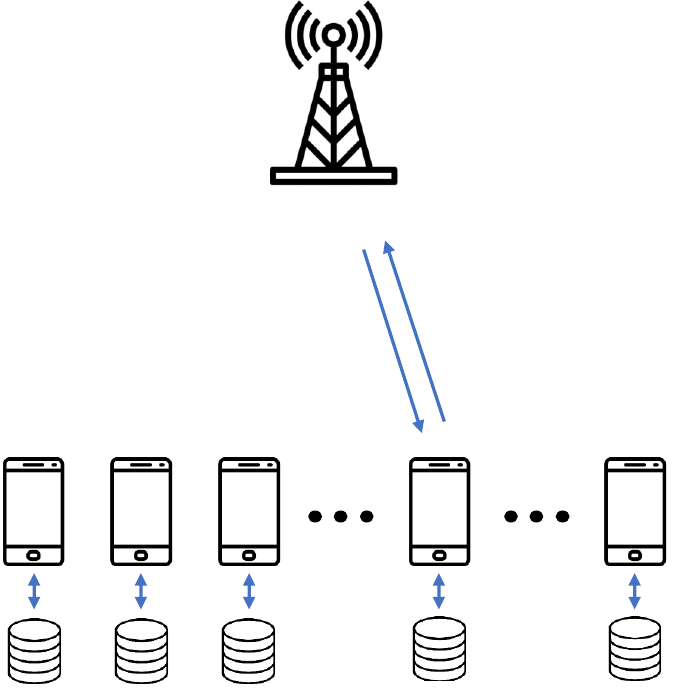}};
        \node[] at (0,1.7) {Parameter Server};
        \node[] at (0.5,-0.15) {$g_k$};
        \node[] at (1.35,-0.15) {$h_k$};
        \node[] at (-3.8,-4.5) {$\mathcal{D}_1$};
        \node[] at (-2.4,-4.5) {$\mathcal{D}_2$};
        \node[] at (-1.1,-4.5) {$\mathcal{D}_3$};
        \node[] at (1.35,-4.5) {$\mathcal{D}_k$};
        \node[] at (3.8,-4.5) {$\mathcal{D}_K$};
    \end{tikzpicture}
    \caption{Illustration of the PS and wireless network setup used throughout this survey. Current wireless communication protocols substantially hinder or completely block distributed training over this setup. The Wireless for \ac{ML} paradigm is an approach to tackle such hinders and blockages.}
    \label{fig:problemdesc}
\end{figure}

All papers that we survey are essentially concerned with the solution to a basic problem, namely the training of a classifier over a wireless communication network constrained by the natural characteristics of the wireless channel. Throughout this survey, we assume a centralized architecture where there is a central controller or \ac{PS} able to make decisions such as user selection, bandwidth allocation, and aggregation frequency control. Such an architecture is representative of most of the wireless networks used today, from large scale mobile to personal area networks. The communication channel is wireless and is thus subject to fading, additive noise, and bandwidth restrictions. The training dataset is always carried by user devices and the training algorithms will always be chosen to minimize a loss based on the global dataset. Unless specified otherwise, the network consists of one \ac{PS}, i.e., the \ac{BS} or the \ac{AP}, and $K$ user devices, e.g., \ac{IoT} devices, \acp{UE}, or other wireless devices. The devices are each carrying a subset $\mathcal{D}_k$ of the global dataset $\mathcal{D}$ and the \ac{PS} carries no data. The global dataset consists of $N$ training samples and corresponds to the union of data available at all the user devices. For communication, the uplink $h_k$ and downlink $g_k$ channel coefficients are of particular importance. Figure~\ref{fig:problemdesc} illustrates the setup, a full list of notation is available in Table~\ref{tab:notation}, and relevant acronyms are available in Table~\ref{tab:acronym}.

The rest of this survey is organized as follows: Section~\ref{sec:distributedml} provides a primer on \ac{DML} and in particular \ac{FL}. In Sections~\ref{sec:ldcomac} and~\ref{sec:orthogonal}, we survey the Wireless for \ac{ML} literature for over-the-air computation and digital communication, respectively. In Section~\ref{sec:future_res}, we discuss the open problems of Wireless for \ac{ML} within both analog over-the-air computation and digital communications.
Then, in Section~\ref{sec:applications}, we discuss applications within Wireless for \ac{ML}.
Finally, we have concluding marks in Section~\ref{sec:conclude}. 

\begin{table}[htbp]
    \caption{Reference list of commonly used variables in this survey. Ordered alphabetically and by case. }
    \label{tab:notation}
    \begin{center}
    \begin{tabular}{|c|l|}
        \hline
        \textbf{Variable} & \textbf{Interpretation} \\
        \hline
        $B$ & Bandwidth available to the learning system \\
        \hline
        $\calD_k$ & Dataset carried by device $k$ \\
        \hline
        $E$ & Number of epochs \\
        \hline
        $K$ & Number of user devices \\
        \hline
        $M$ & Number of antennas at the parameter server \\
        \hline
        $N$ & Number of data samples in the global dataset \\
        \hline
        $N_k$ & Number of data samples stored at device $k$ \\
        \hline
        $\calS^t$ & Set of selected devices at iteration $t$ \\
        \hline
        $T_{\text{round}}$ & Time for federated learning communication round \\
        \hline
        $\beta$ & Learning rate \\
        \hline
        $\eta$ & Post-transmission scalar \\
        \hline
        $\nabla f (\vtW)$ & Gradient of function $f$ evaluated at $\vtW$ \\
        \hline
        $b_k$ & Ratio of total bandwidth allocated to device $k$ \\
        \hline
        $d$ & Number of model parameters in $\vtW$ \\
        \hline
        $f(\vtW)$ & Empirical risk function of the global model $\vtW$ \\
        \hline
        $g_k$ & CSI in downlink direction from server to device $k$ \\
        \hline
        $h_k$ & CSI in uplink direction from device $k$ to server \\
        \hline
        $l(\vtW)$ & Loss function for parameter $\vtW$\\
        \hline
        $p_k$ & Uplink power allocated to device $k$ \\
        \hline
        $v$ & Additive white Gaussian noise \\
        \hline
        $\vtW^t$ & Global model parameters at iteration $t$ \\
        \hline
        $\vtW_k^t$ & Local model parameters for device $k$ at iteration $t$\\
        \hline
        $\vtX$ & Input or feature of data sample\\
        \hline
        $\vtY$ & Output or label of data sample\\
        \hline
    \end{tabular}
    \end{center}
\end{table}

\begin{table}[htbp]
    \caption{Reference list of most acronyms used in this survey.}
    \label{tab:acronym}
    \begin{center}
    \begin{tabular}{|c|l|}
        \hline
        \textbf{Acronym} & \textbf{Phrase} \\
        \hline
        ADMM & Alternating Direction Method of Multipliers \\
        \hline
        AirComp & Over-the-air Computation \\
        \hline
        BAA & Broadband Analog Aggregation \\
        \hline
        BPSK & Binary Phase-Shift Keying \\
        \hline
        BS & Base Station \\
        \hline
        CML & Centralized Machine Learning \\
        \hline
        CoCoA & Comm-efficient distributed dual Coordinate Ascent \\
        \hline
        CoMAC & Computation over Multiple-Access Channels \\
        \hline
        CSI & Channel State Information \\
        \hline
        DML & Distributed Machine Learning \\
        \hline
        DP & Differential Privacy \\
        \hline
        DSGD & Distributed Stochastic Gradient Descent \\
        \hline
        ESN & Echo State Network \\
        \hline
        FD & Federated Distillation \\
        \hline
        FedAvg & Federated Averaging \\
        \hline
        FL & Federated Learning \\
        \hline
        IID & Independent and Identically Distributed \\
        \hline
        IRS & Intelligent Reflective Surface \\
        \hline
        IoT & Internet of Things \\
        \hline
        LTE & Long Term Evolution \\
        \hline
        MIMO & Multiple Input Multiple Output \\
        \hline
        ML & Machine Learning \\
        \hline
        MSE & Mean Square Error \\
        \hline
        OFDMA & Orthogonal Frequency Division Multiple Access \\
        \hline
        PS & Parameter Server \\
        \hline
        RRM & Radio Resource Management \\
        \hline
        SGD & Stochastic Gradient Descent \\
        \hline
        SISO & Single Input Single Output \\
        \hline
        SNR & Signal to Noise Ratio \\
        \hline
        QoE & Quality of Experience \\
        \hline
        UAV & Unmanned Aerial Vehicle \\
        \hline
        VR & Virtual Reality \\
        \hline
        ZF & Zero-Forcing \\
        \hline
    \end{tabular}
    \end{center}
\end{table}


\chapter{Primer on distributed machine learning}
\label{sec:distributedml}
In conventional \ac{ML}, model training is considered to take place in centralized settings, where the processing capability and training datasets are locally available within one computational device. Therefore, \ac{CML} models and algorithms require that all training data must be transmitted from the user devices to the central server. While possible, such an approach has two major practical problems. Firstly, this approach relies on a complete sacrifice of privacy since the all user devices must be willing to reveal their entire datasets to the server. In many cases, this lack of privacy renders training impossible, since the users may not be willing to share their data, it would be considered immoral to collect the data, or the privacy of the users is legally protected. Secondly, the size of training datasets is an important factor in determining the performance of \ac{ML} models, where larger datasets generally generate better results \cite{sun2017revisiting}. This naturally leads to a desire of training with massive datasets, which is very challenging to communicate over a wireless network \cite{huang2018communication}. Recently, \ac{DML} has been proposed as a means to overcome these challenges.
Differently from \ac{CML}, \ac{DML} works over a dataset distributed among many devices, and optionally performs even distributed training.

In \ac{DML} methods, the training can be distributed entirely across the devices, which represents the decentralized architecture; or it can be done jointly by a central \ac{PS} and the devices, which represents the centralized architecture.
In this survey, we focus on the centralized architecture within \ac{DML} because it provides strong guarantees in terms of communication bandwidth usage, latency, parameter update frequency, and desired fault tolerance~\cite{bennun2018survey}.
Figure~\ref{fig:problemdesc} shows the centralized architecture, in which the $K$ devices communicate only with the 
\ac{PS}, which usually has higher computational power than the other devices and is not necessarily represented by a single 
server (see ~\cite[Section 7]{bennun2018survey} for other \ac{PS} infrastructures). 
Notice that the centralized architecture with \ac{PS} is similar to the operation of current cellular networks, Wi-Fi, 
and \ac{IoT} networks with a central controller that could be an app, router, or an \ac{IoT} device.
In \ac{DML}, the training goal is global, i.e, all the participating devices have a common goal. 

The purpose of this section is to introduce the basic concepts in \ac{DML}, which we will use and will refer to often in the rest of the survey, especially for what concerns the mathematical concepts of \ac{ML} and their relation to wireless communication protocols. 
In the following, we discuss the learning goal of \ac{CML} methods before specifically explaining the learning goal of \ac{DML} methods, and then we introduce \ac{FL} methods.

\section{Problem formulation for centralized machine learning}\label{sub:prob_form_cml}
We discuss herein the general \ac{CML} problem of supervised learning, i.e., the problem of labeling unseen 
data based on information from a set of labeled training data~\cite{bottou2018}.
The common learning goal is to represent a prediction function $h:\calX\rightarrow \calY$ from an input space $\calX$ 
to an output space $\calY$ such that, given $\vtX\in\calX$, the value $h(\vtX)$ offers an accurate prediction about the 
true output $y\in\calY$.
Hence, the prediction function $h$ should minimize a risk measure over an adequately selected family of prediction 
functions, termed $\calH$.
Instead of optimizing over a generic family of prediction functions, it is commonly assumed that the prediction 
function $h$ has a fixed form and is parameterized by a real vector $\vtW\in\bbR^d$ with dimension $d$.

Then, for some $h(\cdot;\cdot):\bbR^{d_x}\times\bbR^d\rightarrow\bbR^{d_y}$, the family of prediction functions is 
$\calH\triangleq \{h(\cdot;\vtW): \vtW\in\bbR^d\}$, where $d_x$ and $d_y$ are the dimensions of $\vtX$ and $\vtY$, respectively.

To meet the common learning goal, it is necessary to obtain the prediction function in the family $\calH$ that minimizes
the losses due to inaccurate predictions.
To this end, we assume a loss function $l:\bbR^{d_y}\times\bbR^{d_y}\rightarrow\bbR$ that given an 
input-output pair $(\vtX,\vtY)$, yields the loss $l(h(\vtX;\vtW),\vtY)$~\cite{bottou2018}.
Notice that $h(\vtX;\vtW)$ and $\vtY$ represent the predicted and true outputs, respectively.
The model parameter $\vtW$ is chosen such that the expected loss incurred from any input-output pair is minimized.
The loss functions $l(\cdot;\vtW)$ can be either convex on $\vtW$, such as when used for linear regression or binary 
classification (linear \ac{SVM}), or nonconvex, such as when used for image classification using neural networks with 
several layers.
Let us assume that the losses are measured with respect to a probability distribution $\Pr(\vtX,\vtY)$ in the 
input-output space $\bbR^{d_x}\times\bbR^{d_y}$, i.e., $\Pr:\bbR^{d_x}\times\bbR^{d_y}\rightarrow[0,1]$.
Then, the objective function we want to minimize is
\begin{equation}\label{eq:expected_risk}
R(\vtW) = \int_{\bbR^{d_x}\times\bbR^{d_y}} l(h(\vtX;\vtW),\vtY) d \Pr(\vtX,\vtY) = \bbE[l(h(\vtX;\vtW),\vtY)],
\end{equation}
in which $R: \bbR^d\rightarrow \bbR$ is the expected risk given a parameter vector $\vtW$ with respect to the 
probability distribution $\Pr(\vtX,\vtY)$.
The minimum expected risk, denoted by $R(\vtW^\star)$ with $\vtW^\star \coloneqq \ArgMin{\vtW}{R(\vtW)}$, is also known as the \emph{test} or \emph{generalization error}.
Therefore, the common learning goal in \ac{ML} can be understood as the minimization of the \emph{test error}~\cite{hastie2009}.

To minimize the expected risk in Eq.~\eqref{eq:expected_risk}, it is necessary to have complete information about the 
probability distribution $\Pr(\vtX,\vtY)$ of the input-output pair.
However, such minimization is not possible in most situations because complete information of $\Pr(\vtX,\vtY)$ is not available.
Therefore, the practical learning goal becomes the minimization of an estimation of the expected risk $R$.
To this end, we assume that there are $N\in\bbN$ independently drawn input-output data samples $\calD = \{(\vtX_i,\vtY_i)\}_{i=1}^N 
\subseteq \bbR^{d_x}\times\bbR^{d_y}$, and we define the empirical risk function $R_N:\bbR^d\rightarrow\bbR$ as
\begin{equation}\label{eq:empirical_risk}
R_N(\vtW) = \frac{1}{N} \sum_{i=1}^{N} l( h(\vtX_i;\vtW),\vtY_i ).
\end{equation}
With the empirical risk, the optimization problem is as follows:
\begin{align}\label{eq:problem_emp_risk}
\underset{\vtW}{\text{minimize}}\quad
	& \frac{1}{N} \sum_{i=1}^{N} l( h(\vtX_i;\vtW),\vtY_i ), 
\end{align}
in which the minimization of $R_N$ is the practical optimization problem that needs to be solved when performing supervised learning.
The minimum empirical risk is also known as the \emph{training error} and can be understood as an estimation of the \emph{test error}~\cite{hastie2009}.

To solve optimization problem~\eqref{eq:problem_emp_risk}, several optimization algorithms have been proposed using 
stochastic optimization methods, such as \ac{SGD}, with or without the use of data partition in 
batches~\cite{bottou2018}.
A general \ac{SGD} method solves iteratively optimization problem~\eqref{eq:problem_emp_risk}, with iterations given by
\begin{equation}\label{eq:general_sgd}
\vtW^{t+1} \gets \vtW^t -\beta \sum_{i \in {\cal{S}}^t} \nabla f_{i}(\vtW^t), \forall t\in\bbN
\end{equation}
where $\vtW^t\in\bbR^d$, $\beta$ is the learning rate, $f_i(\vtW)$ is the composition of the loss function $l$ 
and $h$ evaluated at sample $i$, and $\calS^t$ is a set with cardinality $N^t$. 
The sum in~\eqref{eq:general_sgd} depends on the set $\calS^t$ and may represent pure \ac{SGD}, batch gradient descent, or a joint approach with minibatch \ac{SGD}~\cite{bottou2018}.
For $N^t=1$, Eq.~\eqref{eq:general_sgd} 
represents the pure \ac{SGD} method, and the unique element of the set $\calS^t$ corresponds to the seed $\xi^{t}$ of the sample pair $(\vtX^{t},\vtY^{t})$, which is chosen randomly from $\{1,\ldots,N\}$.
For $N^t=N$, Eq.~\eqref{eq:general_sgd} represents the batch gradient descent method, in which the gradient is 
evaluated for all samples $N$ and taken into account at each iteration $t$.
For $1 < N^t < N$, Eq.~\eqref{eq:general_sgd} represents the minibatch \ac{SGD} method, in which $N^t$ is termed batch size and all $N^t$ elements of $\calS^t$ are chosen randomly at each iteration $t$. 
The iterations are evaluated until it reaches a minimizer of the empirical risk $R_N$.

In practice, the training error is evaluated by solving optimization problem~\eqref{eq:problem_emp_risk} with $N$ samples; whereas the test error is evaluated by comparing the prediction function $h(\vtX_j;\vtW)$ using unseen input $\vtX_j\in\calX$ to predict unseen output $\vtY_j\in\calY$.
Specifically to classification problems, the classification accuracy is the ratio between the number of correct predictions and the number of incorrect predictions given by the learning model.
Throughout the survey, the learning performance of \ac{ML} algorithms is related to the training and test errors. 
Specifically to classification problems, we refer to the performance as classification accuracy.

\section{Problem formulation for distributed machine learning}\label{sub:prob_form_dml}
\begin{figure}[t]
\centering
    \begin{tikzpicture}
        \node[inner sep=0pt] (russell) at (0,0)
            {\includegraphics[width=8.5cm]{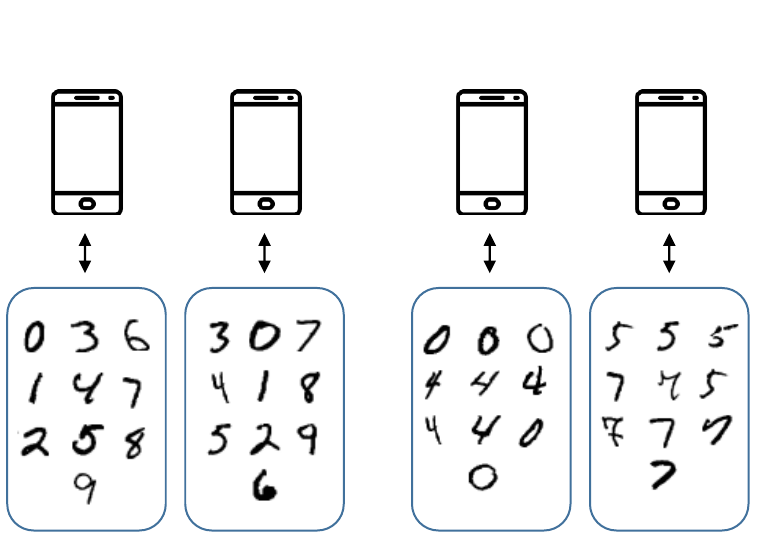}};
        \node[] at (-2.3,2.4) {IID dataset};
        \node[] at (2.2,2.4) {Non-IID dataset};
    \end{tikzpicture}
    \caption{Illustration of \ac{IID} vs. non-\ac{IID} for the MNIST dataset. Non-\ac{IID} data distribution is common when data is generated by user devices, making model convergence of \ac{DML} harder.}
    \label{fig:non-iid}
\end{figure}

Differently from traditional \ac{ML} methods, in \ac{DML} the $N$ samples are either split or generated at different $K$ devices. 
For simplicity, we assume throughout the survey that the samples are generated at $K$ devices.
Let us denote by $\calD_k$ the dataset owned by device $k$ and 
$N_k=|\calD_k|$ as the cardinality of $\calD_k$. 
As a consequence of generating data in a distributed fashion, local data distributions at each device can be skewed in comparison to the global dataset. Consider the classic scenario of digit recognition. In the global MNIST dataset, we have 10\% representation of each digit 0-9~\cite{lecun1998gradient}, which is an \ac{IID} number of digits. If the digits distribution of the local datasets does not match the global one, the distribution is non-\ac{IID}, see Figure~\ref{fig:non-iid}. 

With the splitting of the data across the devices, the empirical risk function can be rewritten as
\begin{equation}\label{eq:dist_emp_risk}
f(\vtW) = \sum_{k=1}^{K} \frac{N_k}{N} F_k(\vtW) = \sum_{k=1}^{K} \frac{N_k}{N} \sum_{i\in\calD_k} f_i(\vtW).
\end{equation}
When the dataset owned by the $K$ devices are \ac{IID}, then $\bbE_{\calD_k}[F_k(\vtW)] = f(\vtW)$, where the 
expectation $\bbE_{\calD_k}[\cdot]$ is taken over the dataset of device $k$.
If the dataset owned by the $K$ devices are non-\ac{IID}, the loss function $F_k(\cdot)$ at device $k$ could be an 
arbitrarily bad approximation of the function $f(\cdot)$~\cite{goodfellow2016book}, thus harming the convergence.

Similar to the traditional \ac{ML} methods, \ac{DML} methods use many optimization techniques to minimize the empirical risk 
in Eq.~\eqref{eq:dist_emp_risk}, such as \ac{DSGD}~\cite{zinkevich2010parallelized}, consensus 
optimization~\cite{nedic2018survey}, and the \ac{ADMM}~\cite{boyd2011distributed}.
For both data distributions, the centralized \ac{DML} architecture needs to exchange information about the parameters between the $K$ devices and the \ac{PS}.
Depending on the optimization technique used, this information, commonly referred to as just \emph{model}, can be the parameter $\vtW$, the gradient $\nabla F_k (\vtW)$, the gradient update $\nabla F_k (\vtW^t) - \nabla F_k (\vtW^{t-1})$, or the parameter update $\vtW^t - \vtW^{t-1}$. 
In this survey, we will use model to refer specifically to the parameter variable $\vtW$, which can be local for each device, $\vtW_k$, or global, $\vtW$.

To improve the applicability of \ac{DML} methods, there are still many challenges for both \ac{DML} architectures and different optimization solvers.
Some of these challenges are the communication efficiency, the systems and statistical heterogeneity, and the privacy 
loss~\cite{tianli2020,konecny2016}. 
The communication efficiency is related to the massive number of messages that need to be exchanged between the PS and a large 
number of devices, which may cause high latency and increase the convergence time.
The systems heterogeneity is related to the different storage, computing, and communication of each device; whereas the statistical heterogeneity is related to the different distribution of the data each device may 
have, which makes the sample distribution among the devices non-\ac{IID}. 
The privacy loss happens when the devices have sensitive data that they do not wish to expose to other devices and/or the \ac{PS}.

Some algorithms to tackle the challenges above have been proposed~\cite{jaggi2014communication,makonecny2017,konecny2016}, including the the \ac{CoCoA} and \ac{CoCoA}$+$ algorithms~\cite{jaggi2014communication,makonecny2017} that address challenges on communication efficiency.
One of these algorithms is \ac{FL}, which has been proposed as a solution aimed at solving all the challenges mentioned and thus 
differing from the \ac{CoCoA} and \ac{CoCoA}$+$ algorithms.

\section{Federated learning}\label{sub:federated_learning}
In \ac{FL} methods~\cite{konecny2016}, a common global model is trained in a distributed manner using the \ac{PS} within the centralized architecture of \ac{DML}.
The common scenario in \ac{FL} is the one in which the number of participating devices is typically large and have slow or unstable connections; the devices do not want to share their raw data with the \ac{PS} or other devices; and there is heterogeneity in 
the data across the devices and in the computation capabilities.
Note that this scenario implies that \ac{FL} methods must address challenges in terms of communication efficiency, privacy, and systems/statistical heterogeneity, which are the challenges common to \ac{DML} methods mentioned in Section~\ref{sub:prob_form_dml}.

\begin{figure*}[t]
	\centering
	\includegraphics[width=0.97\textwidth]{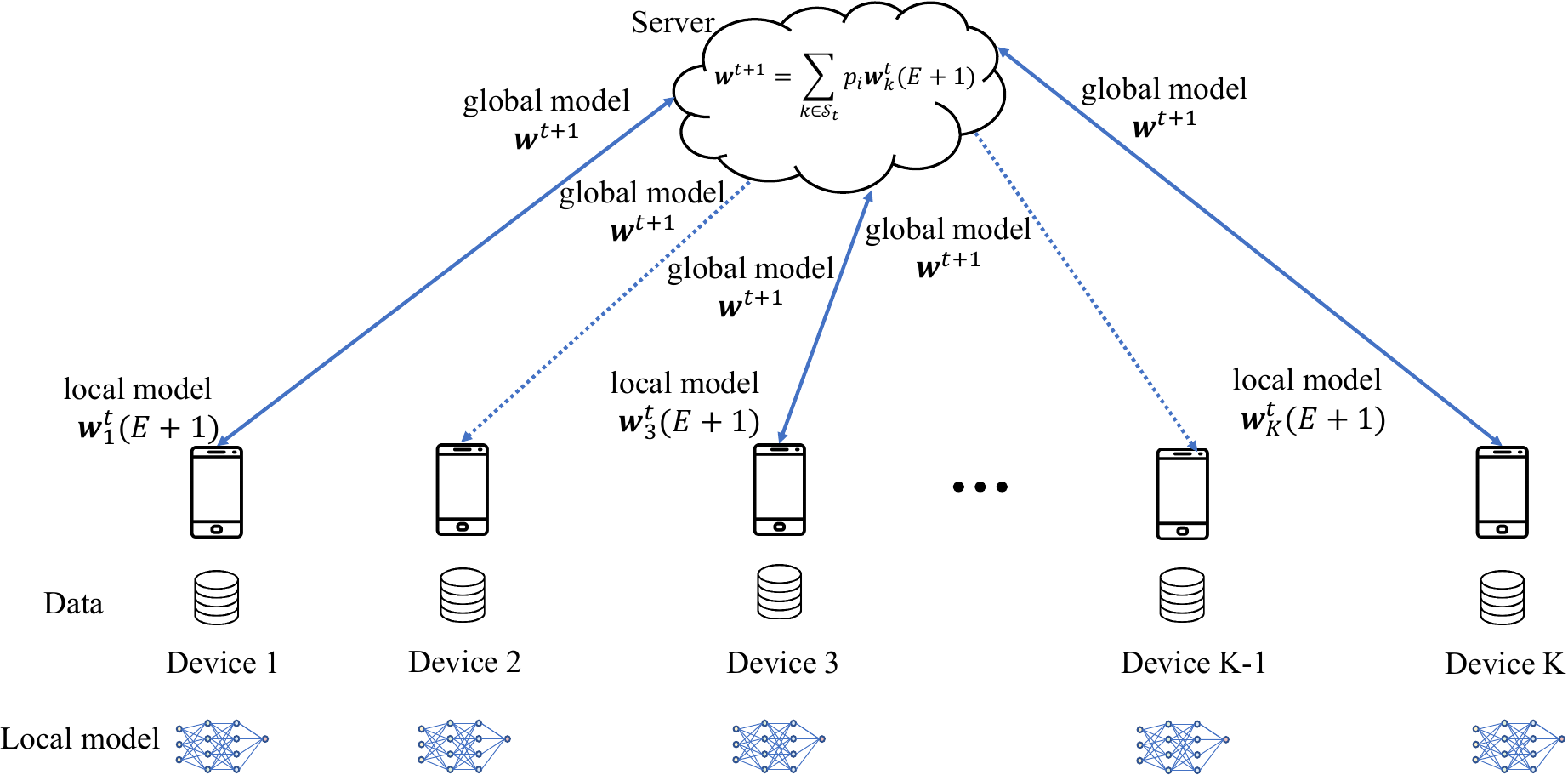}
	\caption{Federated learning scenario with $K$ devices and a \ac{PS}. Only the federated devices, the ones that belong to set $\calS_t$, participate in the learning at communication round $t$ and not all $K$ devices, represented by the solid and dotted lines, respectively. } \label{fig:fed_learning}
\end{figure*}

Figure~\ref{fig:fed_learning} shows a \ac{FL} scenario in which $K$ devices and the \ac{PS} use \ac{FL} towards a common global goal, which is to minimize the empirical risk.
Notice that only the federated devices that belong to the set $\calS^t$ participate in the learning at communication round $t$ and not all the $K$ devices, represented by the solid and dotted lines in Figure~\ref{fig:fed_learning}, respectively.
The raw data is kept locally at each device, and the devices participating in the training minimize their local functions $F_k$, which means having \ac{SGD} updates similar to Eq.~\eqref{eq:general_sgd} for $E$ local iterations.
Then, the devices send to the \ac{PS} their local model $\vtW_k^t(E+1)$ that minimizes the local functions at communication round $t$.
The \ac{PS} aggregates the local models with proper scaling $p_k$ and broadcasts the global model $\vtW^{t+1}$ to all participating devices.
Therefore, \ac{FL} improves the communication efficiency by avoiding many communication rounds with the \ac{PS} due to the transmission of the updated model only after the local iterations; takes into account the devices heterogeneity by the possibility of different number of local iterations at the devices as well as the selection of devices to participate in the training; and finally, it improves privacy by not sending the raw data. 

The first \ac{FL} method proposed was \ac{FedAvg}~\cite[Algorithm 1]{mcmahan2017communication}.
With \ac{FedAvg}, the \ac{PS} randomly selects a fraction $C$, $0 < C \le 1$, of the $K$ devices to participate in the training at 
global iteration $t$, i.e., the set $\calS^t$ has cardinality $\max(\Ceil{CK},1)$.
Each device $k\in\calS^t$ minimizes the local function $F_k$ by computing the gradient $\nabla F_k (\vtW_k^t)$ and iterating $E$ local iterations (named epochs) applying the updates as
\begin{equation}\label{eq:local_fl_iter}
\vtW_{k}^t (i+1) \gets \vtW_{k}^t (i) -\beta \nabla F_k (\vtW_{k}^t (i)), \forall i = 1,\ldots, E.
\end{equation}
Notice that the gradient $\nabla F_k (\vtW_{k}^t (i))$ can be obtained using \ac{SGD} with different batch sizes.
After $E$ epochs, device $k$ sends $\vtW_{k}^t(E+1)$ to the \ac{PS}, which aggregates the local models of the participating 
devices at iteration $t$ to generate the updated global model as
\begin{equation}\label{eq:global_fl_iter}
\vtW^{t+1} \gets \sum_{k\in\calS^t} \frac{N_k}{N} \vtW_{k}^t(E+1).
\end{equation}
Then, the \ac{PS} sends the updated global model $\vtW^{t+1}$ to all participating devices, and the iterative process between the devices and the \ac{PS} continues until global convergence is achieved, at which $\vtW^{t+1} = \vtW^{*}$. To measure the rate of convergence, we use
\begin{equation}
    \label{eq:convergence_rate}
    f(\vtW^{t+1})-f(\vtW^{*}),
\end{equation}
which may be a decreasing function in $t$. For \ac{FL}, there is no closed-form expression for this convergence rate, so model performance cannot be predicted before training. However, for certain scenarios there are theoretical guarantees by upper bounding the convergence rate \cite{chen2020joint}.

Since \ac{FedAvg} was proposed in~\cite{mcmahan2017communication}, many other \ac{FL} methods have been proposed and 
investigated for many scenarios, including sparse and/or quantized \ac{FL} \cite{reisizadeh2020fedpaq}, private \ac{FL} using differential privacy \cite{wei2020federated}, 
fair \ac{FL} \cite{li2019fair}, and FL over wireless communications \cite{tran2019federated}.
For an in-depth overview of recent \ac{FL} methods and applications, we refer the reader 
to~\cite{kairouz2019survey,yang2019book,tianli2020,Yang2020,niknam2020federated}.

\section{Summary}

\ac{DML} methods overcome some challenges from traditional \ac{ML} methods, and similarly, \ac{FL} methods overcome 
some challenges from \ac{DML} methods.
Recently, \ac{FL} has been investigated due to its robustness to a massive number of users participation, privacy-enhancing properties, and both statistical and device heterogeneity.
However, there are still several challenges that \ac{DML} and \ac{FL} still need to overcome when applied to Wireless for \ac{ML}.

In the following sections, we will discuss novel wireless protocols that are specifically designed to address the challenges of Wireless for \ac{ML}. Specifically, we will discuss in Section~\ref{sec:ldcomac} the use of analog over-the-air computation and in Section~\ref{sec:orthogonal} the use of digital \ac{RRM} for \ac{ML}.


\chapter{Analog over-the-air computation}\label{sec:ldcomac}

\section{Primer}\label{sec:comac}
A prominent theme in the Wireless for \ac{ML} literature is a method called either \ac{AirComp} or \ac{CoMAC} \cite{nazer2007computation}. We dedicate this subsection to explain the basics of \ac{CoMAC}. 

In wireless communications, significant attention is put into the avoidance of interference. As an example, \ac{OFDMA} splits the wireless spectrum into small blocks of time and frequency and allocates these blocks to different users in the network. Such a system achieves nearly interference-free communication at the cost of significantly reducing the available transmission time and bandwidth for each user. In contrast, \ac{CoMAC} actively promotes interference. Multiple users are allocated the same time and frequency resources, causing their signals to combine in the air. By carefully designing precoding functions at the transmitting devices, the signal superposition property can be leveraged to calculate functions of the transmitted messages over-the-air~\cite{zhang2006hot}. \ac{CoMAC} addresses applications when the receiver does not need the individual messages, but only some function of them, for example their sum or average.

As \ac{CoMAC} does not allocate orthogonal radio resources, it could be mistaken for the recently proposed \ac{NOMA} schemes. However, unlike \ac{CoMAC}, \ac{NOMA} needs to enable the reconstruction of the individual messages, and thus employs successive interference cancellation to eliminate interfering signals. This cancellation is possible only by introducing diversity in either the power or code domain~\cite{saito2013non,islam2019nonorthogonal}.

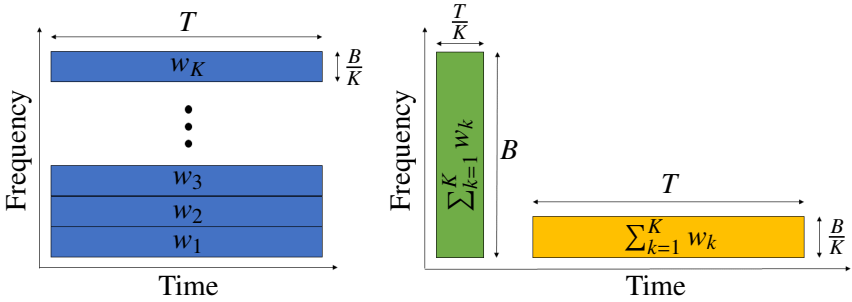
\begin{figure*}[t]
    \begin{tikzpicture}
        \node[inner sep=0pt] (russell) at (0,0)
            {\includegraphics[width=0.95\linewidth]{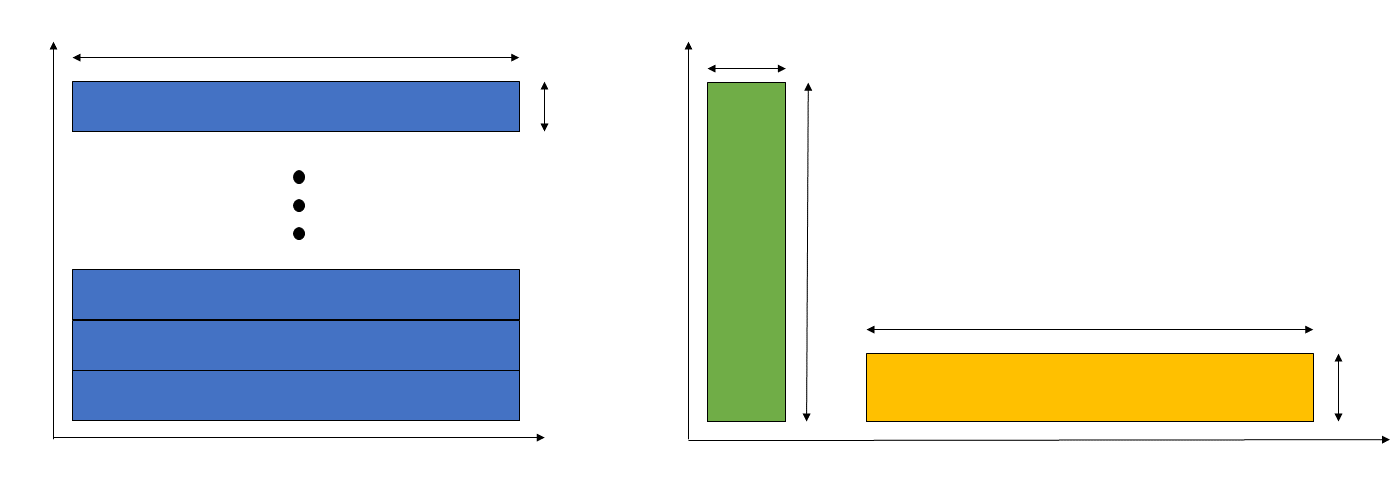}};
        \node[rotate=90] at (-5.45,0) {Frequency};
        \node[] at (-3.2,-1.8) {Time};
        \node[] at (-3.2,-1.25) {$w_1$};
        \node[] at (-3.2,-0.85) {$w_2$};
        \node[] at (-3.2,-0.40) {$w_3$};
        \node[] at (-3.2,1.10) {$w_K$};
        \node[] at (-3.2,1.7) {$T$};
        \node[] at (-1,1.10) {$\frac{B}{K}$};
        \node[rotate=90] at (-0.35,0) {Frequency};
        \node[] at (3,-1.8) {Time};
        \node[rotate=90] at (0.4,-0.2) {$\sum_{k=1}^Kw_k$};
        \node[] at (3.2,-1.15) {$\sum_{k=1}^Kw_k$};
        \node[] at (3.2,-0.45) {$T$};
        \node[] at (1.05,0) {$B$};
        \node[] at (0.4,1.7) {$\frac{T}{K}$};
        \node[] at (5.4,-1.15) {$\frac{B}{K}$};
    \end{tikzpicture}
    \caption{Illustration of communication efficiency for calculating a sum using over-the-air computation. In the left sub-figure, FDMA (blue bars) communicates the $K$ messages over separate frequency bands, and the sum is calculated locally at the parameter server. In contrast, over-the-air computation communicates all $K$ messages jointly by leveraging the interference of simultaneous transmissions. Since over-the-air computation allows complete bandwidth sharing, this results in either $K$ times less latency (green vertical bar) or bandwidth (yellow horizontal bar).}
    \label{fig:comac-bw}
\end{figure*}

Since \ac{CoMAC} does not allocate orthogonal resources or introduce additional diversity, the spectrum efficiency grows linearly with the number of devices in the network~\cite{abari2016over}. Consider the network setup from earlier in Figure~\ref{fig:problemdesc} and that the server wants to calculate a sum of $K$ messages $w_k$ carried by the user devices; over-the-air computation would then require approximately $K$ times less resources to communicate this sum. As illustrated in Figure~\ref{fig:comac-bw}, the protocol designer can choose to crystallize this resource efficiency to reduce latency and/or bandwidth consumption.

\subsection{Sum function example}\label{sec:comacexample}
Herein, we will demonstrate how \ac{CoMAC} calculates a sum function over the air. We follow the system model illustrated in Figure~\ref{fig:problemdesc} where $h_k$ denotes the channel from device $k$ to the \ac{PS}. If all devices transmit simultaneously over the same frequency band, the server will receive a linear combination of these signals due to the additive nature of simultaneously arriving electromagnetic waves. Denote the signal transmitted by device $k$ to be $w_k$. Then, the received signal $r$ at the server is

\begin{equation}
    \label{eq:naivesum}
    r=\sum_{k=1}^Kh_kw_k+v\ ,
\end{equation}
where $v$ is an \ac{AWGN} term. Here we assume that the antenna of the \ac{PS} does not saturate. Because of the fading, the received sum is weighted by different weights for each device, and the server is unable to reconstruct the desired function $\sum_{k=1}^Kw_k$. A possible solution is to let the user devices pre-equalize their channel. So instead of transmitting $w_k$ directly, they transmit $z_k=w_k/h_k$. This way, the server would receive 

\begin{equation}
    \label{eq:preequalizedsum}
    r=\sum_{k=1}^Kh_kz_k+v=\sum_{k=1}^Kw_k+v\ .
\end{equation}
Except for the noise term, this corresponds to the desired function. Considering that the signal strength is the sum of $K$ signals, while the noise $v$ is the same as if a single device transmitted, $r$ is generally a good estimator for the desired sum. While this simple description illustrates the basic idea of \ac{CoMAC}, there are several simplifying assumptions that must be dealt with in practice. We discuss these next.

\emph{1.1) Channel State Information}\\
To pre-equalize the channel in Eq.~\eqref{eq:preequalizedsum}, the user device must know the \ac{CSI} of $h_k$, which cannot be estimated at the device directly. 
If classical channel estimation was employed at device $k$, the estimated value would be $g_k$ in the downlink direction. A naive solution to this problem is to let the server estimate $h_k$ by having the mobile devices transmit individual preamble signals in the uplink direction, and then feedback the \ac{CSI} to the mobile devices. However, the transmission of these preambles would require orthogonal transmission of the uplink signals, negating the benefits of over-the-air computation.

Instead, \cite{abari2016over} presents a solution based on channel reciprocity. The underlying claim is that forward and reverse channels are the same up to a constant multiplier due to differences in hardware between the transmit and receive chains. By introducing a calibration stage in which the up- and downlink channels are measured for each sensor device $k$, this constant multiplier can be found as $c_k = h_k(0)/g_k(0)$, where $h_k(0)$ and $g_k(0)$ are the uplink and downlink channels at time 0, respectively. 
Since the multiplier remains constant, subsequent communication rounds can calculate the uplink channel using the downlink \ac{CSI} measured with the broadcast from the server as $h_k(t)=c_kg_k(t)$. However, this solution was only tested for stationary nodes. For dynamic scenarios, such as cellular and vehicular communications, other solutions have been proposed~\cite{dong2020blind,zhu2018over,jiang2019over}. In the interest of brevity, we refrain from discussing these other methods here, but blind over-the-air computation is discussed in Section \ref{sec:mimo}. Note that this calibration stage is not required in traditional digital communications since the channel can be equalized at the receiver, so the transmitters do not require knowledge of $h_k$.

\emph{1.2) Synchronization}\\
A second problem arises from an inherent assumption in Eq. (\ref{eq:preequalizedsum}), which is that the transmitted signals arrive simultaneously at the server. Even small synchronization errors can lead to major estimation errors because the sum is calculated with an analog signal. Synchronization would be required at a symbol-level, which may be difficult to achieve with traditional synchronization. To overcome such a problem, multiple novel approaches have been proposed, for instance: dedicated hardware that transmits sinusoidal tones~\cite{abari2015airshare}, longer transmission blocks to reduce the synchronization requirement~\cite{goldenbaum2013robust}, or the "timing advance" functionality of LTE networks~\cite{ltetimingadvance}. Additionally, if some devices are far away from the \ac{PS}, there might be a need to estimate the propagation latency and compensate at the transmission.

\emph{1.3) Power control}\\
The pre-equalization scheme from \eqref{eq:preequalizedsum} assumes that the devices have the capability to transmit $z_k = w_k/h_k$. However, if the device is experiencing a deep fade, $h_k$ will be a very small number, thereby requiring a tremendous amount of power for pre-equalization. With practical devices, the peak power is constrained, and such a scheme is unfeasible. To get around this constraint, several researchers have formulated power control problems \cite{cao2020optimized, liu2020over} which introduces a post-transmission scalar $\sqrt{\eta}$. This scalar is applied by the \ac{PS} after receiving the sum, yielding
\begin{equation}
    \label{eq:powercontrolsum}
    r=\sum_{k=1}^K\frac{h_kz_k}{\sqrt{\eta}}+\frac{v}{\sqrt{\eta}}.
\end{equation}
With the newly introduced $\eta$, the amplitude required for pre-equalization of the channel changes to $z_k = \sqrt{\eta}w_k/h_k$. If the post-transmission scalar is selected to be $\eta < 1$, the required transmission power is reduced, enabling more devices to invert their channel. However, a reduction of $\eta$ also leads to an increase in the relative noise power. This tradeoff leads to the power control problem, which aims to optimally select $z_k$ and $\eta$ without exceeding transmission power constraints. We discuss this problem further in Section \ref{sec:power_control}.

\subsection{Summary}
By promoting interference, \ac{CoMAC} allows all devices to share the electromagnetic spectrum without allocating orthogonal radio resources to each user. Such a scheme achieves throughput gains approximately proportional to the number of participating devices, which is a tremendous improvement even with a relatively small number of users. The main drawback of the method is that the individual messages cannot be reconstructed at the receiver, which limits the application to scenarios where a function of the messages is sufficient. In the preceding paragraphs, we gave a simple example that demonstrates how channel pre-equalization \ac{CoMAC} can be used to calculate the sum function. However, this method has several practical issues such as strong demands on \ac{CSI}, stringent synchronization requirements, and limited peak transmission powers at the user devices.

\section{Over-the-air computation for distributed machine learning}
As explained in Section \ref{sub:federated_learning}, the model aggregation step of \ac{FL} consists of transmitting multiple local models from the user devices to the \ac{PS} and then computing a weighted mean of these updates to generate the next iteration of the model, see \eqref{eq:global_fl_iter}. The individual local models are not needed at any point of the \ac{FL} algorithm, only this weighted sum. As such, the sum can be directly computed over-the-air instead of separately transmitting each model vector and then averaging at the \ac{PS}. This basic idea has served as the foundation of a large body of works that explore the impact of \ac{CoMAC} on \ac{DML} and that extend the idea further.

As illustrated in the \ac{CoMAC} example from Section~\ref{sec:comac}, non-uniform fading across the network is a major challenge for estimating the desired function. We presented a power modulation solution based on channel reciprocity and inversion, which is the standard method to overcome this challenge for \ac{SISO} networks. However, for \ac{MIMO} networks, alternative solutions are proposed, such as the blind \ac{CoMAC} which utilizes channel-hardening to avoid the \ac{CSI} acquisition problem. Additionally, the consideration of \ac{MIMO} comes with other interesting \ac{CoMAC}-solutions such as beamforming, cell-free massive \ac{MIMO}, and \ac{IRS}-assisted \ac{CoMAC}. With this in mind, the remainder of this section is split into two parts, \ac{SISO} and \ac{MIMO}. A comprehensive list of papers on \ac{CoMAC} for ML is given in Table~\ref{table:ldma1}, \ref{table:ldma2}, and \ref{table:ldma3}.

\begin{table}[ht]
\caption{Summary of the SISO \ac{AirComp} for \ac{ML} literature. The papers are ordered according to when they are covered in the survey.}
\begin{center}
\begin{tabular}{|l|l|p{17em}|}
    \hline
    \textbf{Topic} & \textbf{Ref.} & \textbf{Summary}\\
    \hline
    \multirow{1}{8em}{Broadband Analog Aggregation} & \cite{zhu2019broadband} & FL using \ac{AirComp} over a broadband channel with truncated channel inversion to handle fading.\\
    \hline
    \multirow{4}{8em}{Gradient Sparsification} & \cite{amiri2020machine} & Sparsification of gradients combined with error accumulation for compression before transmitting.\\
    \cline{2-3}
    & \cite{amiri2019over} & Extension of~\cite{amiri2020machine} to consider fading channels, uses truncated channel inversion.\\
    \cline{2-3}
    & \cite{amiri2020federated} & Performance comparison of~\cite{amiri2019over} scheme, sequential digital transmission, and BAA.\\
    \cline{2-3}
    & \cite{fan2021temporal} & Utilization of temporal structures in the gradient updates to form a Bayesian prior in the gradient estimation step. \\
    \hline
    \multirow{1}{8em}{Federated Distillation} & \cite{ahn2019wireless} & Trains by communicating model outputs instead of model parameters. Over-the-air computation is used to combine model output vectors for each class.\\
    \hline
    \multirow{2}{8em}{Training with Noisy Gradients} & \cite{sery2019sequential} & Proposal of gradient-based multiple-access scheme that does not cancel the fading effect but operates directly with noisy gradients.\\
    \cline{2-3}
    & \cite{sery2020analog} & Convergence rate analysis for gradient-based multiple-access.\\
    \hline
    \multirow{1}{8em}{Data Sharing} & \cite{sun2020energy} & DSGD training using combined gradients. Introduces data redundancy to combat non-IID. data.\\
    \hline
    \multirow{1}{8em}{Analog Federated ADMM} & \cite{elgabli2021harnessing} & Second-order training algorithm with \ac{CoMAC} communication.\\
    \hline
\end{tabular}
\end{center}
\label{table:ldma1}
\end{table}

\begin{table}[ht]
\caption{Continuation of Table \ref{table:ldma1}.}
\begin{center}
\begin{tabular}{|l|l|p{17em}|}
    \hline
    \textbf{Topic} & \textbf{Ref.} & \textbf{Summary}\\
    \hline
    \multirow{2}{8em}{Digital Aggregation} & \cite{zhu2020one} & First digital over-the-air computation method using one-bit quantization of gradients.\\
    \cline{2-3}
    & \cite{jiang2020cluster} & Clustered digital over-the-air computation that minimizes the probability of incorrect gradient sign estimation.\\
    \hline
    \multirow{2}{8em}{Power Control} & \cite{cao2021optimizedfl} & Optimal selection of pre- and post-processing scalars using FL bounds.\\
    \cline{2-3}
    & \cite{zhang2021gradient} &  Estimation of gradient statistics to improve power control for Federated Learning.\\
    \hline
    \multirow{2}{8em}{Retransmissions} & \cite{hellstrom2021retransmit} & Proposal of retransmission-based model update scheme that enables an estimation-communication tradeoff.\\
    \cline{2-3}
    & \cite{hellstrom2021retransmit2} & Development of heuristic to predict the optimal number of retransmissions.\\
    \hline
    \multirow{1}{8em}{Differential Privacy} & \cite{LiuOsvaldoDPFLJSAC2021} & Uses the noise added naturally by the wireless channel to enhance data privacy for free.\\
    \hline
    \multirow{1}{8em}{Byzantine Attacks} & \cite{sifaou2021robust} & Considers the grouping of participating devices to mitigate Byzantine attacks.\\
    \hline
    \multirow{2}{8em}{Device-to-Device Communication} & \cite{xing2020decentralized} & First decentralized machine learning scheme using over-the-air computation.\\
    \cline{2-3}
    & \cite{shi2021over} & Decentralized \ac{SGD} with gradient tracking and variance reduction. \\
    \hline
    \multirow{1}{8em}{Bayesian Learning} & \cite{9562546} & Proposes the channel-driven Monte-Carlo sampling method that leverages channel noise to estimate the posterior distribution of ML parameters.\\
    \hline
\end{tabular}
\end{center}
\label{table:ldma2}
\end{table}

\begin{table}[ht]
    \caption{Summary of the MIMO \ac{AirComp} for \ac{ML} literature. The papers are ordered according to when they are covered in the survey.}
\begin{center}
\begin{tabular}{|l|l|p{17em}|}
    \hline
    \textbf{Topic} & \textbf{Ref.} & \textbf{Summary}\\
    \hline
    \multirow{2}{8em}{Blind Learning} & \cite{amiri2019collaborative} & The assumption of channel knowledge at the user devices is lifted. Instead, multiple antennas at the \ac{PS} is employed to alleviate the fading effect.\\
    \cline{2-3}
    & \cite{mohammadTolgaBlindFLTWC2021} & Extension of \cite{amiri2019collaborative} to consider imperfect channel estimation at the \ac{PS}.\\
    \hline
    \multirow{1}{8em}{Nonlinear Estimator} & \cite{YoSebMohammadMIMOFL} & Recovering the average of local models sent from the devices using their sparsity with a nonlinear estimator.\\
    \hline
    \multirow{1}{8em}{Cell-Free Massive MIMO} & \cite{VuNgoCellFreeFLMIMOTWC2020} & \ac{FL} in a cell-free massive MIMO framework with \ac{CSI} estimation using \ac{CoMAC} pilot transmission.\\
    \hline
    \multirow{1}{8em}{Beamforming and User Selection Co-Design} & \cite{YangJiangFLMIMOTWC2020} & Optimal user scheduling based on tradeoff between maximizing participation and limiting distortion from aggregation error.\\
    \hline
    \multirow{4}{8em}{Intelligent Reflective Surfaces} & \cite{wang2021federated} & Optimized beamforming, user selection, and phase-shift control via intelligent reflective surfaces (IRSs) to maximize device participation.\\
    \cline{2-3}
    & \cite{liu2021reconfigurable} & Optimization over upper bound on FL loss to find proper phase-shift control, device selection, and beamforming for IRS FL.\\
    \cline{2-3}
    & \cite{liu2021csit} & Channel state information free transmission via IRS.\\
    \cline{2-3}
    & \cite{9414785} & Energy minimization with IRS-assisted over-the-air computation.\\
    \hline
\end{tabular}
\end{center}
\label{table:ldma3}
\end{table}

\section{Review of SISO over-the-air computation}
\subsection{Broadband analog aggregation}\label{sec:baa}
The first paper to suggest \ac{CoMAC} as multiple access for \ac{FL} appears to be~\cite{zhu2018towards}. This paper presents a short case study that compares the latency of orthogonal transmission with \ac{CoMAC} under identical conditions. The case study displays a significant reduction in latency, ranging from one to three orders of magnitude, with minor sacrifices in terms of classification accuracy. Later on, the same group presented a fully-fledged scheme called \ac{BAA} in~\cite{zhu2019broadband}. Similar to \ac{LTE}, the \ac{BAA} scheme divides the spectrum into \acp{RB}. However, instead of dedicating each \ac{RB} to a single user, the blocks are dedicated to one element of the model update vector. This way, all $K$ users can transmit their model updates simultaneously over the same \acp{RB} to calculate the weighted sum of model updates from \eqref{eq:global_fl_iter} over-the-air.

As we explained in Section \ref{sec:comacexample}, channel pre-equalization is used to generate the sum function \eqref{eq:preequalizedsum}. As a consequence of this scheme, the receive SNR is identical for every user, because devices with weaker channels compensate by transmitting at higher powers. In \ac{BAA}, devices with sufficiently weak channels are excluded from training, since they are unable to pre-equalize their channels. With this in mind, we consider the inclusion of a post-transmission scalar $\sqrt{\eta}$ as in \eqref{eq:powercontrolsum}. If $\sqrt{\eta}$ is reduced, more devices are able to invert their channels, which increases device participation. In the context of \ac{FL}, higher participation means a larger training dataset. As such, the reduction of $\sqrt{\eta}$ increases data quantity. However, the receive \ac{SNR} is:
\begin{equation}
    \text{SNR} = \eta\frac{\left(\sum_{k=1}^Kw_k\right)^2}{\sigma_z^2},
\end{equation}
which is proportional to $\eta$. Therefore we have a tradeoff between data quantity and receive SNR. In \cite{zhu2019broadband}, they isolate this tradeoff and coin the term communication (SNR)-learning (data quantity) tradeoff. This tradeoff appears in many \ac{CoMAC}-\ac{FL} systems and is important to consider when optimizing such systems.

\subsection{Gradient sparsification}\label{sec:gradsparse}
Although the \ac{BAA} scheme significantly reduces the communication load for \ac{FL}, it does not consider improvements in terms of the \ac{ML} algorithm. In contrast, the next paper we discuss utilizes gradient sparsification together with \ac{CoMAC} to further reduce the communication cost. Gradient sparsification is based on the observation that up to 99.9\% of the gradient exchange in \ac{DSGD} is nearly redundant \cite{lin2017deep}. Therefore, a majority of the gradients can be discarded with minimal reductions to learning accuracy.

In \cite{amiri2020machine}, the combination of gradient sparsification and \ac{CoMAC} appeared for the first time. In this paper, a simple channel model without fading was considered. In \cite{amiri2019over}, the same scheme was extended to consider fading channels, where truncated channel pre-equalization was used to generate the sum. Finally in~\cite{amiri2020federated}, an experimental comparison of three different \ac{FL} approaches (orthogonal transmission, \ac{BAA}, and gradient sparsification with \ac{CoMAC}) is conducted. The study is on training an MNIST classifier, it assumes a limited transmission budget in terms of time slots, and compares the final test accuracy after the transmission budget is out. The results reveal that both \ac{CoMAC} approaches outperform orthogonal communication with up to 40\% better classification accuracy. The study also indicates that the inclusion of gradient sparsification has substantial benefits, with up to 10\% classification accuracy over \ac{BAA}.

\begin{figure}[ht]
\centering
    \begin{tikzpicture}
        \node[inner sep=0pt] (russell) at (0,0)
            {\includegraphics[width=8.5cm]{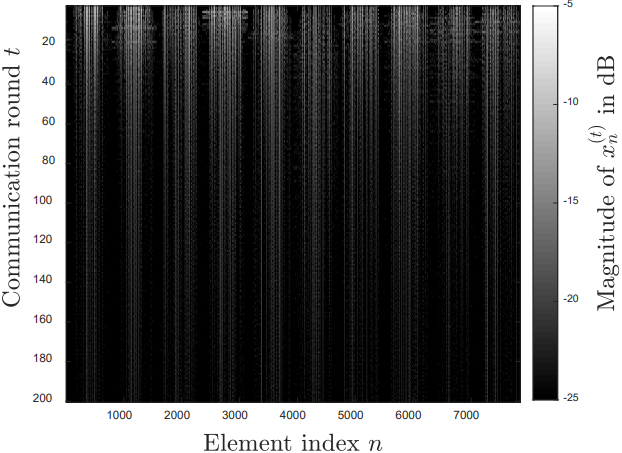}};
    \end{tikzpicture}
    \caption{Illustration of the temporal structure of gradient updates in over-the-air FL. The amplitudes of the gradient elements are encoded in grayscale over time. We can see that the sparsity of the gradient is roughly retained through time, which can be exploited to improve the estimation of the local models. Source: \cite{fan2021temporal}} 
    \label{fig:gradient_sparsity}
\end{figure}

In a more recent work \cite{fan2021temporal}, the authors noticed a predictable structure in the aggregated gradients. From Figure \ref{fig:gradient_sparsity}, we can see that the amplitude of the different gradient elements changes slowly over time, more or less retaining the sparsity structure through the entire training process. To model this structure, \cite{fan2021temporal} uses two independent Markov chains for the support and amplitude. By combining this simple model and the stored gradients from previous communication rounds, a prior belief on the gradient can be formed. As explained in Section \ref{sec:comac}, over-the-air computation always results in noise, therefore the \ac{PS} must estimate the gradient after receiving the uplink signal. 
If there is no prior information, the best estimate is to just directly use the received signal. Instead, this paper uses Bayesian estimation with the prior belief from the Markov chain model to make a better estimation. In the numerical comparisons of \cite{fan2021temporal}, this approach strictly outperforms the results from \cite{amiri2020federated}.

\subsection{Federated distillation}\label{subsec:aircomp_distillation}
As explained in Section~\ref{sec:distributedml}, \ac{FL} achieves consensus by sharing locally trained models with the \ac{PS}. These local models can become enormous when considering deep neural networks with millions of neurons, such as the VGG models that consist of $d$=130-140 model parameters~\cite{simonyan2014very}. With this in mind, there have been attempts to develop an alternative to \ac{FL} called \ac{FD}. In \ac{FD}, model outputs are communicated instead of the model parameters~\cite{jeong2018communication}. In other words $\vtW_k^t$ from Eq.~(\ref{eq:global_fl_iter}) is replaced with the average of the local model outputs, thus communicating an $\bbR^{d_y}$ vector instead of an $\bbR^{d}$ vector. Often classification problems have $d_y\leq100$ labels but millions of parameters $d$, causing a reduction in the number of transmitted bits by many orders of magnitude. Upon receiving these model outputs, the server calculates their average and communicates it back in the downlink. These average model outputs can then be used by the devices to train their \ac{ML} models. As we wish to focus on the communication protocol, we refrain from explaining how these model outputs are used for training and refer the interested reader to~\cite{jeong2018communication}.

In~\cite{ahn2019wireless} \ac{FD} is combined with \ac{CoMAC}. First, each device combines the model outputs over multiple training samples, generating one value for each label. Then, for each label, a global average is calculated over-the-air. This can lead to massive reductions in communication cost but unlike the gradient sparsification schemes, \ac{FD} does have a noticeable drop in classification accuracy. The numerical study conducted in~\cite{ahn2019wireless} suggests it can be between 1-20\% lower than \ac{FL}.

\subsection{Training with noisy gradients}
Unlike all papers we have surveyed so far, which used channel inversion to combat fading, see Eq.~(\ref{eq:preequalizedsum}), the authors of~\cite{sery2019sequential} suggest just transmitting without doing any precoding. Such a scheme has the advantage of not requiring a channel estimate, and a generally simpler implementation. However, since fading is not inverted, the received local models at the \ac{PS} represents a noisy and distorted version of the transmitted local models. This distorted average is then used to perform the \ac{FL} update directly. An important contribution of this paper is an upper bound on the \ac{FL} loss, arguably the first bound that considers \ac{AirComp}. An extended convergence and numerical analysis is given in~\cite{sery2020analog} containing simulation results based on the Million Song Dataset~\cite{bertin2011million}. The results reveal comparable or slightly worse \ac{FL} loss compared to a digital scheme but with significantly reduced energy consumption.

\subsection{Data sharing}
In Section \ref{sub:prob_form_cml}, we explained that there is an important distinction to make between \ac{IID} and non-\ac{IID} training data distribution over the devices. With non-\ac{IID} data, there is no guarantee that the locally trained models resemble the global models, which can significantly harm \ac{FL} performance. In extreme examples, non-\ac{IID} data can harm the classification accuracy by up to 55\% \cite{zhao2018federated}.

Realistically, we should always expect \ac{FL} data distributions to be non-\ac{IID}. For instance, an environmental monitoring device will have a different distribution depending on sensor location, text prediction algorithms depend on user behavior, and body sensor systems depend on the physiology of the host. To combat this,~\cite{sun2020energy} introduces a data sharing phase into \ac{CoMAC} for ML, where each user device shares its dataset with a small number of neighbors before training begins. The study considers the same communication scheme as in \ac{BAA}, but performs data sharing before training begins. Their numerical study on the MNIST dataset considers highly non-\ac{IID} data distributions where each device only carries samples of one digit. They show that classification accuracy goes from 72\% to 82\% by having each user device share its dataset with just one neighbor.

\subsection{Analog federated ADMM}
When over-the-air computation is used to calculate a sum, channel pre-equalization is employed to counteract heterogeneous fading over the network, as explained in Section \ref{sec:comacexample}. Given that all devices perform pre-equalization, the over-the-air computation result in the desired function in expectation. However, some devices are unable to pre-equalize their channel due to limited transmission power. To solve this problem, all papers surveyed up to this point simply exclude those devices from participating. Instead, \cite{elgabli2021harnessing} proposes the first over-the-air computation algorithm that overcomes channel perturbations without pre-equalization, the method is based on a novel \ac{FL} framework rooted in \ac{ADMM}, which they call \ac{A-FADMM}.

For the sake of inclusion, we will not assume that the reader is familiar with \ac{ADMM} and avoid mentioning specifics of the \ac{ADMM} algorithm in this subsection. Instead, we focus on the model update which is communicated by \ac{A-FADMM}, because it differs significantly from what we see in Section \ref{sec:comacexample} and has the interesting property of avoiding channel pre-equalization. For the reader that wants a deeper look into \ac{ADMM}, we refer to \cite{boyd2011distributed}.

We directly state the equation for the update of the global model in \ac{A-FADMM}:
\begin{equation}\label{eq:global_fadmm_iter}
    \vtW^{t+1} \gets \frac{1}{\sum_{k\in{\cal S}^t}|h_{k}|^2}\sum_{k\in\calS^t} \frac{N_k}{N}\left( |h_{k}|^2\vtW_{k}^t(E+1) + h_{k}\vtLambda_k^{t}(E+1)/\rho\right).
\end{equation}
Compared to the standard \ac{FL} update in \eqref{eq:global_fl_iter} there are two major differences. Firstly, the channels $h_k$ have been directly incorporated into the \ac{FL} problem formulation, and secondly there are now two new variables $\vtLambda_k^{t}(E+1)$ and $\rho$ which represents the dual variable of the \ac{ADMM} algorithm and a penalty variable respectively. The semantics of these variables can be ignored for the sake of this discussion. Notice that the channel $h_k$ is a factor both for the local model $\vtW_{k}^t(E+1)$ and the dual variable $\vtLambda_k^{t}(E+1)$. This means that the user devices can transmit
\begin{equation}
    N_k(h_k^*\vtW_{k}^t(E+1)+\vtLambda_k^{t}(E+1)/\rho),
\end{equation}
where $h_k^*$ is the conjugate of $h_k$. Then, using over-the-air computation, the \ac{PS} receives
\begin{equation}
    \sum_{k\in\calS^t}N_k\left( |h_{k}|^2\vtW_{k}^t(E+1) + h_{k}\vtLambda_k^{t}(E+1)/\rho\right) + v.
\end{equation}
If this expression is multiplied by $1/(N\sum_{k\in\calS^t}|h_{k}|^2)$ it generates the desired function from \eqref{eq:global_fadmm_iter} in expectation. Therefore, channel pre-equalization is not required and \ac{A-FADMM} has the advantage of avoiding device exclusion completely. This directly increases the training data quantity, which should improve learning performance. On the other hand, one could argue that the multiplication of $h_k$ leads to weak transmission signals, thereby potentially reducing the \ac{SNR} compared to channel pre-equalization.

In addition to proposing \ac{A-FADMM}, the authors of \cite{elgabli2021harnessing} prove that the algorithm converge for convex functions under time-varying channels. The convergence rate is also evaluated numerically by training with the MNIST dataset. The results suggest that \ac{A-FADMM} converges faster than both traditional \ac{FL} with over-the-air computation as well as digital \ac{ADMM} without over-the-air computation.

\subsection{Digital aggregation}\label{sec:digital_comac}
Current telecommunications infrastructure is almost exclusively designed for digital communications. Because of this, the implementation of analog \ac{CoMAC} in large scale networks becomes problematic. To avoid constructing new analog chipsets at large scale, \cite{zhu2020one} proposes an adaptation to over-the-air aggregation which would be compatible with current transceivers. The proposed protocol is based on 1-bit SGD~\cite{seide20141} which uses single-bit compression of gradient descent updates. Specifically, each element of the user devices' gradient vectors $\nabla F_k(\vtW^t)$ takes one of two values (1 or -1). These binary gradient vectors are combined to form an element-wise majority vote at the \ac{PS}.

The \ac{CoMAC} protocol represents these SignSGD gradients using one of the two \ac{BPSK} symbols. Because the two \ac{BPSK} waveforms are inverted versions of the other, wireless superposition will correctly represent the addition of +1 and -1. In other words, the sum of a +1 \ac{BPSK} waveform and a -1 \ac{BPSK} waveform will be zero, given that their amplitude is identical. Therefore, the \ac{CoMAC} sum function would directly calculate the desired element-wise majority vote over the air.

The performance of one-bit digital \ac{CoMAC} is compared to \ac{BAA}~\cite{zhu2019broadband} by training a classifier for MNIST. The results suggest that the classification accuracy of digital \ac{CoMAC} is nearly identical to \ac{BAA}, with less than 1\% loss of accuracy. This result indicates that \ac{CoMAC} could potentially be implemented in cellular networks without requiring significant change in the hardware. As most \ac{CoMAC} schemes, perfect synchronization is assumed in both theoretical analysis and numerical simulation. One could argue that one-bit digital \ac{CoMAC} is more sensitive to synchronization errors since it relies upon cancellation of two opposite \ac{BPSK} waveforms, unlike analog \ac{CoMAC} which only requires additive powers.

A second digital \ac{CoMAC} scheme was proposed in \cite{jiang2020cluster} which introduces a clustered structure for the majority vote operation of the network. Rather than having all devices communicate directly with the server and thereby casting their vote in a ``direct democracy'' system, they propose intermediate relays that serve as representatives. This breaks the vote into two stages, where all the devices first cast their votes to their closest relay, which uses majority vote to generate a new gradient vector. Then in the second stage, the relays vote to the \ac{PS} in a ``representative democracy'' system. The selection of relays can be done in a smart way so that the relays have similar channel strengths to the server. If the strengths are similar, the channels do not necessarily need to be inverted, which alleviates the need for CSI estimation and improves the probability of success in the final majority vote. Simulation results show improvement over a cluster-free system both in terms of the gradient estimation at the \ac{PS} and the classification accuracy.

\subsection{Power control}\label{sec:power_control}
As explained in Section \ref{sec:comacexample}, there is a power control problem associated with \ac{CoMAC}. The problem arises because the pre-equalization of the channel is restricted by limited transmission power at the user devices. In this section, we discuss the problem of optimal power control.

In all papers mentioned up to this point, sub-optimal power control was used. Specifically, devices with fading below a certain threshold were excluded from participation and the remaining devices perfectly inverted their channels. Instead, \cite{cao2020optimized, zang2020over, liu2020over} study the problem more rigorously to minimize the estimation error under transmission power constraints. They consider problem structure \eqref{eq:power_control} to minimize the mean squared error between the received signal and the desired sum:
\begin{equation}
    \label{eq:power_control}
    \begin{aligned}
    \min_{\vtP, \eta} \quad & \mathbb{E}\left[\left(\sum_{k=1}^K\frac{h_kp_k\Delta \vtW_{n,k}}{\sqrt{\eta}} + \frac{v}{\sqrt{\eta}} - \sum_{k=1}^K\vtW_k\right)^2\right]\\
    \textrm{s.t.} \quad & p_k \leq P_{\text{max}},\ \forall k
    \end{aligned}
\end{equation}
where $p_k$ is the transmission power of device $k$, $P_{\text{max}}$ is the peak power constraint, and the remaining variables are defined in \eqref{eq:powercontrolsum}. There are two sources of error, one is the misalignment error caused by devices being unable to pre-equalize their channel and the second is the noise-induced error by the \ac{AWGN} $v$. The post-processing scalar $\eta$ acts as a tradeoff between the two, where a higher $\eta$ reduces the noise-induced error directly, but indirectly worsens the misalignment error by making it harder to invert the channel. The specific problem posed in \eqref{eq:power_control} is solved to a global minimum in both \cite{cao2020optimized} and \cite{liu2020over}, given certain simplifying assumptions.

In the context of \ac{FL}, the power-control problem affects both the convergence rate and final accuracy of the \ac{ML} model. In \cite{cao2021optimizedfl} (later extended in \cite{cao2022transmission}), a similar set-up to \eqref{eq:power_control} is used, with pre- and post-processing scalars for power control, but with the objective function replaced by an upper bound on \ac{FL} convergence. The proposed scheme vastly outperforms the device-exclusion scheme in terms of prediction accuracy. 

Another work \cite{zhang2021gradient} considers the use of gradient statistics to evaluate the expectation in \eqref{eq:power_control}. For known gradient statistics, they find the optimal solution in closed form using the mean squared norm and the squared multivariate coefficient of variation. In a practical scenario, these statistics would be unknown, but the solution can be used in conjunction with live estimates of the statistics to determine good pre- and post-processing scalars. 

\subsection{Retransmissions}\label{subsec:retransmission}
In digital communications, there is a well-known tradeoff between communication rate and error probability. For example, the modulation order $n$ determines the number of bits $\log_2(n)$ that can be transmitted in a single symbol. Simultaneously, a higher modulation order makes the demodulation problem harder, thereby increasing the probability of error. As such, the modulation order acts as a tradeoff between communication rate and error probability. Similarly, forward error-correcting codes can be used to correct erroneously demodulated bits at the receiver, but simultaneously introduce redundant bits which reduces the rate of communication. In contemporary digital communication protocols, it is common practice to adaptively select the modulation order and coding rate with respect to the estimated channel \cite{goldsmith1998adaptive} but in analog \ac{CoMAC} such a practice does not exist. With this in mind, the authors of \cite{hellstrom2021retransmit} consider a retransmission-based scheme to analyze the tradeoff communication rate and estimation error for over-the-air \ac{FL}.

The scheme presented in \cite{hellstrom2021retransmit} is similar to the power control papers \cite{cao2020optimized, liu2020over} except that the model update $\Delta \vtW_{n,k}$ is transmitted $M$ times in the uplink instead of just once. At the receiver, these $M$ transmissions are collected and the arithmetic mean of them is used to generate the next iteration of the global model update. This way, the signal part of the transmission combines constructively, while the noise part is random and can therefore combine destructively. This scheme is analyzed analytically by proving an upper bound on the \ac{FL} loss, which reveals that the convergence rate is strictly increasing in $M$. To make a fair comparison between transmission with $M=1$ and $M>1$, the authors perform a simulation study in which the uplink transmission budget is fixed to $\overline{C}$, such that only $\overline{C}/M$ communication rounds can be performed. Despite using $M$ times fewer communication rounds, the simulation study indicates that there are scenarios in which $M>1$ achieves higher classification accuracy after consuming the communication budget. Therefore, the performance of Over-the-Air \ac{FL} can be improved by including retransmissions, without incurring additional costs in terms of latency or energy consumption.

In a follow-up study \cite{hellstrom2021retransmit2}, the optimal choice of $M$ is studied further and a heuristic is developed to predict $M^*$ before training begins. Numerical results indicate that the heuristic is generally successful at identifying $M^*$, including the case when $M^*=1$. As such, the system can predict when the conditions are not right for retransmissions and select one-shot uplink transmission.

\subsection{Differential privacy}
Compared to \ac{CML}, \ac{FL} makes a step towards data privacy by keeping the data local at the users.
However, sharing the local models or the gradients may reveal sensitive information about the users data \cite{carlini2019secret,MelisUnintendedLeakage2019}. 
Adding a level of uncertainty to the local models or the gradients computed at the users can enhance the privacy of user data at the cost of lower utility.
Differential privacy (DP) is a privacy measure that quantifies the amount of information leakage about individual data points by measuring the sensitivity of the revealed statistics to a change at a single data point, and it is widely adopted as a promising privacy measure.   

It is shown in \cite{SeifTandonDPFLISIT2020} that the additive nature of the wireless multiple access channel from the user devices to the PS provides local DP guarantees for the devices where the privacy leakage per device is scaled with $1/\sqrt{K}$.  
If the channel noise is not sufficient to satisfy the DP target, a subset of the devices add power constrained artificial noise that benefit all the devices.
Instead, \cite{KodaDPAircompGlobecom2020} introduces an energy efficient differentially private approach for FL over wireless networks by scaling down the transmit power rather than injecting additional noise to the transmit signal at the devices.
In general, a certain level of DP can be achieved for free with the analog transmission from the devices due to the noise added by the wireless multiple access channel which can act as a privacy-inducing mechanism \cite{LiuOsvaldoDPFLJSAC2021}.

\subsection{Byzantine attacks}
An unfortunate consequence of the distributed and privacy-preserving nature of \ac{FL} is that malicious users can transmit modified model updates with the intention of disrupting the training process \cite{vempaty2013distributed}. Even a single client can seriously harm the performance of the end model \cite{blanchard2017machine}. These malicious clients are called Byzantine, and their attacks are called Byzantine attacks. As a countermeasure, a recent idea has emerged for distributed computation among agents called “coded computing”. This idea consists in transforming the client’s information by functions which on the one side hide the client’s information, and on the other side can add robustness to the computation because the \ac{PS} applies another function that attempts to minimize the effect of the Byzantine attacks \cite{bhagoji2019analyzing,fung2018mitigating,bagdasaryan2020backdoor, prakash2020byzantine}. However, these countermeasures generally rely upon detecting anomalies in individual model updates, which is difficult for \ac{AirComp} where the average model updates are calculated directly over the air.

This gap in security for over-the-air \ac{FL} is a serious concern. A first step to address this concern can be found in \cite{sifaou2021robust}. Specifically, \cite{sifaou2021robust} proposes that the participating devices are split into $G$ groups, with $K/G$ devices per group. Each group is allocated its own time slot for over-the-air computation, thereby generating $G$ received model updates at the \ac{PS}. With these $G$ vectors, the \ac{PS} can apply coded computing methods to mitigate potential Byzantine Attacks. In \cite{sifaou2021robust}, the authors prove that the proposed algorithm converges to a neighbourhood of the optimal $\vtW$ when the number of attackers are less than $G/2$. A such, the choice of $G$ acts as a tradeoff between communication efficiency and security.

\subsection{Device-to-device communications}
Up until this point of the survey, we have only considered distributed \ac{ML} over star networks, which can be modeled by the multiple-access channel and therefore leverage \ac{AirComp}. In this subsection, we briefly discuss work on device-to-device communication over more general network topologies. For such topologies, there is no dedicated \ac{PS} and the devices are only able to communicate with their immediate neighbors in a single hop. Therefore, the \ac{FedAvg} algorithm cannot be directly applied for \ac{ML} training. However, there are other methods, such as decentralized \ac{SGD}, which are guaranteed to converge under assumptions of noiseless communication, convexity and connectivity \cite{sundhar2010distributed}. 

In \cite{xing2020decentralized}, the problem of decentralized \ac{SGD} with over-the-air computation was studied for the first time. They consider a connectivity graph model with probibalistic blockages due to shadowing, where unblocked channels are described by Rayliegh fading and \ac{AWGN}. To enable \ac{AirComp} in such a network, they propose a scheduling policy that aims to select as many non-interfering subnetworks with star topologies as possible for each time slot. Once the subnetworks are identified, a two-step iterative procedure is initiated. In the first step, over-the-air computation is leveraged to communicate the average gradient to the center of each star network. In the second step, all centers broadcast the received gradient average to the edge devices. This way, every device in the subnetwork knows the arithmetic mean of the gradients after two time slots. This scheme is evaluated numerically by training an MNIST classifier for $K=8$ devices with randomly generated connectivity graphs. The results suggest that over-the-air computation converges with significantly fewer communication blocks than orthogonal digital communication, but reaches a lower accuracy as the number of communication blocks approach infinity.

In \cite{shi2021over}, a similar setup to \cite{xing2020decentralized} is considered, but with the added consideration of gradient tracking \cite{pu2021distributed} and variance reduction \cite{xin2020decentralized}. Gradient tracking refers to the introduction of an auxiliary variable into the optimization problem of decentralized \ac{SGD} that tracks the average gradient of all devices in the network. With such an auxiliary variable, linear convergence can be guaranteed with a constant step size \cite{pu2021distributed}. With variance reduction, an iterative estimator of the batch gradient is designed, whose variance progressively approaches zero as the parameter vector approaches a local minimizer. With variance reduction, the error floor of \ac{SGD} is eliminated even with a constant step size, which is not possible for vanilla \ac{SGD}. In \cite{shi2021over}, the proposed decentralized scheme is proven to converge linearly under standard convexity assumptions, fully-connected graphs, and bounded gradients.

\subsection{Bayesian learning}
While \ac{ML} has displayed impressive accuracy for many classification tasks, \ac{ML} models are not perfect and will occasionally make mistakes. For certain applications, such mistakes could have unwanted consequences that limit the applicability of \ac{ML}. To mitigate the harm caused by \ac{ML} mistakes, it is desirable to consider models with the ability of assessing the certainty of its predictions. Consider the application of fall detection among elderly. Multiple accelerometers can be attached to a patient's body with the intent of detecting falls and alerting the patient's medical assistant \cite{wu2017autonomous}. A common problem with these systems is that alerts are communicated to assistants for normal, healthy activity which causes unnecessary and unwanted visits \cite{broadley2018methods}. If the alerts were sent to the assistant together with a measure of the model's uncertainty, the assistant could make a better decision on whether they should intervene. In statistics, the term \textit{predictive uncertainty} is used to describe this virtue, where many state-of-the-art \ac{ML} methods, such as neural networks, are poor at quantifying predictive uncertainty, and tend to produce overconfident predictions \cite{lakshminarayanan2017simple}.

Bayesian learning is a popular method to quantify the predictive uncertainty of neural networks, in which a prior distribution is specified upon the parameters of a neural network and then, given the training data, the posterior distribution over the parameters is computed. If we compare this to traditional \ac{ML}, we can say that traditional \ac{ML} generates a point estimate of the parameters, i.e., one instantiation of the weights and biases of the neural network, while bayesian learning attempts to generate a full distribution over the parameters, i.e., the posterior distribution. Exact calculation of the posterior distribution is in general intractable, so approximate methods are used to generate an estimate of the distribution, such as Monte-Carlo sampling \cite{lakshminarayanan2017simple}. Once estimated, the posterior distribution is leveraged to quantify the uncertainty of any given prediction. The interested reader can refer to \cite{wilson2020bayesian} for a detailed description of the uncertainty quantification.

In \cite{9562546}, distributed bayesian learning is brought into the wireless setting using over-the-air computation. The main contribution of the paper is the introduction of an idea called \textit{channel-driven Monte-Carlo sampling} where the channel noise is utilized as an integral part of the sampling for estimating the posterior distribution. If accounted for, the channel noise combined with the analog transmissions in over-the-air computation may not cause harm to the performance of the learning. This is in contrast to \ac{FL}, where the noise generally slows down convergence and should be compensated for, as discussed in Sections \ref{sec:power_control} and \ref{subsec:retransmission}. In \cite{9562546}, the channel-driven Monte-Carlo method is analyzed analytically by means of a convergence proof and numerically by extensive simulations.

\section{Review of MIMO over-the-air computation}\label{sec:mimo}

\subsection{Blind learning}\label{subsubsec:blind}
Similar to traditional \ac{MIMO} communications, the channel estimation effort of \ac{CoMAC} systems is in the opposite direction of traditional \ac{SISO} communication, since equalization is performed at the transmitter instead of at the receiver. This is problematic, because while the downlink channel can be estimated using the model broadcast of \ac{FL}, the uplink channel can not. To solve this problem, one can use channel reciprocity together with a calibration factor to estimate the uplink channel \cite{abari2016over} but this is both more expensive (requires calibartion stage) and less precise than downlink channel estimation. In \ac{CoMAC}, this problem is exacerbated since the \ac{CSI} knowledge is used to achieve signal alignment, and poor channel estimation will result in distorted function computation \cite{zhu2021over}. With this in mind, the channel hardening phenomenon of \ac{MIMO} communications carries particular importance for \ac{CoMAC}. In~\cite{goldenbaum2014channel}, channel hardening is leveraged to perform over-the-air computation without deterministic channel knowledge at any node in the network. Specifically, the authors quantify the gap in performance between a system with full \ac{CSI} knowledge and one with only statistical knowledge at the \ac{PS} and no \ac{CSI} knowledge at the user devices. For a network with $M > 1$ antennas at the \ac{PS} and $K > 1$ single-antenna sensor devices, they prove that this performance gap approaches zero as $KM \rightarrow \infty$.

In the previous section, we highlighted~\cite{amiri2020machine} that introduces gradient sparsification to over-the-air \ac{FL}. In ~\cite{amiri2019collaborative}, this scheme is extended to consider blind learning. The main contributions of this work are to propose a \ac{CoMAC}-based \ac{FL} technique that requires no transmit \ac{CSI} from the devices and to provide insights into how the number of antennas affect learning accuracy. The numerical results show that for $M=2K^2$, the accuracy nearly matches a non-fading channel. For a lower number of antennas $M=2K$ the accuracy drop compared to the non-fading channel is about 5\%.

The work in \cite{amiri2019collaborative} is then further extended in \cite{mohammadTolgaBlindFLTWC2021} to consider imperfect \ac{CSI} at the \ac{PS}. The authors show that the lack of perfect \ac{CSI} results in an additional zero-mean interference term with a variance proportional to $1/M$. Similarly, worst-case analysis shows that the imperfect \ac{CSI} results in slower convergence but that the effect is inversely proportional to the number of antennas. Finally, numerical analysis on the MNIST and CIFAR-10 datasets reveal significant performance improvement as $M$ increases with a more pronounced effect when channel estimation is not perfect.

\subsection{Nonlinear estimator}
One challenge in the FL over wireless network is the presence of a noisy shared wireless medium from the typically abundant users to the PS, over which the users transmit their local models or gradients. 
The goal is to deliver users' signals  to the PS as accurately as possible.
Equipping the PS with multiple antennas can improve communication reliability between the users and the PS, where multi-antenna transmission and/or reception beamforming techniques can be employed \cite{mohammadTolgaBlindFLTWC2021,YangJiangFLMIMOTWC2020,VuNgoCellFreeFLMIMOTWC2020}.
However, the above works consider only linear beamforming techniques at the multi-antenna PS to estimate the signals transmitted from the users. 

In general, a linear beamforming technique at a multi-antenna receiver does not lead to any optimal estimation performance \cite{YoSebMohammadMIMOFL}.
Instead, the authors in \cite{YoSebMohammadMIMOFL} design an estimator based on the sparsity of the gradient vectors computed at the users. 
Motivated by this sparsity, a compressive sensing approach in the user domain is employed, where the gradient vectors at different users are permuted using different patterns such that only a small subset of the users transmit non-zero entries at each dimension.
This results in a sparse transmitted signal from the users, and using this sparsity, the PS employs a nonlinear estimator to recover the average of the gradients almost accurately. 
This approach is extended in \cite{YoSebMohammadCommEffiMIMOFL} by employing the gradient compression technique introduced in \cite{amiri2020machine,amiri2020federated} to reduce the transmission bandwidth over the wireless multiple access channel from the users to the PS.   

\subsection{Cell-free massive MIMO}
Recently, a new architecture for multi-user \ac{MIMO}, called cell-free massive \ac{MIMO}, has emerged. In cell-free massive \ac{MIMO}, a large number of \acp{AP} collaboratively serve users over the same time/frequency resources~\cite{ngo2017cell}. All \acp{AP} collaborate through a backhaul network, enabling fine synchronization that can be used for conjugate beamforming in the downlink and matched filtering in the uplink. The main advantage of the cell-free architecture is the broad coverage due to the high number of \ac{AP}s. This is especially important for over-the-air \ac{FL} since the communication quality of \ac{CoMAC} for Machine Learning is determined by the device with the worst channel \cite{zhu2019broadband}.

In~\cite{VuNgoCellFreeFLMIMOTWC2020}, a comprehensive scheme combining cell-free massive \ac{MIMO} and \ac{FL} was proposed. The \ac{FL} process is divided into four steps, starting with \ac{CSI} acquisition and ending in global model aggregation at the centralized \ac{PS}. Unlike the previous subsection, the proposed scheme does not utilize blind transmission but it is able to estimate the channel using non-orthogonal transmission. By making all sensor devices transmit their pilot sequence simultaneously over the same bandwidth, the channels can be estimated using multiple measurements received by the large number of \acp{AP}. Numerical results show that cell-free massive \ac{MIMO} can reduce training time by up to 33\% when compared to massive \ac{MIMO} with collocated antennas.

\subsection{Beamforming and user selection co-design}\label{subsub:bf_user_sel}
Due to the communication-learning tradeoff, see Section \ref{sec:baa}, user selection should be made to strike a balance between receive \ac{SNR} and data quantity. The solution to this problem in the SISO case was to set a fading threshold based on a power constraint and only include users below that threshold. By introducing multiple antennas at the \ac{AP},~\cite{YangJiangFLMIMOTWC2020} instead proposes receive beamforming to maximize the participating users while ensuring that the aggregation error is constrained. The proposed user selection and beamforming scheme is compared to a semidefinite relaxation baseline and a global optimization approach with exponential time complexity. In terms of probability of feasibility, the proposed approach was significantly better than semidefinite relaxation and nearly identical to the global optimum. Additionally, the approach was used to train on the CIFAR-10 dataset \cite{krizhevsky2009learning} and the proposed approach achieved nearly double the relative classification accuracy of semidefinite relaxation.

\subsection{Intelligent reflective surfaces}\label{section:IRS}
\begin{figure}[ht]
\centering
    \begin{tikzpicture}
        \node[inner sep=0pt] (russell) at (0,0)
            {\includegraphics[width=8.5cm]{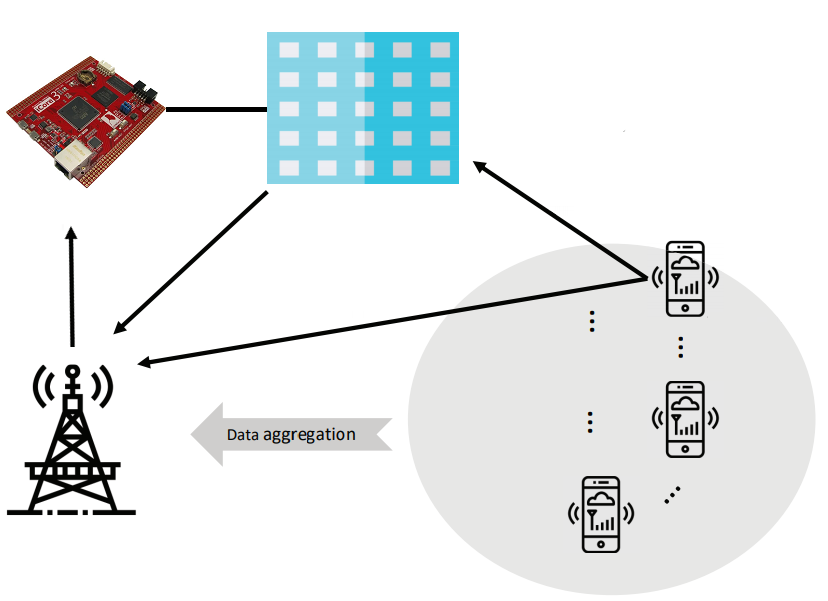}};
        \node[] at (2.1,-3) {User devices};
        \node[] at (-3.5,-2.4) {Access Point};
        \node[] at (-0.55,1) {IRS};
        \node[] at (-3.4, 1.1) {Controller};
    \end{tikzpicture}
    \caption{Illustration of an IRS for over-the-air computation. In this figure, the uplink data transmission is assisted by the reflective surface to improve the wireless channel. To control the phase shifts of the reflected signals, the \ac{AP} communicates with a controller attached to the reflecting surface.}
    \label{fig:irs}
\end{figure}
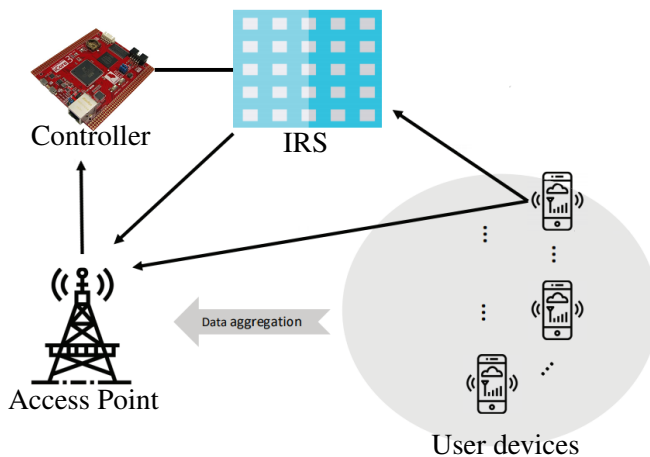
The \ac{IRS} is a recent technological development that has received strong interest from both academia and industry \cite{zhao2019survey}. The purpose of an \ac{IRS} is to introduce a "mirror" for electromagnetic waves that can be tuned to reflect incident signals toward the intended receiver. The surfaces consist of passive reflective elements which adjust the phase shift of the incoming signal, effectively creating a \ac{MIMO} effect. In addition to the reflecting elements, a controller is installed that allows for \acp{AP} to configure the phase shifts, illustrated in Figure \ref{fig:irs}. For the case of \ac{CoMAC}, \cite{jiang2019over} was the first paper to propose a joint beamforming and \ac{IRS} phase shift design to minimize the aggregation error. The paper showed incredible potential with up to 4 orders of magnitude lower estimation error than an \ac{IRS}-free propagation environment.

For the case of \ac{FL}, the \ac{IRS} can be used to enable higher user participation. As we know from the previous Subsection~\ref{subsub:bf_user_sel}, it is natural to constrain the number of participating devices to ensure that the aggregation error falls below an acceptable level. In \cite{wang2021federated}, the authors proposed a joint beamforming, user selection, and \ac{IRS} phase shift design to maximize the number of participating devices. The resulting scheme was able to approximately double the number of participating users compared to an equivalent system without \ac{IRS}, which can improve the test accuracy by up to 20\% under the right conditions. 

The maximization of user participation clearly has a positive impact on learning, but it is a rough proxy for the classification accuracy, which is the metric of interest. To address this issue, \cite{liu2021reconfigurable} found an upper bound on the \ac{FL} loss under the \ac{IRS} over-the-air setup and proposed an optimization problem that incorporates the loss function, thereby targeting the accuracy more directly. The simulation results of \cite{liu2021reconfigurable} reached nearly the same test accuracy as training over an error-free channel, outperforming \cite{jiang2019over} by 3\%-30\% depending on the experimental setup.

Besides improving the classification accuracy, \ac{IRS} can also be used to enable blind transmissions without having a large antenna array \cite{liu2021csit}. When using \ac{IRS}, blind transmission can be achieved even with single-antenna devices and single-antenna \acp{AP}. However, the system is not completely blind, as it still requires receive \ac{CSI} at the \ac{PS}. Since there is no \ac{CSI} at the transmitter, the devices cannot invert their channel before transmitting. Instead, \cite{liu2021csit} proposes that the devices transmit with maximum power, and the \ac{PS} configures the \ac{IRS} phase shift vector to achieve the desired function over-the-air. Such an approach achieves a significantly worse aggregation error than a system with \ac{CSI} at the transmitter, but the error is still sufficiently low to achieve a comparable classification accuracy. Since \ac{FL} works well with some level of noisy updates, the 4 orders of magnitude reduction from the \ac{IRS} design can be excessive, opening up for designs that are less efficient in terms of \ac{MSE}.

In \cite{9414785}, the authors investigated the use of multiple \acp{IRS} and over-the-air computation to support the deployment of FL. In their considered model, the devices can directly transmit FL models to the BS or using IRS. The authors jointly optimized the device selection, phase shift matrix, decoding vector, and power control so as to minimize the energy that the devices use to transmit and train FL models. Simulation results comparing communication with and without an \ac{IRS} reveal that the energy consumption of the \ac{FL} training can be reduced by approximately an order of magnitude by transmitting via an \ac{IRS}.

\chapter{Digital communications}\label{sec:orthogonal}

\section{Primer}\label{subsec:digital_primer}

The \ac{CoMAC} systems discussed in the previous section provide an attractive solution to the \ac{DML} problem. However, the technology is dependent on prerequisites that can be difficult to realize in practical scenarios, such as very stringent synchronization and customized hardware. Due to the challenges with \ac{CoMAC}, digital communications still has to be considered as a basis for \ac{DML}. Within digital communications, we consider orthogonal communication methods that leave the physical layer as it is. Then, the attention is placed on the data link and network layer, with a particular emphasis on \ac{RRM} protocols for \ac{DML}.

As explained in Section \ref{sec:introduction}, the problem of \ac{DML} differs in several ways from that of general data communication. These differences result in new constraints in terms of computational complexity, training time, training data, and more. In this setting, general data communication protocols perform poorly, motivating the design of digital communication protocols tailored to support \ac{DML}. In this primer, we will discuss some of these differences in more detail to better understand why new digital protocols are needed.

\subsection{Fairness}
In traditional \ac{RRM}, the well-known water-filling method~\cite{yu2001constant} allocates more transmission power to users experiencing a good channel. This method leads to very efficient spectrum utilization, but generally leads to some users with no allocated power. Therefore, despite utilizing spectrum less efficiently, max-min-fairness protocols are often used to ensure a minimum level of service for all users in the network~\cite{xue2009max}. This sacrifice is not reasonable for \ac{FL} since the participation of every user is not necessary to train a good model. In fact, if our goal is to maximize the classification accuracy of the \ac{ML} model, the data-importance discussion in the previous section indicates that we should be deliberately treating users in a discriminatory manner, contradicting the demands on user fairness. Even if data-importance is not considered, there is no reason to sacrifice spectrum utilization to ensure user fairness for FL.

\subsection{Training data}
It is well-known that supervised \ac{ML} performance is intricately connected to the quality and size of the training dataset. Therefore, we would ideally utilize every collected data point to train machine learning models. However, over resource-constrained wireless networks, this is not always possible. Therefore, we are posed with the problem of optimally selecting which data points to utilize. In centralized machine learning, this problem is related to which data points are communicated to the server, and in \ac{DML}, the problem is related to which devices should participate (and thereby their datasets). One useful metric to guide such a selection is data importance (discussed further in Section \ref{sec:importance}), which can be utilized to value one data sample over another.

\subsection{Computational capability}
Since \ac{FL} is traditionally a synchronous algorithm, it suffers from a problem known as the \textit{straggler effect}, i.e., the effect where the slowest device acts as a bottleneck while remaining users idly wait for the next communication round~\cite{ha2019coded}. Therefore, the heterogeneity of communication and computational capabilities becomes an important factor to consider for device scheduling and \ac{RRM}. As an example, more bandwidth could be allocated to slow devices, thereby helping them to compensate for their slow training by communicating their local models quicker.

\subsection{Energy}
Most \ac{DML} algorithms rely on multiple rounds of communication to reach convergence in the model training process, each of which consumes a significant amount of energy. Additionally, each communication round is associated with a computational task of training the model, which leads to further energy costs. To maintain an acceptable battery level at the training devices, the energy-efficiency of this process is of critical importance. There are specific properties of the \ac{FL} algorithm which can be utilized to either consume less energy or transfer power from the base station to the user devices. As an example, there is a period of naturally occurring radio silence in \ac{FL}, when the user devices are doing their local training. During this time it is possible to perform power transfer from the \ac{BS} to the devices.

\section{Digital communications for distributed machine learning}
In this section, we have divided the digital \ac{DML} literature into two categories: importance-aware communication and \ac{RRM} for \ac{FL}. 
The first category considers prioritization schemes that select users based on how valuable their training data is to the \ac{ML} model. The second category tries to optimize \ac{RRM} algorithms for \ac{FL}. A comprehensive list of papers for digital \ac{DML} methods can be found in Table~\ref{table:ld-orthogonal1}, \ref{table:ld-orthogonal2}, and \ref{table:ld-orthogonal3}.

\begin{table}[ht]
    \caption{Summary of the Importance-Aware Communications literature. Papers that consider both importance-aware communications and radio resource management is covered in table \ref{table:ld-orthogonal3}.The papers are ordered according to when they are covered in the survey.}
\begin{center}
\begin{tabular}{|l|l|p{17em}|}
    \hline
    \textbf{Topic} & \textbf{Ref.} & \textbf{Summary}\\
    \hline
    \multirow{3}{8em}{Centralized Learning} & \cite{liu2019wireless} & Retransmission protocol with data-sample prioritization.\\
    \cline{2-3}
    & \cite{liu2020wireless} & Extension of~\cite{liu2019wireless} to consider more advanced \ac{ML} models such as convolutional neural networks.\\
    \cline{2-3}
    & \cite{liu2020data} & User selection protocol.\\
    \hline
    \multirow{2}{8em}{Federated Learning} & \cite{goetz2019active} & Importance-aware user selection step.\\
    \cline{2-3}
    & \cite{leng2022client} & Comparison of different data importance metrics for user selection step.\\
    \hline
\end{tabular}
\end{center}
\label{table:ld-orthogonal1}
\end{table}

\begin{table}[ht]
    \caption{Summary of the Radio Resource Management for Machine Learning literature. The papers are ordered according to when they are covered in the survey.}
\begin{center}
\begin{tabular}{|l|l|p{17em}|}
    \hline
    \textbf{Topic} & \textbf{Ref.} & \textbf{Summary}\\
    \hline
    \multirow{3}{8em}{Participation Maximization} & \cite{nishio2019client} & Client selection scheme that aims to maximize the number of participants in the Federated Learning training step.\\
    \cline{2-3}
    & \cite{Xu2021} & Joint client selection and bandwidth allocation considering the \emph{later-is-better} phenomenon of \ac{FL}.\\
    \cline{2-3}
    & \cite{Xu2021b} & Joint time slot and bandwidth allocation with multiple co-existing \ac{FL} services that share wireless resources.\\
    \hline
    \multirow{3}{8em}{Energy Efficiency} & \cite{zeng2020energy} & Joint client selection and bandwidth allocation scheme that aims to minimize the energy consumed for FL training.\\
    \cline{2-3}
    & \cite{9264742} & Joint time slot allocation, bandwidth allocation, and transmit power allocation.\\
    \cline{2-3}
    & \cite{Dinh2021} & Joint time slot allocation, clock frequency optimization, and local accuracy optimization.\\
    \hline
    \multirow{3}{8em}{Packet Error Impact} & \cite{chen2020joint} & Performs convergence analysis on the impact of packet errors in FL training. Utilizes the resulting upper bound to perform client selection, resource block allocation, and power allocation.\\
    \cline{2-3}
    & \cite{Salehi2021} & Client selection scheme that weighs the ML model update contribution of individual devices based on their probability of successful transmission. \\
    \cline{2-3}
    & \cite{jin2022communication} & Analyzes the convergence of SignSGD-based distributed learning.\\
    \hline
    \multirow{2}{8em}{Total Time Minimization} & \cite{shi2020device} & Joint client selection and bandwidth allocation to minimize the total time spent training the \ac{ML} model.\\
    \cline{2-3}
    & \cite{Chen2021} & Joint client selection and resource block allocation. \\
    \hline
\end{tabular}
\end{center}
\label{table:ld-orthogonal2}
\end{table}

\begin{table}[ht]
    \caption{Continuation of Table \ref{table:ld-orthogonal2}.}
\begin{center}
\begin{tabular}{|l|l|p{17em}|}
    \hline
    \textbf{Topic} & \textbf{Ref.} & \textbf{Summary}\\
    \hline
    \multirow{1}{8em}{Empirical Classification Error} & \cite{wang2020machine} & Attempts to estimate the classification error empirically and uses this estimate to guide power allocation.\\
    \hline
    \multirow{1}{8em}{Federated Distillation} & \cite{9121290} & Combines Federated Distillation in the uplink with Federated Learning in the downlink. Also employs data sample mixing to enhance user privacy. \\
    \hline
    \multirow{1}{8em}{Batch Size Selection} & \cite{ren2020accelerating} & 
    Treats hyperparameters of the machine learning algorithm as decision variables for the RRM problem. Specifically, a joint batch size selection and time-slot allocation scheme is developed.\\
    \hline
    \multirow{2}{8em}{Importance RRM} & \cite{ren2020scheduling} & Combines importance-aware communication and RRM for FL by considering a client selection scheme. Specifically, the gradient divergence is used to guide the selection of participating devices.\\
    \cline{2-3}
    & \cite{Chen2021} & Considers update staleness and update drift to develop a joint client selection and resource block allocation scheme. \\
    \hline
    \multirow{2}{8em}{Energy Harvesting/Power Transfer} & \cite{Zeng2021} & Joint batch size selection, clock frequency optimization, and learning-wireless power transfer tradeoff.\\
    \cline{2-3}
    & \cite{Silva2021} & Joint local number of iterations optimization and time slot allocation to transmit, compute and harvest energy. \\
    \hline
    \multirow{1}{8em}{Noisy Downlink} & \cite{MDSVNoisyDownlinkJournal2020} & Digital downlink transmission of the global model is compared to analog transmission.\\
    \hline
    \multirow{1}{8em}{Federated Meta-Learning} & \cite{9681911} & The combination of meta-learning and \ac{FL} is considered in a wireless network, where users are scheduled based on a convergence bound.\\
    \hline
\end{tabular}
\end{center}
\label{table:ld-orthogonal3}
\end{table}

\section{Review of importance-aware communications}\label{sec:importance}

\begin{figure}[ht]
\centering
    \begin{tikzpicture}
        \node[inner sep=0pt] (russell) at (0,0)
            {\includegraphics[width=8.5cm]{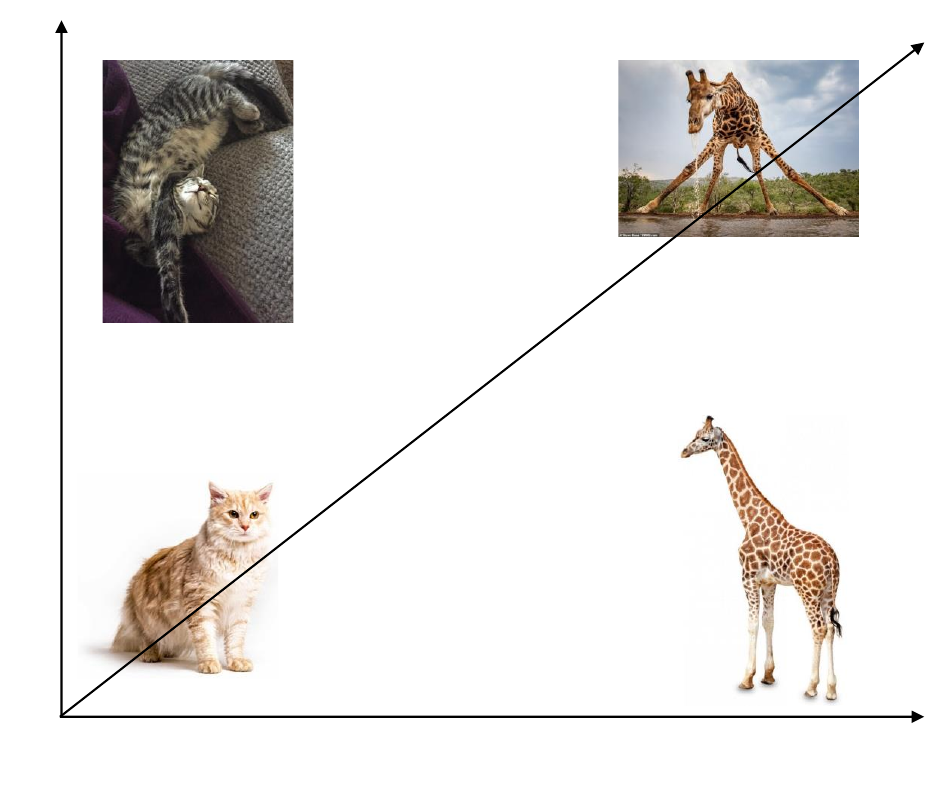}};
        \node[] at (0,-3.2) {Diversity};
        \node[rotate=90] at (-4,0) {Uncertainty};
        \node[rotate=40] at (0,0.5) {Data Importance};
    \end{tikzpicture}
    \caption{Data-importance illustration for image classification. Uncertainty measures how difficult a data sample is to classify, while diversity measures the rarity of the data sample label. The importance of training data is a good metric for prioritizing data samples when communication resources are limited.}
    \label{fig:importance}
\end{figure}
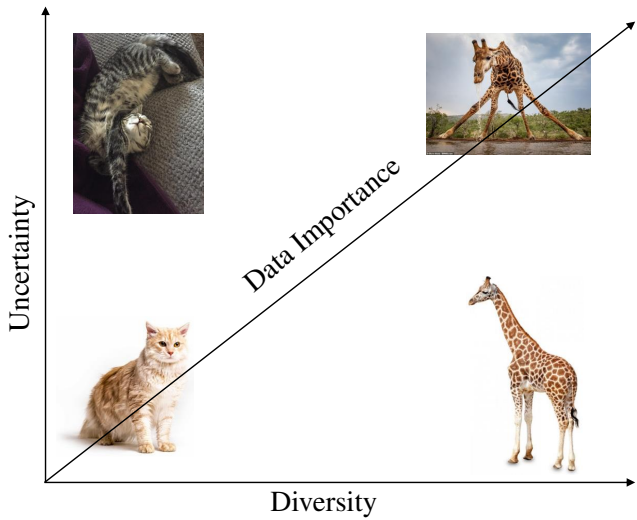

When traditional communication algorithms are designed to maximize data rate, they are implicitly assigning equal worth to each bit regardless of their information content. This makes sense in classical packet-switched networks since the abstraction of information in the OSI model prohibits the controller from interpreting the payload. However, in \ac{DML}, data-importance is not uniform~\cite{settles2012active}, thus if we consider that each bit has the same worth, resources are wasted to transmit low-importance data. The non-uniform data-importance for \ac{ML} stems from two qualities: uncertainty and diversity~\cite{huang2013active}. Uncertainty refers to the confidence level with which the current model can classify a data sample, and diversity refers to the rarity of the label compared to the remaining training data set. Consider an image classification system for animals in Figure~\ref{fig:importance}. Low data-importance images would correspond to something that is easy to classify, such as a simple white background, a common animal, and a natural pose. As either the diversity (rarity of the animal) or uncertainty (difficult pose/background) increases, so does the data importance. By prioritizing samples with high data-importance, \ac{ML} training is accelerated~\cite{wen2019overview}. Since non-uniform data-importance is common, communication algorithms concerned with learning performance should incorporate uncertainty and diversity in their design by prioritizing high-importance data.

This idea of evaluating data samples based on their importance during the training of the classifier model comes from a branch of \ac{ML} called Active Learning~\cite{settles2009active}. The problem considered in Active Learning is with regards to the cost of labeling. Using a speech recognition example, the cost would come from having a human interpreter listening to recorded samples and transcribing labels to be used for the \ac{ML} algorithm. In Wireless for \ac{ML}, each sample is instead associated with a cost related to transmission, and since we consider supervised learning, the label is already available at the device. Although the fundamental goal is different, the metrics developed in Active Learning for evaluating data samples have been tested for the communication problem and have been shown to reduce the communication cost~\cite{liu2020wireless, goetz2019active}.

\subsection{Centralized learning}
In this section, we consider centralized learning using data distributed over multiple devices in a network. In other words, training only occurs at the \ac{PS} but wireless communication is still used to collect the data. This scenario is relevant when the user devices do not have sufficient computational resources to perform local training but are still carrying data relevant for learning. In~\cite{liu2019wireless} and~\cite{liu2020wireless}, the problem of developing an \ac{ARQ} protocol for \ac{ML} is considered. Despite being orthogonal, the communication is with analog transmission, so there is always some distortion of the received sample. This distortion can be reduced by taking the mean of multiple transmissions of the same signal, thus improving effective receive \ac{SNR}. Given a time slot budget, the goal is to maximize the final learning accuracy. Given that the time slot budget is not sufficient to upload every sample to the server, additional retransmissions reduce the total number of samples uploaded for training. This problem gives rise to a communication-learning tradeoff like earlier but based on retransmissions instead of participation. 

On top of finding a balance between data quantity and quality, the protocols are designed to prioritize samples with higher importance. Three solutions are suggested in~\cite{liu2019wireless}, as an example, we discuss "Importance \ac{ARQ} for binary \ac{SVM} classification" in detail. The protocol considers the acquisition of a data sample $x$ from a user device. Using the first transmission, the \ac{PS} estimates the data-importance and then \ac{PS} repeatedly requests the device to retransmit $x$ until the effective receive \ac{SNR} satisfies

\begin{equation}
\label{eq:iaARQ}
    \text{SNR}(T)>\min (\theta_0 \mathcal{U}_d(\hat{x}(T)),\theta_{\text{SNR}})\ ,
\end{equation}
where $T$ is the number of retransmissions, $\theta_0$ is a scaling factor, $\mathcal{U}_d(\hat{x}(T))$ is the uncertainty measure, and $\theta_{\text{SNR}}$ is the maximum \ac{SNR}. The maximum \ac{SNR} is there to prevent one sample from consuming too many transmissions, and $\mathcal{U}_d(\hat{x}(T))$ is defined as the distance to the \ac{SVM} boundary, which is an uncertainty measure. The protocol based on Eq. (\ref{eq:iaARQ}) will allocate sufficiently many retransmissions for each device to reach their guaranteed minimum \ac{SNR}. Devices carrying low-importance data are guaranteed lower minimum \acp{SNR} and are therefore given fewer retransmissions even if the channel is poor. 

Apart from binary \ac{SVM} classification,~\cite{liu2019wireless} contains extensions to multi-class \ac{SVM}, generic classifiers, and \acp{CNN}. According to experimental studies, the protocol outperforms purely channel-aware retransmission protocols in terms of classification accuracy by around 2-3\% when training on the MNIST dataset.

In~\cite{liu2020data}, importance-aware user selection is addressed. The devices are scheduled in a time-division manner and take turns to upload a data sample in each time slot. Once again, the radio resources are limited and the problem is to schedule devices in a manner that maximizes the final test accuracy. User selection is based on two factors, the channel quality of each user and the importance of their data. Devices experiencing lower fading are prioritized so that higher data rates are achieved, but only if their data is sufficiently important.

Unlike the retransmission case, it is not obvious how data-importance should be communicated to the \ac{PS}. The problem lies in that both the model and the data samples are required to measure importance, and they are not present in the same entity. To solve this problem,~\cite{liu2020data} suggests using popular model compression methods~\cite{cheng2017survey,zhu2017prune} to transmit a lighter version of the \ac{ML} model to the user devices. This would reduce the size $d$ of the local model $\vtW\in\bbR^d$. This way, the user device can evaluate their importance locally, and then inform the \ac{PS}.

\subsection{Federated learning}
Since \ac{FL} communicates local models or gradients instead of data samples, there is a need for data importance metrics that can be applied to gradients. In~\cite{goetz2019active}, the loss function is proposed as an importance metric. This metric is an uncertainty metric, as it directly describes how difficult a sample is to classify. Additionally, it is cheap to compute by performing inference on the already locally available \ac{ML} model.

Using the loss function as a metric, the model importance is defined as
\begin{equation}
    \label{eq:valuefunction}
    I_k = \frac{1}{\sqrt{N_k}}\sum_{i=1}^{N_k} l( h(\vtX_k^i;\vtW_k),\vtY_k^i )\ ,
\end{equation}
where $I_k$ is the importance of device $k$'s gradient to the global model, $\vtX_k$ is the vector of data samples, $\vtY_k$ are the labels for those samples, and $N_k$ is the number of samples. The importance is evaluated locally at each user and is transmitted in the uplink together with the local model, illustrated in Figure~\ref{fig:activebd}. The \ac{PS} takes advantage of $I_k$ to determine which users to schedule for the upcoming communication round.

\tikzstyle{plain}=[draw=none]
\tikzstyle{rect}=[draw, fill=blue!20, minimum size=2em]
\tikzstyle{circ}=[draw, ellipse, fill=blue!20, minimum size=2em]
\tikzstyle{init} = [pin edge={to-,thin,black}]
\pgfdeclarelayer{background}
\pgfsetlayers{background,main}

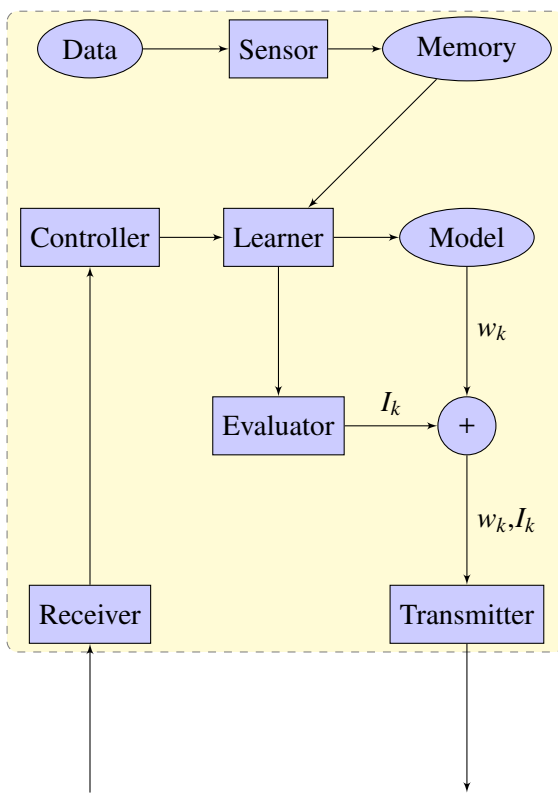
\begin{figure}[ht]
    \centering
    \begin{tikzpicture}[node distance=2.5cm,auto,>=latex']
        \node [circ] (a) {Data};
        \node [rect] (b) [right of=a] {Sensor};
        \node [circ] (c) [right of=b] {Memory};
        \node [circ] (d) [below of=c] {Model};
        \node [rect] (e) [left of=d] {Learner};
        \node [circ] (f) [below of=d] {+};
        \node [rect] (g) [left of=f] {Evaluator};
        \node [rect] (h) [below of=f] {Transmitter};
        \node [plain] (i) [below of=h] {}; 
        \node [plain] (j) [left of=h] {};
        \node [rect] (k) [left of=j] {Receiver};
        \node [rect] (l) [left of=e] {Controller};
        \node [plain] (m) [below of=k] {}; 
        
        \path[->] (a) edge node {} (b);
        \path[->] (b) edge node {} (c);
        \path[->] (c) edge node {} (e);
        \path[->] (e) edge node {} (d);
        \path[->] (d) edge node {$w_k$} (f);
        \path[->] (e) edge node {} (g);
        \path[->] (f) edge node {$w_k$,$I_k$} (h);
        \path[->] (g) edge node {$I_k$} (f);
        \path[->] (h) edge node {} (i);
        \path[->] (k) edge node {} (l);
        \path[->] (l) edge node {} (e);
        \path[->] (m) edge node {} (k);
        
        Background
        \begin{pgfonlayer}{background}
            \path (a)+(-1.1,+0.5) node (nw) {};
            \path (h)+(+1.3,-0.5) node (se) {};
            \path[fill=yellow!20,rounded corners, draw=black!50, dashed]
                (nw) rectangle (se);
        \end{pgfonlayer}
    \end{tikzpicture}
    \caption{User device block diagram for importance-aware \ac{FL} system. The importance measurement is calculated by an evaluator block whose output is transmitted together with the local model to the \ac{PS}.}
    \label{fig:activebd}
\end{figure}

Unlike the user selection case for \ac{CML}, the \ac{ML} model is now naturally present at the user device and the data-importance can be evaluated without the need for transmitting a compressed model to the user devices. As illustrated in Figure~\ref{fig:activebd}, the user device evaluates the importance locally and appends the data importance to the uplink packet containing the local model. At the start of each communication round, the \ac{PS} selects a fixed number of users for participation. In vanilla \ac{FL} the choice would be randomized, this scheme proposes to select the users with the highest $I_k$. Using active \ac{FL}, the proposal in~\cite{goetz2019active} achieves the same performance as vanilla \ac{FL} using 20-70\% fewer epochs. Since the data size of the importance evaluation is small in comparison to the local model, this method also has a negligible overhead.

Rather than using just the loss as the data-importance metric, \cite{leng2022client} opts to use a combination of the information entropy and the loss value. Specifically, the elements of the gradient vectors are assumed to follow a random distribution, and the entropy gradient elements is used as the data-importance metric. This way of quantifying gradient information originally comes from researchers aiming to perform completely different tasks such as fast tree approximation, community discovery \cite{ding2019gradient}, and autoencoding \cite{elkhalil2021fisher}, similarly to how data-importance measures from Active Learning had completely different original purposes. In a simulation study, the authors of \cite{leng2022client} compares user scheduling based on the gradient norm and gradient divergence to that of gradient entropy. The simulations indicate that the gradient entropy is superior to the norm and divergence when the dataset is non-IID.

\section{Review of radio resource management for federated learning}\label{sec:wirelessFL}
Because of the differences in objective between Wireless for \ac{ML} and traditional data communications, direct application of the \ac{FL} protocol without consideration of practical constraints in wireless communication systems, makes the overall training process inefficient~\cite{nishio2019client, chen2020joint, qin2021federated}. Instead, \ac{RRM} protocols should be customized for \ac{FL} to enable efficient training of \ac{ML} models using distributed data. Within \ac{RRM}, we include the allocation of transmission power, bandwidth, time slots, and user scheduling. The objective of \ac{RRM} for \ac{ML} is a learning goal, such as the classification accuracy of a model, rather than a general data communication goal, such as data-rate maximization. This difference shapes Wireless for \ac{ML} \ac{RRM} in ways that might seem contradictory compared to traditional \ac{RRM}.

\begin{figure}[t]
\centering
    \begin{tikzpicture}
        \node[inner sep=0pt] (russell) at (0,0)
            {\includegraphics[width=8.5cm]{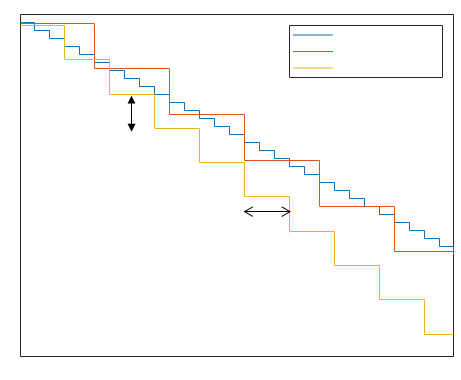}};
        \node[] at (0.2,-3.4) {Time};
        \node[rotate=90] at (-4.2,0) {Loss};
        \node[] at (-2.2,1.4) {$\Delta l$};
        \node[] at (0.7,-0.8) {$T_{\text{round}}$};
        \node[] at (2.73,2.85) {\scriptsize{Short iterations}};
        \node[] at (2.72,2.54) {\scriptsize{Long iterations}};
        \node[] at (2.88,2.25) {\scriptsize{Balanced iterations}};
    \end{tikzpicture}
    \caption{Qualitative plot illustrating \ac{FL} convergence. There is a tradeoff between communication and computation, where slower iterations means more computation and less communication. For optimal convergence, the \ac{RRM} protocol should be able to adapt to available computational resources and channel qualities.}
    \label{fig:FL-tradeoff}
\end{figure}
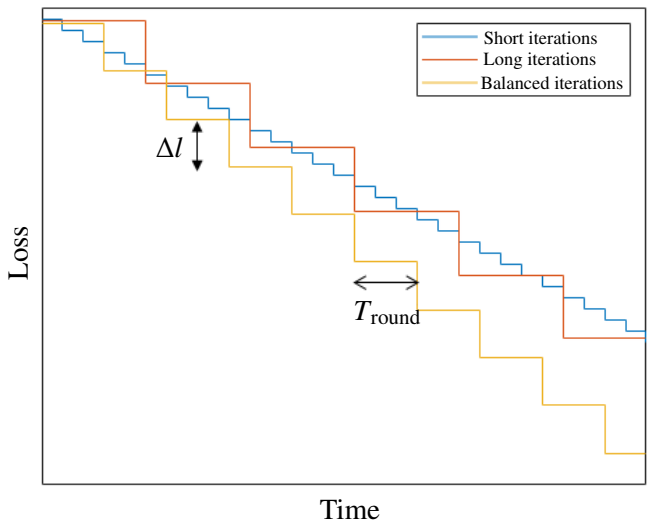

In \ac{FL}, multiple communication rounds have to be performed until the desired accuracy is reached. As is generally true for iterative algorithms, there is a tradeoff in \ac{FL} between the computational complexity of each communication round versus the total number of rounds. Specifically, the time per communication round ($T_{\text{round}}$) and the loss decay per round ($\Delta l$) must be carefully balanced, as illustrated in Figure~\ref{fig:FL-tradeoff}. Both $T_{\text{round}}$ and $\Delta l$ are impacted by the \ac{RRM} decisions. Additionally, this need for balance leads to new decisions to be made by the \ac{PS} such as: 
\begin{enumerate}
    \item Deciding how many users will participate in each round;
    \item Performing aggregation frequency control, which means to decide how many local training iterations each device performs before communicating their update;
    \item Selecting the batch size of each user device's training algorithm.
\end{enumerate}

In the following, we discuss important topics within \ac{RRM} for \ac{FL} that the literature has addressed.

\subsection{Participation maximization}\label{subsub:partic_max}
The first paper written about \ac{RRM} for \ac{FL} is on the topic of client selection~\cite{nishio2019client}. In the original \ac{FL} protocol, \mbox{FedAvg}, each communication round begins by the \ac{PS} selecting a random fraction of clients and sending them the global classifier model~\cite{mcmahan2017communication}. The authors of~\cite{nishio2019client} demonstrated the inefficiency of this selection over wireless networks, due to the heterogeneity of channel conditions in the network. If clients with poor channels are selected, the uplink transmission is slow and the straggler problem will significantly slow down the training process. Alongside channel heterogeneity, computational resource heterogeneity will lead to the same problem. 

To find a better client scheduling policy, an optimization problem is formed. Ideally, the objective function would be the classification accuracy on the test data, but there is currently no closed-form expression for this, see Eq.~(\ref{eq:convergence_rate}). Instead, the number of participants is maximized, which can serve as a rough proxy for the convergence rate~\cite{mcmahan2017communication}. The problem is constrained to exclude slow users, by introducing a deadline for the entire \ac{FL} algorithm $T_{\text{round}} \in \mathbb{R}^+$. The \ac{PS} is then subjected to a tradeoff in client selection between the number of participants in each round, and the time required to complete each round. A good $T_{\text{round}}$ value is found experimentally, and then the following problem is solved:

\begin{equation}
    \label{eq:nishio}
    \begin{aligned}
    \max_{\calS} \quad & |\calS|\\
    \textrm{s.t.} \quad & T_{\text{round}} \geq T_{\text{cs}} + T_\calS^d+T_c+T_\calS^u+T_{\text{agg}}\ ,
    \end{aligned}
\end{equation}
where $\calS$ is the set of clients selected for the round, $T_{\text{cs}}$ is the time to select clients, $T_\calS^d$ is the downlink transmission time, $T_c$ is the computational time to train the local model, $T_\calS^u$ is the uplink transmission time, and $T_{\text{agg}}$ is the time required to aggregate the local models at the \ac{PS}.
Using the proposed scheme, an extensive experimental study is conducted based on \ac{LTE} networks in a mobile edge computing context. The studies indicate that the proposed solution consistently converges faster than out-of-the-box FL regardless of the choice of dataset (Fashion-MNIST or CIFAR-10), and the distribution of the data (IID or Non-IID).

In~\cite{Xu2021}, the authors highlight a phenomenon in \ac{FL}, termed \emph{later-is-better}, in which the learning 
rounds are temporally interdependent and have varying significance towards the desired learning outcome.
When using \ac{FL} over the wireless network, the authors show that it is important to take into account this 
phenomenon when designing resource management methods to support the \ac{FL} task.
To make use of these findings, the authors formulate a stochastic client selection and bandwidth allocation problem for 
a finite number of communication rounds while considering finite energy constraints on the clients.
The problem aims to maximize the weighted sum of selected clients for a fixed number of communication rounds, whose 
weights depend on a temporal parameter to capture the significance of selecting more clients in different communication rounds.
The authors show that an increasing sequence of these temporal parameters often results in better \ac{FL} performance due to a higher number of clients being selected in later rounds of the learning convergence.
The constraints include a long-term energy budget on individual clients and feasibility constraints on the bandwidth 
allocation.
Due to the time-varying and unpredictable wireless channel conditions, the authors use Lyapunov optimization to solve 
the optimization problem and propose an algorithm, named \mbox{OCEAN}, for online client selection and bandwidth 
allocation.
In the results, the authors show that the \mbox{OCEAN} algorithm is adaptive to changing network environments and 
outperforms greatly other benchmarks that ignore the \emph{later-is-better} effect of \ac{FL}.

Differently from~\cite{nishio2019client,Xu2021}, the authors in~\cite{Xu2021b} consider a scenario with multiple 
\ac{FL} services co-existing and sharing resources in a wireless network and propose bandwidth allocation to ensure 
sufficient client participation for each \ac{FL} service.
Specifically, they propose a two-level resource management framework comprising of intra- and inter-service resource
allocation. 
The intra-service resource management problem aims to minimize the \ac{FL} communication round time by optimizing the 
bandwidth allocation among the clients within each \ac{FL} service. 
Subsequently, the inter-service resource management problem aims to distribute bandwidth resources among multiple
simultaneous \ac{FL} services.
For both problems, the authors analyse both cooperative and non-cooperative \ac{FL} service providers.
For cooperative providers, they propose a distributed bandwidth allocation solution to optimize the overall performance 
of multiple \ac{FL} services while considering the fairness among \ac{FL} services and the privacy of clients and 
providers. 
For non-cooperative providers, they propose a new auction scheme with the \ac{FL} providers as the bidders and the 
wireless server as the auctioneer, which is able to balance learning accuracy and fairness among the \ac{FL} services. 
The bid is based on the bandwidth requested by the \ac{FL} provider and the price it is willing to pay to get the requested bandwidth.
The results show that the proposed solutions outperform other benchmarks, such as equal bandwidth allocation among 
clients or services, and bandwidth allocation proportional to the number of clients for each service, for various 
wireless network conditions.

\subsection{Energy efficiency}\label{subsubsec:energy_efficiency}
Since \ac{FL} over wireless networks are mostly concerned with either mobile or sensor devices, low energy consumption is critical. In~\cite{zeng2020energy}, this topic is investigated in a joint bandwidth allocation and client selection scheme. Specifically, the energy consumption of transmitting the local model in the uplink is considered as

\begin{equation}
    E_k^{up}=b_k B p_k t_k\ ,
\end{equation}
where $b_k$ is the bandwidth allocation ratio, $B$ is the total bandwidth, $p_k$ is the power allocation in Watt/Hz, and $t_k$ is the model uploading time. The joint bandwidth allocation and user selection scheme is then found by solving:

\begin{equation}
    \label{eq:energy_opt}
    \begin{aligned}
    \min_{b_k, t_k, I_k} \quad & \sum_{k=0}^{K-1}E_k^{up} - \lambda \sum_{k=0}^{K-1} I_k\\
    \textrm{s.t.} \quad & \beta_k \in\{0,1\}, \\
    & \sum_{k=0}^{K-1}b_k = 1, \\
    & 0 \leq t_k \leq T_k
    \end{aligned}
\end{equation}
where $I_k$ is an indicator function that is 1 if device $k$ is selected, and $T_k$ is a maximum time budget for each device. Similarly to the client selection scheme of the previous section, the number of participating devices has been used as a proxy for the convergence rate of the FL model. A numerical study on the MNIST dataset suggests that the proposed scheme outperforms a baseline of selecting every possible client in energy consumption by up to 25\% with a 1-2\% loss in classification accuracy.

In~\cite{9264742}, the energy consumption for computation is considered in addition to transmission. The energy for computing the local model updates at device $k$ is
\begin{equation}
    E_k^c = \kappa A_k \log_2\left(\frac{1}{a}\right)f_c^2\ ,
\end{equation}
where $\kappa$ is the effective switch capacitance that depends on the chip architecture, $A_k$ is an approximation of the energy consumption per training iteration, $a$ is the local classification accuracy, and $f_c$ is the computation capacity of device $k$ measured in CPU cycles per second. To minimize this energy, the proposed scheme allows the \ac{PS} to control the local classification accuracy by selecting the number of local iterations per communication round and the computation capacity of user devices (presumably by giving the training task higher priority on their CPUs). It is worth to note that this paper does not consider \ac{FedAvg} but uses the distributed approximate Newton-type method (DANE) \cite{shamir2014communication} for training, in which the user devices implicitly uses the local Hessian to compute their computes. In \cite{9264742}, upper bounds on DANE convergence is used to determine the number of local iterations per communication round, thereby leading to different constraints of the radio resource management problem than for \ac{FedAvg}. Simulation results suggest that the proposed scheme significantly outperforms baseline schemes of equal bandwidth allocation, fixed CPU frequency allocation, and fixed target accuracy allocation. 

In~\cite{Dinh2021}, the authors propose a novel \ac{FL} method, named \mbox{FEDL}, to handle heterogeneous user data 
and physical resource, and employ the proposed \ac{FL} model to a resource management problem focused on the energy consumption and the communication round time. 
For the proposed \mbox{FEDL} model, the local model updates at the users minimize a surrogate function of the local objective function 
using the previous averaged global model and global gradient estimate.
The authors provide the convergence analysis and establish the convergence rate of \mbox{FEDL}, which depends on the number of epochs and global iterations.
For the resource management problem, the objective is to minimize the energy consumption and the communication round time while 
considering as variables the computation capacity of the users, the \acl{UL} communication time, the desired accuracy 
for the \mbox{FEDL} method, the controllable parameter for the local surrogate function, the communication time in one 
global round, and the time to compute one epoch.
The proposed problem is non-convex and the authors provide a solution by decomposing the original problem into three 
subproblems.
The numerical results indicate that \mbox{FEDL} outperforms \mbox{FedAvg} in various learning and wireless 
communication settings.

\subsection{Packet errors}
The papers discussed so far considered perfect \ac{CSI} and error-free transmission. In~\cite{chen2020joint} instead, the authors consider an outage model where packet-errors can happen, with error probability dependent on the allocated bandwidth and transmission power. The \ac{FL} averaging step is updated using the outage model to consider potential packet losses. With this new averaging step, an upper bound on the learning convergence is derived, that reveals the impact of packet errors on the training loss. Using this upper bound, the authors design a joint user selection and bandwidth/power allocation scheme, which converges despite the errors, but after convergence, the following optimality gap remains
\begin{equation}
    \mathbb{E}\left[l(\vtW^t)\right]-\mathbb{E}\left[l(\vtW^\star)\right] = C\sum_{i=1}^KN_k(1-I_k+I_kq_k(b_k, p_k))\ ,
\end{equation}
where $l(\vtW^\star)$ is the loss of the optimal model, $C$ is a constant depending on the number of training samples and the Lipschitz parameter of the loss function, $N_k$ is the number of training samples at device $k$, $I_k$ is an indicator that is 1 if device $k$ is scheduled, and $q_k(b_k, p_k)$ is the probability of packet error given the bandwidth and power allocation. This result shows that proper bandwidth and power allocation reduces the optimality gap, leading to better results after convergence. 

Similar to~\cite{chen2020joint}, the authors in~\cite{Salehi2021} consider a transmission success probability, complementary to the probability of error, which impacts the client scheduling policy and convergence analysis.
The \ac{FL} averaging step uses the success probability together with the scheduling policy and sends in the 
uplink the difference between the local model after $E$ epochs, $\vtW_{k}^t(E)$, and the global model of the current communication round, $\vtW^t$.
The transmission success probability for each device is derived using stochastic geometry tools in a cellular wireless 
network considering a fixed number of transmission attempts in the \acl{UL}.
The authors study two scheduling policies to allocate $M$ resource blocks: the first using uniform sampling of devices without 
replacement, and the second using a sampling of devices with predefined probability $\{\hat{q}_k\}$ with replacement.
Subsequently, they also propose a suboptimal scheduling policy to improve the convergence rate.
The authors derive the convergence analysis via an upper bound on the learning convergence and show that unsuccessful 
transmissions do not affect the convergence rate significantly after proper adjustment of the averaging step. 
They also show the impact of the number of local epochs, communication rounds, and transmission 
attempts on the convergence rate. 
Among the interesting results of~\cite{Salehi2021}, the authors prove and show numerical results that other schemes, which 
do not include the transmission success probability in the global model update step, may converge to the solution of a 
different \ac{FL} problem, specifically biased towards the model of devices with high success probabilities. To avoid such a bias, \cite{Salehi2021} proposes to weigh the model update contribution of devices based on their probability of packet loss.

To improve the communication efficiency of \cite{chen2020joint}, \cite{jin2022communication} adopts the idea of SignSGD over a lossy wireless network. This is similar to the \ac{DML} algorithm that was considered in \ref{sec:digital_comac}, but rather than to enable digital \ac{AirComp} it is used to increase communication efficiency. Since only one bit per element of the gradient vector need to be transmitted, SignSGD is over an order of magnitude more communication efficient than standard 32-bit elements. This efficiency comes at a cost of representing the gradient more coarsely, which intuitively should slow down convergence. However, such intuition is not always right. In fact, SignSGD has been proven to converge with a theoretical rate similar to or in some circumstances even better than standard \ac{SGD} \cite{bernstein2018signsgd}. With this SignSGD scheme, the authors of \cite{jin2022communication} attempt to minimize the outage probabilities and maximize the number of communication rounds, while maintaining an energy consumption constraint. Simulation results show that the proposed scheme can achieve both higher classification accuracy (1-3\%) and lower energy consumption (10-50\%) than vanilla FedAVG.

\subsection{Total time minimization}
In the previous papers, the proposed \ac{RRM} schemes were greedy algorithms in the sense that they only optimized for the current communication round. Instead,~\cite{shi2020device} proposes to minimize the total time of the entire \ac{FL} process, from the first communication round until convergence. 

The proposed solution is a joint bandwidth allocation and client scheduling protocol which is formed by minimizing the product of the total number of communication rounds and $T_{\text{round}}$. The problem is solved by decomposing the problem into one client scheduling sub-problem and one bandwidth allocation sub-problem. The reason for the decomposition is that the client scheduling problem is a combinatorial optimization problem, which is infeasible to solve exactly. Experimental results on the MNIST dataset compare the scheme to~\cite{nishio2019client} and show that the classification accuracy can be significantly improved.

Differently, the authors in~\cite{Chen2021} aim to minimize the total convergence time that depends on the FL parameter transmission delay per iteration and the number of iterations that FL requires to converge.
In this problem, the authors consider a user selection matrix and a resource block allocation matrix as variables, 
which directly impacts the users participating or not in the training.
The authors propose a probabilistic user selection, to schedule users that have a high impact 
in the global \ac{FL} model, and an uplink resource block allocation, given the user selection.
To further reduce the total convergence time, the authors use a neural network to estimate the local \ac{FL} models of 
users that did not receive a resource block and use these estimated models to improve the convergence speed.
The numerical results indicate a reduction in the \ac{FL} convergence time of $56\%$ and improvement in the 
accuracy of $20\%$ when compared to an \ac{FL} algorithm that randomly determines the subset of selected users and resource blocks allocated to each user for FL parameter transmission.

\subsection{Empirical classification error}
Although the ultimate goal of these \ac{RRM} algorithms is to reach the highest possible classification accuracy under communication constraints, none of the protocols maximize the accuracy directly. There still exists a gap in \ac{ML} theory, which is a closed-form expression for the relationship between the number of training samples and the classification accuracy. In this survey, we have seen multiple examples of getting around this gap by using other metrics as proxies for classification accuracy. In~\cite{wang2020machine} instead, the use of an empirical function is proposed to model how the accuracy depends on the sample size. The empirical function of the classification error $\Theta(N)$ with respect to the number of training samples $N$ is designed to satisfy three properties:
\begin{itemize}
    \item The classification error is a percentage that must lie within $0 \leq \Theta(N) \leq 1$;
    \item More data provides more information, and thus $\Theta(N)$ should be a monotonically decreasing function of $N$;
    \item As $N$ increases, the magnitude of the derivative $\partial \Theta(N)/(\partial N)$ should gradually decrease and eventually go to zero, since infinitely increasing the sample size should not improve the classification error.
\end{itemize}
Based on these properties, the function $\Theta(N) = a \cdot N^{-b}$ is chosen, where $a$ and $b$ are tuning parameters. The function is then trained with a limited number of training samples, the classification accuracy is tested, and this data point is used to fit the tuning parameters. By repeating this process, samples on classification accuracy are gathered, and the parameters are found via nonlinear least-squares fitting. After fitting, the function is used as the objective for an optimization problem to find an \ac{RRM} scheme. In a numerical study, the resulting \ac{RRM} scheme is compared to the classic max-min fairness and sum-rate maximization protocols. When training a classifier for the MNIST dataset, the proposed scheme outperformed both baselines by a classification accuracy of about 1-2\%. If the same classification accuracy is targeted, the proposed scheme saves at least 30\% transmission time compared to both baselines.

\subsection{Federated Distillation}
In Section \ref{subsec:aircomp_distillation}, we discuss a \ac{DML} scheme known as \ac{FD}. There, the model outputs are combined in the uplink direction via \ac{AirComp} to reach exceptional communication efficiency. In this section, we are instead considering a novel Federated Distillation approach that uses digital communication. In \cite{9121290}, the authors consider uplink-downlink asymmetric channels, where the uplink channel capacity is more limited than the downlink. Since the downlink channels are limited, the cheap communication of model outputs in \ac{FD} makes sense. However, in the more powerful downlink channel, it would be better to communicate more information than what is contained in model updates, considering that pure \ac{FD} sacrifices accuracy to pay for the communication efficiency.

Therefore, \cite{9121290} proposes a scheme that communicates model outputs in the uplink (as in \ac{FD}) and model parameters in the downlink (as in \ac{FL}). To achieve this, a method known as \ac{FL} after distillation \cite{park2019distilling} is utilized. Specifically, this means that the server converts uploaded model outputs to \ac{ML} model parameters, using a process known as knowledge distillation \cite{hinton2015distilling}. In addition to model outputs, this process requires additional training samples from the user devices, which violates user privacy. Therefore, \cite{9121290} utilizes a mixup scheme to obscure the original samples, in which the idea is to create locally superpositioned samples using the mixup algorithm \cite{zhang2017mixup} which provides realistic synthetic samples for the knowledge distillation process without sacrificing too much privacy. In a numerical study, the proposed mixup scheme achieves 42.4x smaller payload size than \ac{FL}, which leads to significantly more communication rounds for a fixed period of time. As a result, their proposal achieves up to 16.7\% higher classification accuracy than \ac{FL}.

\subsection{Batch size selection}
In~\cite{ren2020accelerating}, the authors decided to include the selection of batch size among the decision variables for the \ac{RRM}. The motivation is based on the aforementioned straggler effect (see Section~\ref{subsec:digital_primer}), which causes the slowest device to act as a bottleneck. By giving the \ac{RRM} control over the batch size, this situation can be improved in two ways. The fastest devices of the network can be asked to train with a larger batch size ($N^t$ in Eq.~(\ref{eq:general_sgd})), thus increasing the accuracy of their gradients without decelerating the \ac{FL} process. Similarly, the batch size of the slowest devices can be decreased, sacrificing some of their performance to accelerate the \ac{FL} process. The scheme improved the classification accuracy by approximately 2\% compared to both random selection of batch sizes and uniform selection of batch sizes.

\subsection{Importance-aware radio resource management}\label{subsub:import_rrm}
The proposal in~\cite{ren2020scheduling} is a user selection scheme taking both channel fading and data importance into account. Similar to how~\cite{nishio2019client} uses the number of scheduled users as a proxy for convergence rate,~\cite{ren2020scheduling} uses data importance. The optimal user selection is found by the following optimization problem:
\begin{equation}
    \label{eq:importancerrm}
    \begin{aligned}
    \min_{p_1,p_2,...,p_K} \quad & \sum_{k=1}^Kp_k\left(\rho (-I_k)+(1-\rho)T_k\right)\\
    \textrm{s.t.} \quad & \sum_{k=1}^Kp_k = 1\ ,
    \end{aligned}
\end{equation}
where $p_k$ is the probability that $k$ is scheduled, $I_k$ is the importance of the gradient at device $k$, and $T_k$ is the time for device $k$ to upload its gradient. In this case, they use the gradient divergence as the importance measurement, i.e., $I_k = ||\nabla F_k(\vtW^t)-\nabla f(\vtW^t)||^2$. Note that the importance is negative since we want to maximize importance but minimize latency. The solution to this problem strikes a balance between data importance and channel quality, where the weight between the two is controlled by $\rho\in[0,1]$. When training an MNIST classifier, the channel and importance aware user scheduler outperformed a channel-based scheduler both in convergence rate and final classification accuracy. The simulation results suggest a decrease of less than half the convergence time and an improvement of up to 2\% higher in the final accuracy.

Differently from~\cite{nishio2019client,ren2020scheduling}, the authors in~\cite{Xia2021} introduce scheduling policies 
that use novel update importance and latency policies for client scheduling to reduce the required number of 
communication rounds and the total time in a communication round.
The update importance policy is based on two sub-metrics: update staleness and update drift.
The update staleness 
measures the staleness associated with the local updates of each client and aims to keep the local updates as fresh as possible.
The age of update rule on client $k$ for communication round $t+1$ is defined as $a_k(t+1) = (a_k(t) + 1)(1-s_k(t))$, 
where $a_k(t)$ is the age of the local update in round $t$, and $s_k(t)$ is a binary indicator that equals $1$ if 
client $k$ receives the global model in round $t$, i.e., if the wireless channel is above a predefined threshold for the signal detection, and $0$ otherwise.
The update drift 
is based on the distance, either the Manhattan or the Euclidean distance, between the local model and the global model.

For the latency-based policy, it considers a long-term fairness constraint to allow fair participation among clients 
that may have important data while having a bad channel condition.
The results show that the proposed scheduling policies achieve a higher accuracy than \mbox{FedAvg} with random 
scheduling and that different policies are recommended for different goals. 
To reduce the number of communication rounds, a scheduling using update importance metrics is recommended; whereas to 
shorten the total time in a communication round, a schedule using latency metrics is recommended.

In~\cite{MohammadUpdateAwareISIT2020}, the authors analyse the significance of the local models and the quality of the channels over the wireless multiple access channel from the users to the PS as user scheduling metrics. 
The main idea is to share the limited wireless resources with the users that have significant contribution to the model, rather than all the users.
As a result, users with more significant updates can have more resources and can transmit their updates more accurately.
On the other hand, users with very bad channels may not be able to communicate their updates accurately unless they are allocated a relatively significant portion of the resources; it is irrelevant if the updates are significant or not.   
It is shown numerically in \cite{MohammadUpdateAwareISIT2020} that considering both these metrics in user scheduling results in a better performance than considering each metric individually. 
The authors extend this result by deriving a convergence rate in \cite{MohammadConvUpdateAwareTWC2021} that corroborates the experimental results.

\subsection{Energy harvesting and power transfer}\label{subsubsec:energy_harvesting}
One promising solution to overcome the energy limitations in \ac{IoT} is energy harvesting, which allows devices to 
harvest \acl{RF} energy when communicating with a \ac{PS}~\cite{Clerckx2021}.
In \ac{FL} over wireless \ac{IoT}, the dowlink transmission of the aggregated model parameters from the \ac{PS} to the \ac{IoT} devices could be used to provide energy to the devices. 
Hence, the use of energy harvesting for \ac{IoT} devices with \ac{FL} would be a perfect combination.
However, how to allow the devices to harvest sufficient energy to train a \ac{FL} model while not substantially increasing the communication round time is largely an open question.
The use of energy harvesting for \ac{FL} is highly novel and to the best of our knowledge, there are only two works in the literature~\cite{Zeng2021,Silva2021}.

In~\cite{Zeng2021}, the authors consider a \ac{FL} application in which a wireless network uses power-beacons to transfer \acl{RF} energy.
The key components of the work are the distributed gradient estimation, local-computation optimization, and optimal 
learning-wireless power transfer tradeoff.
The distributed gradient estimation is related to the convergence of the \ac{FL} method based on the mini-batch size of devices, number of active devices, and computation-outage probability. 
The computation-outage is an event in which a device does not harvest more energy than the necessary to transmit, thus not being to able to transmit.
The local-computation optimization aims to minimize the local gradient deviation present in the expected convergence rate expression, 
whose solution is accomplished through the optimization of the mini-batch size and processor clock frequency.
Then, the authors derive an optimal learning-wireless power transfer tradeoff, which shows that a higher density of power beacons improves the learning convergence and the local gradient deviation.
Moreover, it provides scaling laws of the convergence rate with respect to the transferred energy and the devices’ computational capacities.

\begin{figure}[t]
\centering
\includegraphics[width=0.75\linewidth,trim=0mm 0mm 0mm 0mm,clip]{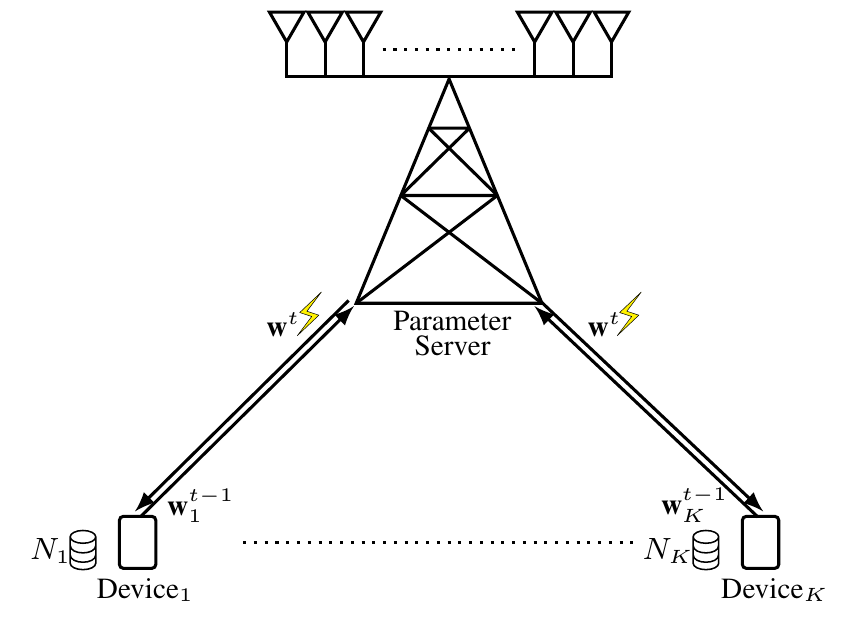}
\caption{An example of a network employing \ac{FL} and \acl{SWIPT} with $K$ devices. The devices send the local model 
$\vtW_k^{t-1}$ in the \acl{UL} in one time slot, while the edge server sends the global model $\vtW^t$ and energy in the \acl{DL} on the subsequent time slot.}
\label{fig:harv_federated}
\end{figure}

The authors in~\cite{Silva2021} analyse a multi-antenna \ac{PS} using the \acl{SWIPT} technology for \ac{IoT} devices.
The scenario considers \ac{FL} simultaneously training a learning model while communicating with a \ac{PS} 
(see Figure~\ref{fig:harv_federated}). 
The authors consider the use of \mbox{FedProx}~\cite{Li2020b}, a recent generalization of \ac{FL} that allows to 
optimize the number of local iterations at each device, while guaranteeing convergence to (non-)convex learning tasks.
The work aims to minimize the number of communication rounds and communication round time while optimizing the 
number of local iterations, the time to transmit/receive, and to harvest a percentage of the total energy spent at each 
round and device.
From the energy harvesting literature~\cite{Timotheou2014}, the \ac{MRT} beamforming is better at harvesting energy than the \ac{ZF} beamforming while \ac{ZF} is better at providing higher rates than \ac{MRT} due to the interference cancellation.
Hence, it is non-trivial to decide the beamforming method due to a possible increase in the communication round time if the devices do not have sufficient energy to harvest or do not have sufficient rate to transmit the model parameters.
Due to this reason, the authors consider \ac{MRT} and \ac{ZF}, and analyse which method is more suitable for energy harvesting within \ac{FL}.
The results indicate that the test accuracy using either \ac{MRT} or \ac{ZF} with the optimization of the local number of iterations outperform a solution without such optimization.
Moreover, it shows that \ac{MRT} vastly outperforms \ac{ZF} in terms of minimum communication round time for all the 
percentage of the energy harvesting required.

\subsection{Noisy downlink}
Although the PS typically has access to more resources than the edge users, it is essential to consider imperfect transmission over wireless networks, where the PS shares the global model with the users for local training.
In this case, users may not receive the global model available at the PS accurately, and the analysis of the convergence behaviour of FL should account for noisy version of the global model at the users.  
Digital transmission of the model over a bandwidth-limited noisy downlink leads to a relatively coarse estimation of the global model at the devices since the model vector has a high empirical variance, and quantizing the model itself does not provide an accurate estimate.
Therefore, it is suggested in \cite{ExpandingRedClientResFL} to project the model vector linearly using a random matrix before quantizing it.
This random linear projection spreads the information in the model vector more evenly across its dimensions, and leads to a smaller empirical variance.
Then, the PS quantizes the projected model and broadcasts the quantized vector over the downlink, where the users recover the actual model from the quantized vector having knowledge about the random matrix employed at the PS.
In a follow-up work, \cite{DoubleSqueezeTangConf} suggests to compress the model itself while accumulating and compensating the quantization error.
This may lead to a coarse estimate of the model in the users if the downlink capacity is not large enough, in which case the model is compressed with a relatively low quatization level.  

It is shown in \cite{MDSV_NIPS2020} that the global model update, with respect to the last global model estimate available at the devices, has significantly less empirical variance than the global model.
As a result, quantizing the global model update provides a more accurate estimate rather than quantizing the global model itself for the same quantization level.
The authors in \cite{MDSV_NIPS2020} introduce quantizing the global model updates at the PS with respect to the last model available at the users.
This approach provides a significant improvement over the ones introduced in \cite{ExpandingRedClientResFL,DoubleSqueezeTangConf}, which is due to the availability of a more accurate estimate of the global model at the users.
This approach is extended in \cite{MSV_DownUserSel_Conf2021} by considering broadcasting different global model descriptions to different users based on the broadcasting capacity region such that the users with better capacities receive a more accurate estimate of the global model.
This introduces a new user scheduling metric, which is based on the downlink capacity, through which at each iteration only the devices with relatively good channels, i.e., better global model estimates, can be selected to participate in the training.   

It is worth highlighting that, as studied in \cite{MDSVNoisyDownlinkJournal2020}, analog transmission of the model from the PS allows different devices to receive different noisy copies of the model, where less noisy devices receive a better version of the model.
As such, devices with more accurate estimates of the model can compensate the lack of accurate estimates of the model at noisier devices when averaging the local models transmitted over uplink. 
This may lead to performance improvement compared to digital transmission of the global model from the PS \cite{MDSVNoisyDownlinkJournal2020,JinHyunAhnFLNoisyDownlink}. 

\subsection{Federated meta-learning}
Within certain narrow fields of \ac{ML}, state-of-the-art systems are in parity with or even beyond human capabilities, such as playing the game of Chess and Go \cite{bory2019deep}. However, to reach such capabilities, state-of-the-art \ac{ML} systems require significantly more exposure to data than a human. For instance, the training process of AlphaGo included approximately 600 billion moves of Go to train the value network \cite{lee2016human}. If a human plays for 8 hours every day of its life, spending an average of 10 seconds per move, it would take the human more than 500,000 years to play 600 billion moves.

To address this efficiency gap, the field of meta learning was born \cite{finn2017model}. In meta learning, the goal is to train a parameterized algorithm that, in turn, is used to train \ac{ML} models, i.e., the parameterized algorithm is learning to learn (meta learning). Practically, the fundamental difference between meta learning and standard \ac{ML} can be expressed as the division of testing and training cases. In standard \ac{ML}, the dataset is divided into training data and testing data, where the training data is used to train the model and the testing data is used to evaluate its performance. In meta learning, there is instead a collection of tasks, which are divided into training tasks and testing tasks, where the tasks are generally non-overlapping, e.g., one task might be to classify different animals and another to classify plants. The idea is that the training tasks are used to train the parameterized learning algorithm, which learns to detect common structures among the non-overlapping tasks. The testing tasks are then used to evaluate how well the learning algorithm trains \ac{ML} models on the previously unseen tasks using just a few data samples.

In the space of \ac{DML}, \ac{FML} is a recently proposed framework for achieving fast learning with distributed data \cite{chen2018federated}. The \ac{FML} framework leverages the data of multiple devices to train the parameterized learning algorithm. This algorithm can then be used by the participating devices to train an \ac{ML} model but more importantly, new devices can be given the parameterized learning algorithm upon joining the network so that they can quickly train an \ac{ML} model using just a few data points. The underlying assumption here is that the devices carry data for a similar class of tasks, e.g., image classification, but with non-overlapping tasks within that class. In \cite{9681911}, the \ac{FML} framework is brought into the wireless setting. First, the authors claim that the uniform selection of devices in each round (which is part of vanilla \ac{FML}) leads to slow convergence rates. Then, they propose a non-uniform device selection scheme that maximizes a lower bound on the convergence speed of \ac{FML}. In the same paper, the model is also extended to a joint user device selection and \ac{RRM} problem. The paper contains both theoretical insights in terms of convergence bounds and numerical results that reveal strictly lower losses for the proposed system compared to a greedy and random \ac{RRM} baseline.

\chapter{Open problems}\label{sec:future_res}
The current literature on Wireless for ML has demonstrated that many critical metrics can be substantially improved by tailoring wireless network protocols to support \ac{ML}, including latency, classification accuracy, energy consumption, and spectrum efficiency. However, the literature is still young and there are fundamental problems that remain unsolved. In this section, we give a brief insight into these open problems to inspire future research.

\section{Over-the-air computation}
\ac{CoMAC} for \ac{ML} is an exciting area of research since it offers a radically new way to think about wireless protocol design. However, the divergence from digital communications poses challenges of incompatibility with standard hardware and lack of prior experience. There is a need for careful investigation of assumptions in the theory and extensive testing in practice. If these challenges are overcome, great bounties await in the form of massively improved spectrum efficiency, approximately proportional to the number of participating devices.

\subsection{Digital over-the-air computation}
As explained in Section \ref{sec:digital_comac}, a recent work proposed a digital \ac{CoMAC} protocol, based on one-bit quantization of gradient elements and \ac{BPSK} modulation~\cite{zhu2020one}. The proposal carries great importance for the practical implementation of \ac{CoMAC} since it is compatible with the digital wireless transceivers we are using today. However, there are two potential issues with the scheme that should be investigated further.

First, \ac{BPSK} demands more precise synchronization than comparable analog schemes. For example, \cite{goldenbaum2013robust} showed that analog \ac{CoMAC} can be achieved with just coarse block-synchronization by encoding its real-valued message in the transmit power of a series of random signal pulses. In contrast, the \ac{BPSK}-based scheme is dependent upon constructive and destructive interference of phase modulated signals to represent the transmission of "+1" and "-1". Such a scheme requires very precise alignment of the analog waveforms, which may be unreasonably difficult or expensive to achieve in practice \cite{goldenbaum2013robust}.

Second, the restriction of using one-bit quantization of the gradient elements could pose problems. The numerical study in \cite{zhu2020one} found that the classification accuracy of one-bit quantization was comparable to analog communication, but this could easily change depending on the properties of the wireless network. As learning bounds on over-the-air \ac{FL} demonstrates, noisy estimations of the local models slows down convergence and harms the final accuracy of the model \cite{sery2020analog}, and the combination of quantization noise and channel noise can yield undesirable results.

\subsection{Channel state information}
As we have seen in Section~\ref{sec:comac} and~\cite{goldenbaum2014channel,dong2020blind, abari2016over}, the \ac{CSI} acquisition effort is greater for over-the-air computation than for digital communications. Multiple solutions have been devised to solve the issue, such as blind estimation using either \ac{MIMO} or \acp{IRS}. However, there are still open questions related to \ac{CE} that remain unaddressed. In particular, the current literature assumes the availability of perfect \ac{CSI}, which allows for perfect inversion of the channel. In reality, noisy \ac{CSI} will lead to distorted sums. Instead of the channel inversion in Eq.~(\ref{eq:preequalizedsum}), the received vector will be
\begin{equation}
    \label{eq:imperfectCSI}
    \sum_{k=1}^Kh_kz_k+v=\sum_{k=1}^K\frac{w_kh_k}{\hat{h}_k}+v,
\end{equation}
where the estimated channels $\hat{h}_k$ do not cancel out $h_k$. To understand the effect of imperfect \ac{CSI} on learning performance, this needs to be studied. Additionally, the performance comparisons of \ac{CoMAC} and digital communications have not considered the cost of \ac{CSI} acquisition, which could be a non-negligible difference due to the increased channel estimation effort.

\subsection{Security}
A fundamental consequence of \ac{CoMAC}, is that it is impossible to see who is transmitting model updates in the uplink. This can be seen as a blessing or a curse. The upside is that user privacy is guaranteed, stopping the potential for model inversion at the \ac{PS}~\cite{fredrikson2015model}. The downside is that it opens up for potential adversaries to corrupt the training process. Because of the inherent anonymity of \ac{CoMAC}, it is easy for an adversary to send malicious model updates and harm training. This process is known as model poisoning and has received some attention from the \ac{FL} community~\cite{bhagoji2019analyzing,fung2018mitigating,bagdasaryan2020backdoor}. However, the defense strategies proposed in the literature depend on detecting anomalies in individual model updates, which is impossible for \ac{CoMAC}. Hence, there is a need to find new strategies against model poisoning that work without seeing individual model updates. One possible countermeasure is the consideration of coded computing, but so far there is only one paper which would be applicable to \ac{CoMAC} \cite{sifaou2021robust}. Another idea is briefly mentioned in~\cite{zhu2018broadband} where all legitimate devices are assigned a common secret spreading code. Consequently, the \ac{PS} can exclude devices that are not using the secret code. However, despite these initial steps, the security problem of over-the-air \ac{FL} is far from solved.

\subsection{Self-aware power control}
The power control schemes developed for \ac{CoMAC} are all reliant upon an assumption of random messages being transmitted by the devices \cite{cao2020optimized, cao2021optimizedfl, zang2020over, liu2020over}. Such an assumption is made to reflect that the transmitting devices are unaware of the messages to be sent by other devices in the network. However, for mathematical simplicity, these schemes are not only assuming that other devices' messages are unknown but also the message of the transmitting device itself. In practice, each device of course knows the message it is about to transmit, therefore there is room to improve the power control by taking this information into account. Since these schemes use analog modulation, the strength of the transmitted signal depends on the value being sent, and therefore the knowledge of this value should change the optimal transmission power.

\section{Digital communications}
Today's digital communication systems are optimized for communication metrics such as data rate, packet error rate, latency, or fairness. These metrics are in some way beneficial for the goals of \ac{ML} but are not completely aligned. Instead, digital Wireless for ML systems should optimize metrics such as classification accuracy, data importance, or training time. In the current Wireless for ML literature, we have seen that customized retransmission and \ac{RRM} protocols generate significantly better \ac{ML} models than generic communication protocols. However, since machine learning performance is difficult to predict ahead of training, it is not clear what the correct objective of these protocols should be, leaving us with proxies for classification accuracy, such as data importance, user participation, or bounds on the learning loss. A deeper theory of these objectives and the interplay between communication and learning is needed.

\subsection{Data-importance metrics}
In most Wireless for \ac{ML} scenarios, the acquisition of data from user devices is the bottleneck of training. Therefore, the selection of which data points to collect or which devices to schedule is of critical importance to efficiently train an \ac{ML} model. In much of the current literature \cite{liu2019wireless,liu2020data,inagaki2019prioritization,shinkuma2019data,ren2020scheduling}, this selection is based on data-importance metrics from the field of Active Learning. The original problem studied in Active Learning was that of labeling data samples but there are important differences between the problem of labeling data samples and communicating them, which opens up for new research directions. Specifically, we have listed two such differences below:
\begin{itemize}
    \item In most Wireless for \ac{ML} scenarios, the labels are available at the user devices. By using importance metrics from Active Learning as-is, potentially valuable information (the labels) is completely unutilized. This calls for the investigation of new importance metrics which incorporates the label;
    \item In \ac{DML}, the devices do not communicate data samples but local models, model updates, or gradients. However, for the sake of device scheduling, we are still interested in the importance of the update. In one paper, the local loss was proposed as a measure of gradient-importance \cite{goetz2019active} but no more work has been done in this direction. This measure could potentially be used to improve \ac{RRM} for \ac{ML} and other metrics for gradient-importance could be developed.
\end{itemize}

\subsection{Data-importance staleness}
In several importance-aware \ac{RRM} schemes, the data-importance is not updated in every communication round. For instance in~\cite{goetz2019active}, the data importance is measured on a user basis and is calculated locally during training to be transmitted in conjunction with the local model on the uplink. However, only a subset of users is selected for any given round, leaving the \ac{PS} with a mix of old and fresh data importance measurements. As the global model is trained, the importance of a user's data could change substantially. This calls for further studies on the effect of data importance staleness on learning convergence, and eventually solutions to combat this effect.

\subsection{Channel uncertainty}
Despite the strong progress on developing \ac{RRM} schemes for \ac{FL}, there are still fundamental questions that are unanswered. One example is the impact of channel uncertainty on the learning convergence. In practical systems, the \ac{RRM} decisions will always be based on an imperfect estimate of the wireless channel and the impact of this uncertainty on these systems is still unexplored. Despite affecting the \ac{RRM} decision, channel uncertainty will also have an impact on the packet error rates, which will thus worsen the optimality gap~\cite{chen2020joint}. A recent work~\cite{wadu2020federated} has taken a first step to address imperfect \ac{CSI} but more work is needed.

\subsection{Energy harvesting for federated learning}
With the increasing use of IoT devices for monitoring applications, the importance of energy harvesting for \ac{FL} is 
quickly increasing. 
The works we discussed~\cite{Zeng2021,Silva2021} are the first attempts to analyse this emerging field, but substantial 
work is still necessary.
Specifically, the impact of its application with bandwidth limited transmissions, such as narrowband IoT, which limits 
the transmission rate for the IoT devices. 
Moreover, the impact of \ac{CSI} errors in the process also needs to be considered given that the errors will impact 
the learning accuracy and may imply the need for retransmissions. If retransmissions are needed, this may also be 
beneficial for the energy harvesting of the devices, but will impact the ultimate convergence time of the process.
Hence, there is a tradeoff in terms of retransmissions, in case of \ac{CSI} errors, energy harvesting, and learning 
accuracy.

\section{Problems relevant to analog and digital communications}
An important missing piece of analytical performance evaluation of FL over wireless networks is its gap to the centralized learning, where the entire data is available at a single server carrying out all the processing. 
FL over wireless networks suffers from unreliable communications between the nodes in addition to the various heterogeneity aspects that exist with the FL framework.
This gap should capture the impact of various factors that exist with the FL framework due to its distributed nature and communications over noisy channels.
It would be particularly interesting to analyze the impact of noisy communications on the performance gap to the centralized learning.

\chapter{Applications}\label{sec:applications}
The term Wireless for ML is meant to capture any wireless technology tailored to solve a machine learning problem, including model training, data collection, and inference. However, the current Wireless for ML literature is almost exclusively focused on supervised learning using a distributed data set. Therefore, the work we have surveyed in this article applies to any application that falls within that domain, given that the data-collecting devices are connected via a wireless link. There are already a number of such applications envisioned or used in practice, such as Vehicular Internet of Things \cite{DuWu2020}, \ac{FL} for wireless \cite{niknam2020federated}, environmental monitoring \cite{park2021large}, mobile keyboard prediction \cite{hard2018federated}, and Industrial IoT \cite{nguyen2021federated}. Besides the current applications, Wireless for ML argue for the creation of foundations of an infrastructure for \ac{DML}. Such an infrastructure will be able to support many upcoming applications that we cannot envision today. In this section we expose a few current applications to discuss the challenges they pose and how Wireless for ML addresses those challenges.

\section{Smart city}
\begin{figure}
\centering
\includegraphics[width=0.80\textwidth]{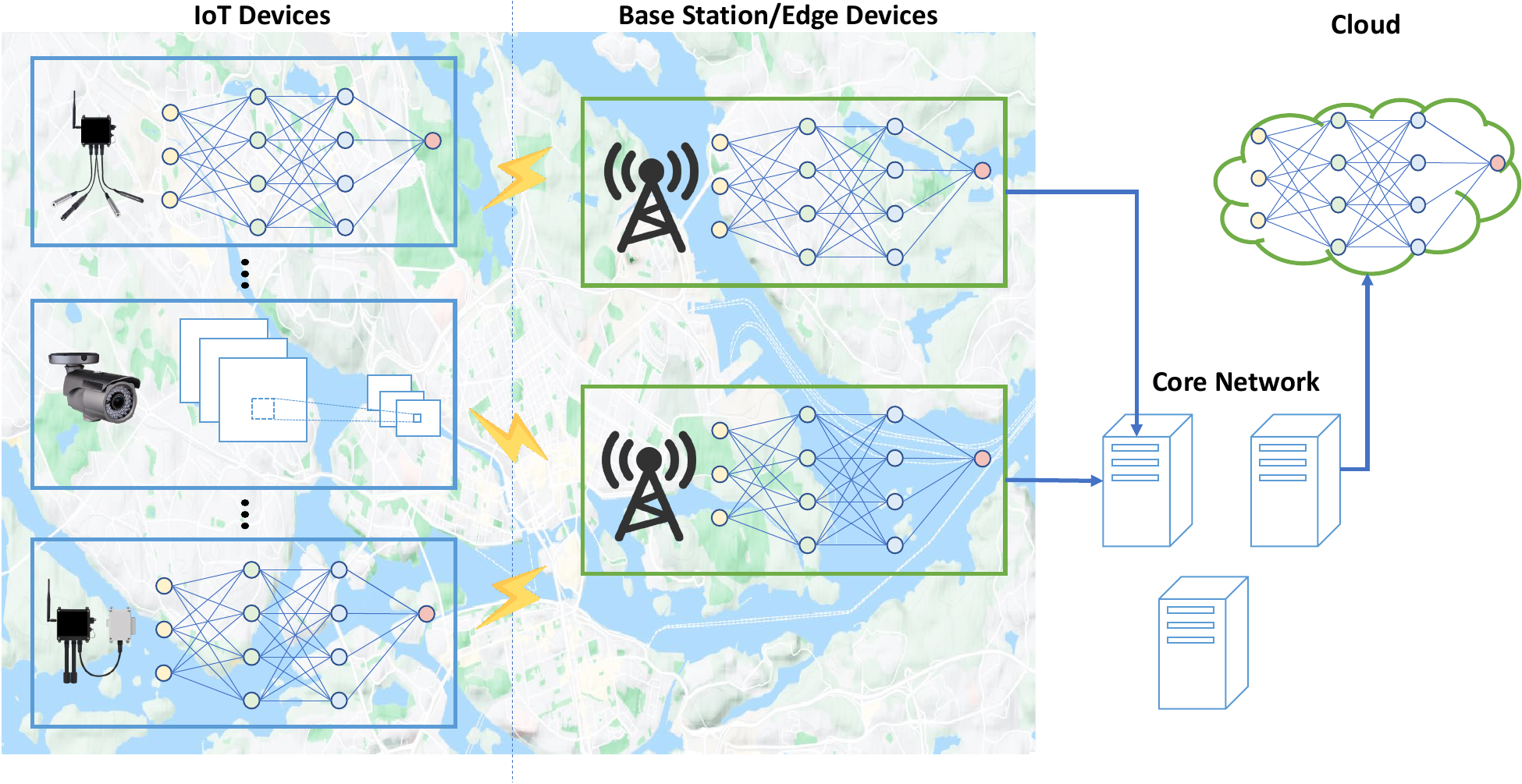}
\caption{An example of wireless IoT for ML monitoring in smart cities, including IoT devices for water monitoring, 
security surveillance, and mobile monitoring.} 
\label{fig:smart_cities}
\end{figure}

The future smart cities critically depend on the reliable monitoring of large civil infrastructures such as roads, 
tunnels, bridges, water networks, renewable energy sources, or smart electrical grids.
The denser we can measure relevant information in space and time, the higher is the potential to perform an accurate 
monitoring. 
Recently, IoT is becoming instrumental to perform such fine-grained monitoring and is opening the potential for 
several new monitoring services. 
Although IoT devices can collect a large amount of data, it is challenging to have sustainable, secure, and reliable 
monitoring services. 
To overcome such challenges, a key promising solution is the use of \ac{ML} over the IoT devices in a distributed manner across the wireless network, as illustrated in Figure \ref{fig:smart_cities}. 

Using data-driven and model-based solutions to perform reliable data analysis, it is possible to establish a 
methodology for scalable, resource-efficient learning and decision making under physical, communication, and security 
constraints. 
With the increase in the computation capacity of sensors, it is now possible to consider a scenario in which the IoT 
devices perform part of the learning and/or prediction tasks locally and at the cloud or edge server. 
Using \ac{DML} across the wireless network, the IoT devices may reduce the need to transmit a large 
amount of data to the network, alleviate the storage and energy consumption due to less intensive transmission needs, 
and enhance privacy by not transmitting the raw data over the network. 
For example, the authors in~\cite{Du2020} propose model compression for IoT devices monitoring water conditions in 
Sweden. 
The proposed model compression shows a degradation of $2.5\%$ in test accuracy while saving $96\%$ in transmissions 
compared to a scheme that sends all raw data.

Many Smart City \ac{IoT} nodes will be placed in inaccessible or remote locations, such as chimneys, water pipes, lakes, and underground. As such, there is a large cost associated with performing maintenance on these devices, including charging or replacing the battery. The results from Section \ref{subsubsec:energy_efficiency} suggest that \ac{RRM} for energy-efficient learning can significantly prolong the battery life of such devices. While Section \ref{subsubsec:energy_harvesting} suggests that energy harvesting can be leveraged to completely compensate for the consumed energy by increasing the communication round time. Therefore, the use of Wireless for ML can help to learn and predict relevant phenomena in critical infrastructures 
of smart cities, such as water leakage in water distribution networks and structural problems in the road infrastructure.

\section{Vehicular communication}
\begin{figure}
\centering
\includegraphics[width=0.80\textwidth]{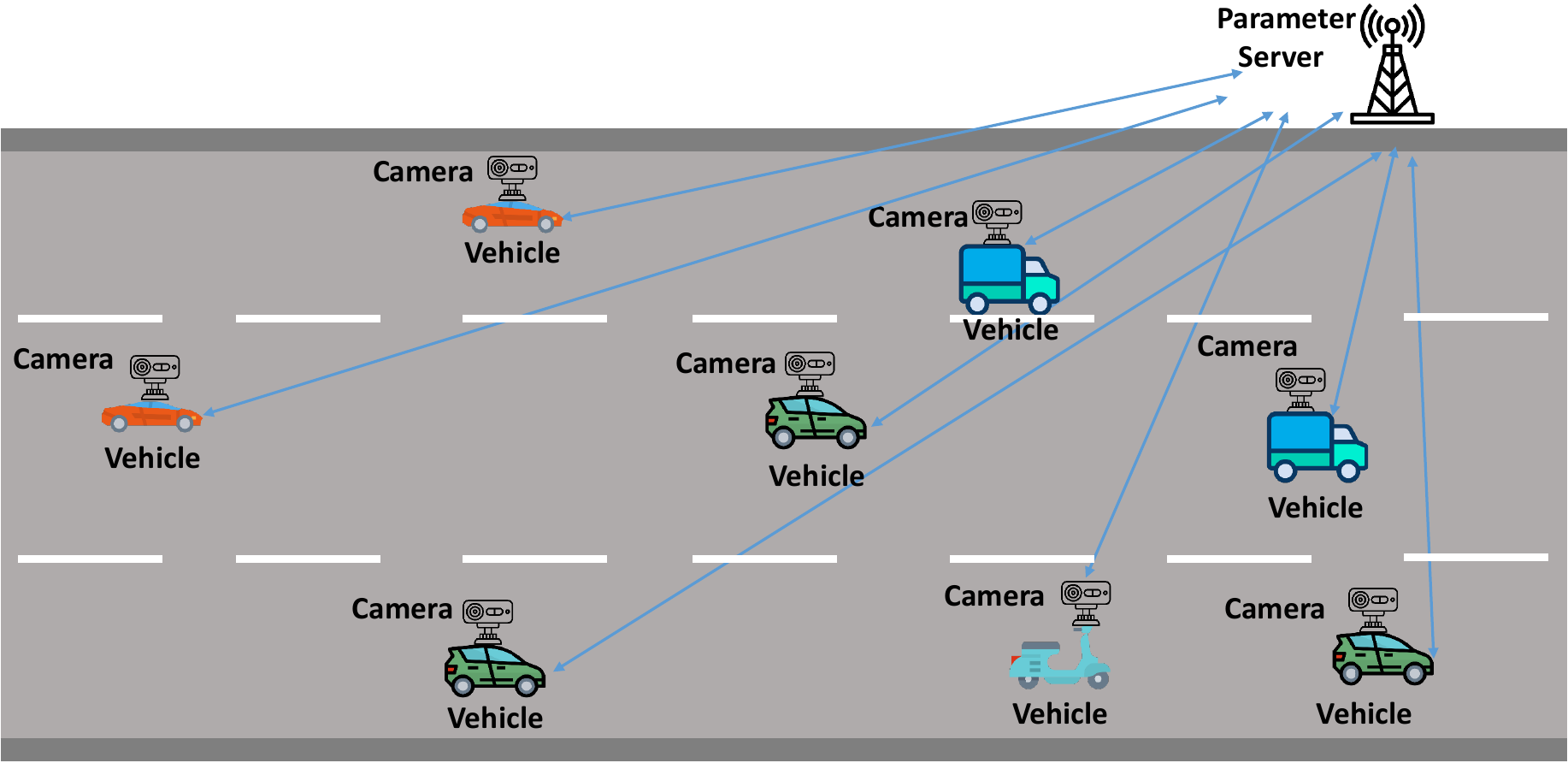}
\caption{An example of vehicular networks using distributed machine learning for training. The vehicles exchange 
their learning models with a parameter server towards a global common goal, such as traffic sign recognition.} 
\label{fig:vehicular_fed}
\end{figure}

To enable intelligent transportation systems, such as autonomous driving and advanced driver assistance systems, it is 
necessary to integrate vehicular communications and machine learning.
Vehicular communications provide communications between vehicles, pedestrians, road infrastructure, and the Internet, 
and has severe requirements in terms of low latency, high reliability and high rates~\cite{DuWu2020}.
Due to the advantages of \ac{FL} in terms of distributed computation, communication efficiency, and privacy by not 
sending raw data, its use in vehicular communications has started to get momentum (see Figure~\ref{fig:vehicular_fed}). 

The literature has recently considered \ac{FL} methods in learning tasks at the vehicles, such as collision 
avoidance, and traffic sign recognition, which can be considered as \ac{FL} in vehicular applications but 
without tailoring wireless methods for ML.
Specifically, the authors in~\cite{Elbir2020} investigate \ac{FL} applications for vehicular communications in the 
literature, including autonomous driving, road safety prediction, and vehicular object detection, and highlight some of 
the challenges and research directions for \ac{FL} in vehicular communications.
Conversely, \ac{FL} methods have been applied to resource management problems in vehicular communications, such as 
power control.
For instance, the authors in~\cite{Samarakoon2020} address wireless resource management problems in vehicular 
communications by using \ac{FL} to estimate the tail distribution of the network-wide queue lengths.

However, we are interested in this survey on the joint design of vehicular communications and \ac{FL}, or \ac{RRM} in vehicular 
communications for \ac{FL}, in which a \ac{PS} and vehicles jointly optimize their learning goals together with the communication requirements.
These three directions are illustrated in a recent survey~\cite{DuWu2020}, in which the authors discuss mainly the 
communication and learning aspects, while briefly mentioning the challenges of joint learning and communication of 
\ac{FL} in vehicular communications.

To the best of our knowledge, there is only one paper that fits our criteria~\cite{Zeng2021b}.
The authors in~\cite{Zeng2021b} consider the problem of learning and optimizing their autonomous controller design, 
which allows the vehicle to execute near real-time decisions, in the presence of wireless uncertainties and 
environmental dynamics.
To this end, this work proposes a dynamic federated proximal algorithm to account for the varying participation of 
vehicles due to mobility and wireless channels.
To improve the convergence of the proposed \ac{FL} algorithm, the authors design an incentive mechanism for the device 
participation using contract theory.
The incentive mechanism acts as a device participation and importance \ac{RRM}, such as the ones in 
Sections~\ref{subsub:partic_max} and~\ref{subsub:import_rrm}, by taking into account the data quality and devising a 
power allocation mechanism to maximize the convergence gain between two consecutive rounds.
The results show substantial improvements in the convergence speed compared to the other \ac{FL} algorithms, such as 
\mbox{FedAvg}, and baselines of their own proposed \ac{FL} algorithm using maximum and random power allocations.

The work in~\cite{Zeng2021b} jointly analyzed some of the control and learning challenges, but the communication 
challenges are still open.
Specifically, the impact of severe requirements in terms of low latency, high reliability, and high rates in a joint 
communication and learning approach needs to be considered.
Moreover, the impact of quick channel variations need to be analysed together with the learning convergence of the 
\ac{FL} method.
Therefore, research for this application is still quite open and there are many challenges ahead.

\section{Augmented and virtual reality}
For augmented and virtual reality (AR) and (VR) services provided by wireless networks, any sudden drop in the data rate or increase in the delay can negatively affect the quality of experience (QoE) of VR users. Although 5G beyond networks support operation at high frequency bands as well as flexible frame structure to minimize latency, the performance of communication links at high frequencies is highly prone to blockage thus reducing the QoE of VR users. 

\begin{figure}
    \centering
    \includegraphics[width=0.80\textwidth]{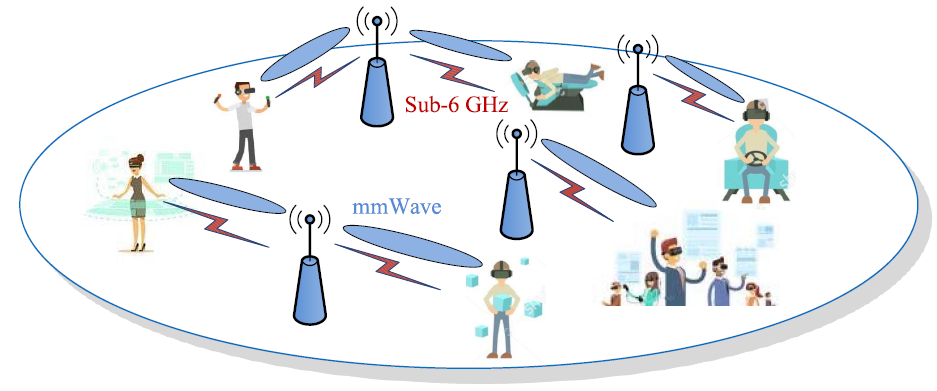}
    \caption{The architecture of a wireless virtual reality network. Virtual reality applications impose stringent throughput requirements. Therefore, millimeter wave communication is employed in the downlink.} 
    \label{fig:virtual_reality}
\end{figure}

One key application of using FL for improving QoE of wireless VR users is presented in \cite{8851408}. In the considered model, each \ac{BS} serves several VR users. Each user will transmit tracking information to the BS. Then, the BS will generate VR images according to the received tracking information and transmit the generated VR images to the VR users over millimeter wave frequency, which can be seen in Figure \ref{fig:virtual_reality}. Since VR images are transmitted over millimeter wave links, user movement such as mobility and orientation will introduce blockages to the millimeter wave transmission links thus decreasing the QoE of VR users.    

The goal of \cite{8851408} is to minimize the breaks in presence (BIP) of all VR users via optimizing user association. Since user association depends on the user mobility patterns and orientation, it is necessary to design a novel learning method to analyze the mobility patterns and orientation of each VR user. Meanwhile, since user association changes over time, each user may connect to different BSs at different time slots and hence, each BS can collect partial information related to user mobility patterns and orientation. Hence, traditional centralized learning algorithms that are implemented by a given BS cannot predict the entire VR user's locations and orientations without knowing the user's data collected by other BSs. To minimize the BIP of all users, an echo state network (ESN) based FL algorithm is designed, which enables the BSs to collaboratively generate a global ESN model to predict the whole set of locations and orientations for each user without transmitting the collected data to other BSs. Meanwhile, different from traditional FL algorithms that need to transmit the entire FL model, ESN based FL only needs to transmit the parameters of the output layer which can significantly reduce the size of data transmitted over wireless links thus improving convergence speed. In many envisioned VR and AR applications, co-located users share the same virtual world, for example in Smart Campus \cite{yagol2018new} and the Metaverse \cite{ning2021survey}. When many co-located users share an \ac{ML} task, over-the-air FL offers radical communication-efficiency improvements over orthogonal communications, as discussed in Section \ref{sec:ldcomac}. Therefore, Wireless for ML can assist in meeting the heavy communication demands imposed by AR and VR.

\section{Edge caching}
Caching of popular content at the network edge has been introduced as a promising approach to push the network traffic closer to the edge and reduce data traffic on backhaul networks \cite{MaddahAliCentralized}. 
Popular content is stored close to the edge terminals, at small \acp{BS}, \acp{AP}, or edge devices, proactively, such that it can be accessed more easily by the edge users.
This is particularly appealing for applications with stringent delay and bandwidth requirements. 
One of the challenges in edge caching is determining popularity of the content which is stored in the cache memories. 
Static and dynamic models have been introduced to capture the content popularity, where static models do not consider the time varying nature of the real-time content. 
On the other hand, dynamic models require accessing data for content differentiation.
This is not desirable in wireless systems since sharing data with other nodes may violate the privacy of users.

\begin{figure}[t]
\centering
    \begin{tikzpicture}
        \node[inner sep=0pt] (russell) at (0,0)
            {\includegraphics[width=0.8\linewidth]{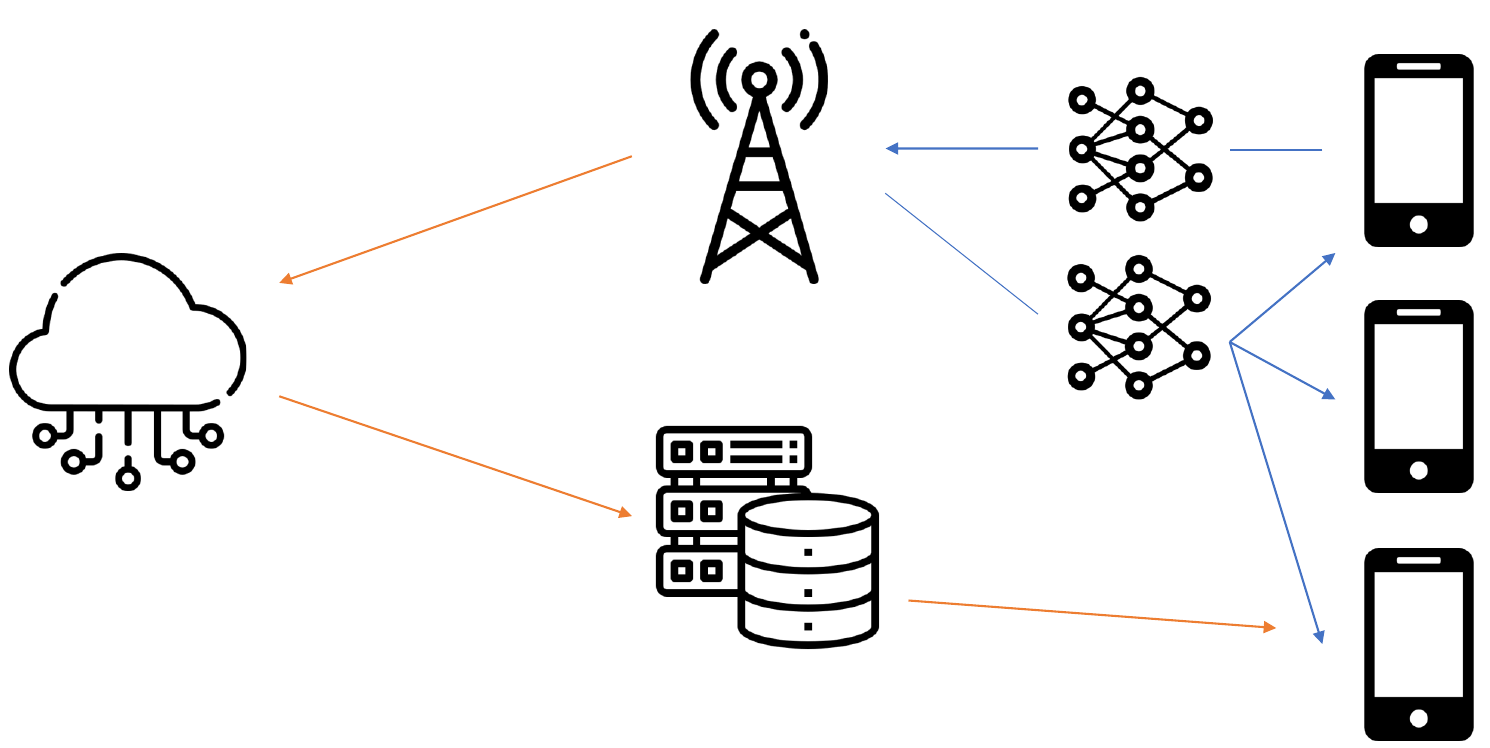}};
        \node[rotate=0] at (4.2,-3) {User devices};
        \node[rotate=0] at (0.2,-3) {Base station};
        \node[rotate=0] at (-3.9,-3) {Cloud};
    \end{tikzpicture}
    
    \caption{Illustration of federated learning for content popularity prediction. The blue lines illustrate how the user devices are collaboratively training the prediction model, using the base station as a \ac{PS}. The red lines illustrate the communication of content. The base station requests content based on the output of the popularity prediction model.}
    
    \label{fig:caching}
\end{figure}
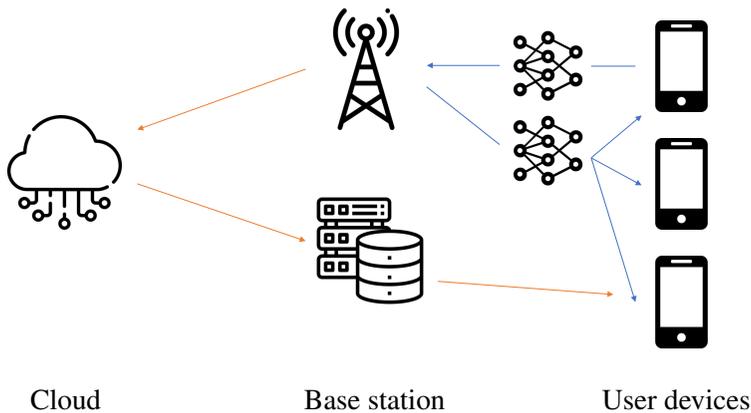

A distributed ML framework is a perfect fit to learn content popularity for edge caching by utilizing processing capabilities of edge devices. 
In this approach, local data at the users can be used to train a global model that is shared with all the users in order to learn the content popularity, see Figure \ref{fig:caching}. 
Therefore, the entire data across the network is used to determine the popularity of the content while data never leaves the users. 
The popular content is then stored close to the users to reduce the network traffic. 
For example, in augmented reality local data at the users can be used to learn popular elements, and the information about these elements can be cached proactively close to the users to reduce the latency and improve users experience.
Furthermore, in an autonomous driving example, information about the traffic, which can be learned collaboratively using the data collected by different vehicles, can be pre-fetched into the road side units.

Since the \ac{BS} is often both the arbiter of \ac{RRM} decisions and the host of the cache, it is natural to consider \ac{RRM} tailored to learn content popularity, as discussed in Section \ref{sec:wirelessFL}. Such dedicated wireless methods could improve the communication efficiency of training the content prediction model as well as reduce training and communication energy costs. Since trends in popular content changes regularly at a moments notice, the prediction model should be retrained continuously, which further emphasizes the importance of communication and energy efficiency.

\section{Unmanned aerial vehicles}
The low-altitude airspace of contemporary cities is generally empty or dominated by urban wildlife. In the upcoming decades, this underutilized real estate is predicted to be populated by search-and-rescue drones, delivery vehicles, and aerial \acp{BS} \cite{hayat2016survey, polka2017use, brik2020federated}. These applications are enabled by the \ac{UAV} technology, which provides cheap, easy to deploy, and highly maneuverable drones. However, there are many communication challenges associated with flying devices. First, there are stringent energy constraints as the weight of the battery increases the cost of flying. Secondly, the \ac{UAV} air-to-ground channel is more susceptible to fading, path loss, and delay spread because of the 3D movement of the vehicles \cite{brik2020federated}. Finally, \acp{UAV}
are never completely still, generating continual fast-fading.

\begin{figure}[t]
\centering
    \begin{tikzpicture}
        \node[inner sep=0pt] (russell) at (0,0)
            {\includegraphics[width=0.6\linewidth]{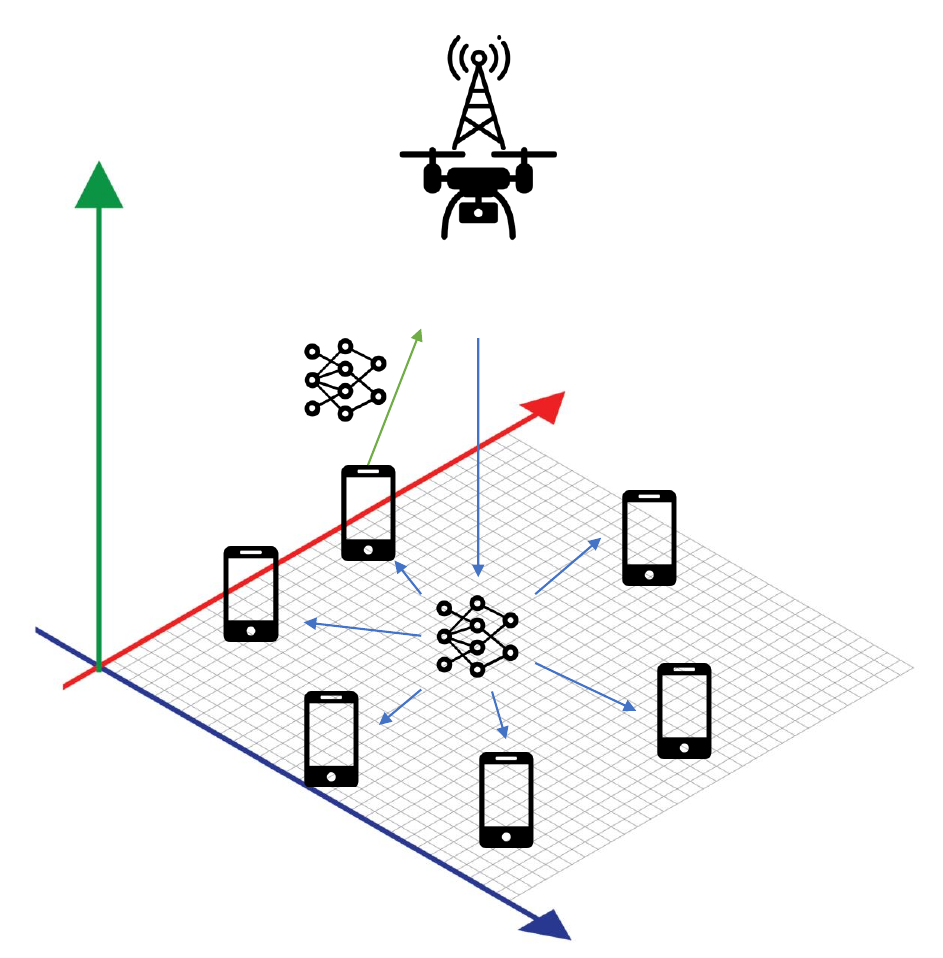}};
        \node[rotate=0] at (1.7,-2.5) {User devices};
        \node[rotate=0] at (0.2,1.5) {Flying base station};
    \end{tikzpicture}
    
    \caption{Illustration of federated learning for flying base stations. The goal is to predict the optimal 3D flying base station location that provides maximum channel gains to the user devices. Unlike a traditional base station, the flying base station can adjust its location dynamically.}
    
    \label{fig:uav}
\end{figure}

One interesting use-case of \acp{UAV} are the deployment of flying \acp{BS}, especially in geographical zones with lacking cellular infrastructure or as temporary deployment to increase cellular capacity during large events. Unlike a traditional \ac{BS}, these would be able to dynamically adjust their location to improve channel quality. The prediction of the correct location is a challenging problem that depends on the propagation environment, the number of users, and their mobility patterns. \ac{FL} is a perfect fit for training such a prediction model using the distributed data collected by the \ac{UAV} \ac{BS} and mobile devices \cite{brik2020federated}, see Figure \ref{fig:uav}. In this case, the training data is channel state information collected by the \ac{UAV} \acp{BS}. As such the data distribution changes quickly, and it is important to retrain continuously, which calls for efficient wireless protocols. Since the channels are changing quickly, the communication method must offer low latency to cope with the short channel coherence time. The Over-the-air computation methods discussed in Section \ref{sec:ldcomac} can offer low latencies that scale inversly with the number of users, which is perfect for a flying \ac{BS} deployed to a large event. Additionally, the blind methods discussed in Section \ref{subsubsec:blind} offer CSI-free over-the-air computation which is helpful when the channels are changing quickly.

Another critical application of \acp{UAV} are search and rescue missions at disaster locations. In these missions, the terrain is often unknown, since disasters such as floods, explosions, and earthquakes can change the known map completely \cite{hayat2016survey}. The time to locate victims is critical since survival is often heavily dependent on quick retrieval. Unfortunately, the cellular infrastructure easily gets destroyed by the disaster, leaving rescue workers in an unknown environment, with strict time constraints, and without connectivity. \acp{UAV} could be helpful in these scenarios to quickly set up multi-hop ad-hoc networks as a replacement for the damaged cellular infrastructure and to map out the environment. However, the highly mobile environment results in uncertain channel conditions that make routing difficult. A possible solution is a \ac{ML}-based model to predict the channels of potential next-hop nodes \cite{brik2020federated}. The inference of such models would be used to dynamically update the \acp{UAV} routing tables. Additionally, by training with the rescue team's devices, the \acp{UAV} can predict areas of poor coverage and adjust their locations to compensate.

\chapter{Conclusions}\label{sec:conclude}

Given the continuous growth of \ac{IoT} and mobile devices, the demand for \ac{ML} over wireless networks is expected to grow significantly. However, traditional communication protocols have been shown to be greatly inefficient for carrying \ac{ML} related data, creating a demand for new wireless solutions. In this survey, we have reviewed the most important contributions in this area, specifically focusing on analog over-the-air computation and digital \ac{RRM} for \ac{DML}. 

Analog over-the-air computation offers the most radical improvements in communication efficiency, exhibiting a throughput improvement approximately proportional to the number of participating devices. However, for contemporary communication, digital transmission is the de-facto standard. Within digital \ac{RRM} for \ac{DML}, significant performance improvements are achieved by considering data-importance and tailored \ac{RRM} protocols for \ac{FL}. However, this field is still in its infancy and several fundamental problems remain. For analog over-the-air computation, the main concerns are with integration into contemporary wireless infrastructure and functionality in dynamic wireless environments. Within digital \ac{RRM} for \ac{DML}, there are still many open questions relating to data-importance, such as choice of metrics and staleness of importance updates. 

It is highly relevant to find answers to these open questions, since efficient Wireless for \ac{ML} solutions could have profound effects on society, which we demonstrate by discussing five application areas: Smart City, Vehicular Communication, Virtual Reality, Edge Caching, and Unmanned Aerial Vehicles. The development of wireless methods specifically for \ac{ML} is a fertile area of research that could provide significant benefits in terms of energy efficiency, spectrum efficiency, and latency.

\backmatter  

\printbibliography

\end{document}

%% file: math.tex
\DeclareMathAlphabet{\mathppl}{T1}{ppl}{m}{it}
\DeclareMathAlphabet{\mathphv}{T1}{phv}{m}{it}
\DeclareMathAlphabet{\mathpzc}{T1}{pzc}{m}{it}
\newlength{\norlen} 

\newcommand{\ArgMin}[2]{\underset{#1}{\arg\min}\left\{#2\right\}}

\newcommand{\Ceil}[1]{\left\lceil #1 \right\rceil}

\newcommand{\Vt}[1]{\mathbf{\lowercase{#1}}}

\newcommand{\vtP}{\Vt{P}}

\newcommand{\vtW}{\Vt{W}}
\newcommand{\vtX}{\Vt{X}}
\newcommand{\vtY}{\Vt{Y}}

\newcommand{\vtLambda}{\Vt{\boldsymbol{\lambda}}}

\newcommand{\bbE}{\mathbb{E}}

\newcommand{\bbN}{\mathbb{N}}

\newcommand{\bbR}{\mathbb{R}}

\newcommand{\calD}{\mathcal{D}}

\newcommand{\calH}{\mathcal{H}}

\newcommand{\calS}{\mathcal{S}}

\newcommand{\calX}{\mathcal{X}}
\newcommand{\calY}{\mathcal{Y}}



%% file: acronyms.tex

\begin{acronym}[LTE-Advanced]
  \acro{2G}{Second Generation}
  \acro{3-DAP}{3-Dimensional Assignment Problem}
  \acro{3G}{3$^\text{rd}$~Generation}
  \acro{3GPP}{3$^\text{rd}$~Generation Partnership Project}
  \acro{4G}{4$^\text{th}$~Generation}
  \acro{5G}{5$^\text{th}$~Generation}
  \acro{A-FADMM}{analog federated alternating direction method of multipliers}
  \acro{AA}{Antenna Array}
  \acro{AC}{Admission Control}
  \acro{AD}{Attack-Decay}
  \acro{ADC}{analog-to-digital converter}
  \acro{ADMM}{alternating direction method of multipliers}
  \acro{ADSL}{Asymmetric Digital Subscriber Line}
  \acro{AHW}{Alternate Hop-and-Wait}
  \acro{AirComp}{\LU{O}{o}ver-the-air \LU{C}{c}omputation}
  \acro{AMC}{Adaptive Modulation and Coding}
  \acro{AP}{\LU{A}{a}ccess \LU{P}{p}oint}
  \acro{APA}{Adaptive Power Allocation}
  \acro{ARMA}{Autoregressive Moving Average}
  \acro{ARQ}{\LU{A}{a}utomatic \LU{R}{r}epeat \LU{R}{r}equest}
  \acro{ATES}{Adaptive Throughput-based Efficiency-Satisfaction Trade-Off}
  \acro{AWGN}{additive white Gaussian noise}
  \acro{BAA}{\LU{B}{b}roadband \LU{A}{a}nalog \LU{A}{a}ggregation}
  \acro{BB}{Branch and Bound}
  \acro{BCD}{block coordinate descent}
  \acro{BD}{Block Diagonalization}
  \acro{BER}{Bit Error Rate}
  \acro{BF}{Best Fit}
  \acro{BFD}{bidirectional full duplex}
  \acro{BLER}{BLock Error Rate}
  \acro{BPC}{Binary Power Control}
  \acro{BPSK}{Binary Phase-Shift Keying}
  \acro{BRA}{Balanced Random Allocation}
  \acro{BS}{base station}
  \acro{BSUM}{block successive upper-bound minimization}
  \acro{CAP}{Combinatorial Allocation Problem}
  \acro{CAPEX}{Capital Expenditure}
  \acro{CBF}{Coordinated Beamforming}
  \acro{CBR}{Constant Bit Rate}
  \acro{CBS}{Class Based Scheduling}
  \acro{CC}{Congestion Control}
  \acro{CDF}{Cumulative Distribution Function}
  \acro{CDMA}{Code-Division Multiple Access}
  \acro{CE}{\LU{C}{c}hannel \LU{E}{e}stimation}
  \acro{CL}{Closed Loop}
  \acro{CLPC}{Closed Loop Power Control}
  \acro{CML}{centralized machine learning}
  \acro{CNR}{Channel-to-Noise Ratio}
  \acro{CNN}{\LU{C}{c}onvolutional \LU{N}{n}eural \LU{N}{n}etwork}
  \acro{CPA}{Cellular Protection Algorithm}
  \acro{CPICH}{Common Pilot Channel}
  \acro{CoCoA}{\LU{C}{c}ommunication efficient distributed dual \LU{C}{c}oordinate \LU{A}{a}scent}
  \acro{CoMAC}{\LU{C}{c}omputation over \LU{M}{m}ultiple-\LU{A}{a}ccess \LU{C}{c}hannels}
  \acro{CoMP}{Coordinated Multi-Point}
  \acro{CQI}{Channel Quality Indicator}
  \acro{CRM}{Constrained Rate Maximization}
	\acro{CRN}{Cognitive Radio Network}
  \acro{CS}{Coordinated Scheduling}
  \acro{CSI}{\LU{C}{c}hannel \LU{S}{s}tate \LU{I}{i}nformation}
  \acro{CSMA}{\LU{C}{c}arrier \LU{S}{s}ense \LU{M}{m}ultiple \LU{A}{a}ccess}
  \acro{CUE}{Cellular User Equipment}
  \acro{D2D}{device-to-device}
  \acro{DAC}{digital-to-analog converter}
  \acro{DC}{direct current}
  \acro{DCA}{Dynamic Channel Allocation}
  \acro{DE}{Differential Evolution}
  \acro{DFT}{Discrete Fourier Transform}
  \acro{DIST}{Distance}
  \acro{DL}{downlink}
  \acro{DMA}{Double Moving Average}
  \acro{DML}{Distributed ML}
  \acro{DMRS}{demodulation reference signal}
  \acro{D2DM}{D2D Mode}
  \acro{DMS}{D2D Mode Selection}
  \acro{DPC}{Dirty Paper Coding}
  \acro{DRA}{Dynamic Resource Assignment}
  \acro{DSA}{Dynamic Spectrum Access}
  \acro{DSGD}{\LU{D}{d}istributed \LU{S}{s}tochastic \LU{G}{g}radient \LU{D}{d}escent}
  \acro{DSM}{Delay-based Satisfaction Maximization}
  \acro{ECC}{Electronic Communications Committee}
  \acro{EFLC}{Error Feedback Based Load Control}
  \acro{EI}{Efficiency Indicator}
  \acro{eNB}{Evolved Node B}
  \acro{EPA}{Equal Power Allocation}
  \acro{EPC}{Evolved Packet Core}
  \acro{EPS}{Evolved Packet System}
  \acro{E-UTRAN}{Evolved Universal Terrestrial Radio Access Network}
  \acro{ES}{Exhaustive Search}
  \acro{FD}{\LU{F}{f}ederated \LU{D}{d}istillation}
  \acro{FDD}{frequency division duplex}
  \acro{FDM}{Frequency Division Multiplexing}
  \acro{FDMA}{\LU{F}{f}requency \LU{D}{d}ivision \LU{M}{m}ultiple \LU{A}{a}ccess}
  \acro{FedAvg}{\LU{F}{f}ederated \LU{A}{a}veraging}
  \acro{FER}{Frame Erasure Rate}
  \acro{FF}{Fast Fading}
  \acro{FL}{Federated Learning}
  \acro{FML}{Federated Meta Learning}
  \acro{FSB}{Fixed Switched Beamforming}
  \acro{FST}{Fixed SNR Target}
  \acro{FTP}{File Transfer Protocol}
  \acro{GA}{Genetic Algorithm}
  \acro{GBR}{Guaranteed Bit Rate}
  \acro{GLR}{Gain to Leakage Ratio}
  \acro{GOS}{Generated Orthogonal Sequence}
  \acro{GPL}{GNU General Public License}
  \acro{GRP}{Grouping}
  \acro{HARQ}{Hybrid Automatic Repeat Request}
  \acro{HD}{half-duplex}
  \acro{HMS}{Harmonic Mode Selection}
  \acro{HOL}{Head Of Line}
  \acro{HSDPA}{High-Speed Downlink Packet Access}
  \acro{HSPA}{High Speed Packet Access}
  \acro{HTTP}{HyperText Transfer Protocol}
  \acro{ICMP}{Internet Control Message Protocol}
  \acro{ICI}{Intercell Interference}
  \acro{ID}{Identification}
  \acro{IETF}{Internet Engineering Task Force}
  \acro{ILP}{Integer Linear Program}
  \acro{JRAPAP}{Joint RB Assignment and Power Allocation Problem}
  \acro{UID}{Unique Identification}
  \acro{IID}{\LU{I}{i}ndependent and \LU{I}{i}dentically \LU{D}{d}istributed}
  \acro{IIR}{Infinite Impulse Response}
  \acro{ILP}{Integer Linear Problem}
  \acro{IMT}{International Mobile Telecommunications}
  \acro{INV}{Inverted Norm-based Grouping}
  \acro{IoT}{Internet of Things}
  \acro{IP}{Integer Programming}
  \acro{IPv6}{Internet Protocol Version 6}
  \acro{IRS}{intelligent reflective surface}
  \acro{ISD}{Inter-Site Distance}
  \acro{ISI}{Inter Symbol Interference}
  \acro{ITU}{International Telecommunication Union}
  \acro{JAFM}{joint assignment and fairness maximization}
  \acro{JAFMA}{joint assignment and fairness maximization algorithm}
  \acro{JOAS}{Joint Opportunistic Assignment and Scheduling}
  \acro{JOS}{Joint Opportunistic Scheduling}
  \acro{JP}{Joint Processing}
	\acro{JS}{Jump-Stay}
  \acro{KKT}{Karush-Kuhn-Tucker}
  \acro{L3}{Layer-3}
  \acro{LAC}{Link Admission Control}
  \acro{LA}{Link Adaptation}
  \acro{LC}{Load Control}
  \acro{LDC}{\LU{L}{l}earning-\LU{D}{d}riven \LU{C}{c}ommunication}
  \acro{LOS}{line of sight}
  \acro{LP}{Linear Programming}
  \acro{LTE}{Long Term Evolution}
	\acro{LTE-A}{\ac{LTE}-Advanced}
  \acro{LTE-Advanced}{Long Term Evolution Advanced}
  \acro{M2M}{Machine-to-Machine}
  \acro{MAC}{medium access control}
  \acro{MANET}{Mobile Ad hoc Network}
  \acro{MC}{Modular Clock}
  \acro{MCS}{Modulation and Coding Scheme}
  \acro{MDB}{Measured Delay Based}
  \acro{MDI}{Minimum D2D Interference}
  \acro{MF}{Matched Filter}
  \acro{MG}{Maximum Gain}
  \acro{MH}{Multi-Hop}
  \acro{MIMO}{\LU{M}{m}ultiple \LU{I}{i}nput \LU{M}{m}ultiple \LU{O}{o}utput}
  \acro{MINLP}{mixed integer nonlinear programming}
  \acro{MIP}{Mixed Integer Programming}
  \acro{MISO}{multiple input single output}
  \acro{ML}{Machine Learning}
  \acro{MLWDF}{Modified Largest Weighted Delay First}
  \acro{MME}{Mobility Management Entity}
  \acro{MMSE}{minimum mean squared error}
  \acro{MOS}{Mean Opinion Score}
  \acro{MPF}{Multicarrier Proportional Fair}
  \acro{MRA}{Maximum Rate Allocation}
  \acro{MR}{Maximum Rate}
  \acro{MRC}{Maximum Ratio Combining}
  \acro{MRT}{maximum ratio transmission}
  \acro{MRUS}{Maximum Rate with User Satisfaction}
  \acro{MS}{Mode Selection}
  \acro{MSE}{\LU{M}{m}ean \LU{S}{s}quared \LU{E}{e}rror}
  \acro{MSI}{Multi-Stream Interference}
  \acro{MTC}{Machine-Type Communication}
  \acro{MTSI}{Multimedia Telephony Services over IMS}
  \acro{MTSM}{Modified Throughput-based Satisfaction Maximization}
  \acro{MU-MIMO}{Multi-User Multiple Input Multiple Output}
  \acro{MU}{Multi-User}
  \acro{NAS}{Non-Access Stratum}
  \acro{NB}{Node B}
	\acro{NCL}{Neighbor Cell List}
  \acro{NLP}{Nonlinear Programming}
  \acro{NLOS}{non-line of sight}
  \acro{NMSE}{Normalized Mean Square Error}
  \acro{NOMA}{\LU{N}{n}on-\LU{O}{o}rthogonal \LU{M}{m}ultiple \LU{A}{a}ccess}
  \acro{NORM}{Normalized Projection-based Grouping}
  \acro{NP}{non-polynomial time}
  \acro{NRT}{Non-Real Time}
  \acro{NSPS}{National Security and Public Safety Services}
  \acro{O2I}{Outdoor to Indoor}
  \acro{OFDMA}{\LU{O}{o}rthogonal \LU{F}{f}requency \LU{D}{d}ivision \LU{M}{m}ultiple \LU{A}{a}ccess}
  \acro{OFDM}{Orthogonal Frequency Division Multiplexing}
  \acro{OFPC}{Open Loop with Fractional Path Loss Compensation}
	\acro{O2I}{Outdoor-to-Indoor}
  \acro{OL}{Open Loop}
  \acro{OLPC}{Open-Loop Power Control}
  \acro{OL-PC}{Open-Loop Power Control}
  \acro{OPEX}{Operational Expenditure}
  \acro{ORB}{Orthogonal Random Beamforming}
  \acro{JO-PF}{Joint Opportunistic Proportional Fair}
  \acro{OSI}{Open Systems Interconnection}
  \acro{PAIR}{D2D Pair Gain-based Grouping}
  \acro{PAPR}{Peak-to-Average Power Ratio}
  \acro{P2P}{Peer-to-Peer}
  \acro{PC}{Power Control}
  \acro{PCI}{Physical Cell ID}
  \acro{PDCCH}{physical downlink control channel}
  \acro{PDD}{penalty dual decomposition}
  \acro{PDF}{Probability Density Function}
  \acro{PER}{Packet Error Rate}
  \acro{PF}{Proportional Fair}
  \acro{P-GW}{Packet Data Network Gateway}
  \acro{PL}{Pathloss}
  \acro{PRB}{Physical Resource Block}
  \acro{PROJ}{Projection-based Grouping}
  \acro{ProSe}{Proximity Services}
  \acro{PS}{\LU{P}{p}arameter \LU{S}{s}erver}
  \acro{PSO}{Particle Swarm Optimization}
  \acro{PUCCH}{physical uplink control channel}
  \acro{PZF}{Projected Zero-Forcing}
  \acro{QAM}{Quadrature Amplitude Modulation}
  \acro{QoS}{quality of service}
  \acro{QPSK}{Quadri-Phase Shift Keying}
  \acro{RAISES}{Reallocation-based Assignment for Improved Spectral Efficiency and Satisfaction}
  \acro{RAN}{Radio Access Network}
  \acro{RA}{Resource Allocation}
  \acro{RAT}{Radio Access Technology}
  \acro{RATE}{Rate-based}
  \acro{RB}{resource block}
  \acro{RBG}{Resource Block Group}
  \acro{REF}{Reference Grouping}
  \acro{RF}{radio frequency}
  \acro{RLC}{Radio Link Control}
  \acro{RM}{Rate Maximization}
  \acro{RNC}{Radio Network Controller}
  \acro{RND}{Random Grouping}
  \acro{RRA}{Radio Resource Allocation}
  \acro{RRM}{\LU{R}{r}adio \LU{R}{r}esource \LU{M}{m}anagement}
  \acro{RSCP}{Received Signal Code Power}
  \acro{RSRP}{reference signal receive power}
  \acro{RSRQ}{Reference Signal Receive Quality}
  \acro{RR}{Round Robin}
  \acro{RRC}{Radio Resource Control}
  \acro{RSSI}{received signal strength indicator}
  \acro{RT}{Real Time}
  \acro{RU}{Resource Unit}
  \acro{RUNE}{RUdimentary Network Emulator}
  \acro{RV}{Random Variable}
  \acro{SAC}{Session Admission Control}
  \acro{SCM}{Spatial Channel Model}
  \acro{SC-FDMA}{Single Carrier - Frequency Division Multiple Access}
  \acro{SD}{Soft Dropping}
  \acro{S-D}{Source-Destination}
  \acro{SDPC}{Soft Dropping Power Control}
  \acro{SDMA}{Space-Division Multiple Access}
  \acro{SDR}{software-defined radio}
  \acro{SDP}{semidefinite programming}
  \acro{SER}{Symbol Error Rate}
  \acro{SES}{Simple Exponential Smoothing}
  \acro{S-GW}{Serving Gateway}
  \acro{SGD}{\LU{S}{s}tochastic \LU{G}{g}radient \LU{D}{d}escent}  
  \acro{SINR}{signal-to-interference-plus-noise ratio}
  \acro{SI}{self-interference}
  \acro{SIP}{Session Initiation Protocol}
  \acro{SISO}{\LU{S}{s}ingle \LU{I}{i}nput \LU{S}{s}ingle \LU{O}{o}utput}
  \acro{SIMO}{Single Input Multiple Output}
  \acro{SIR}{Signal to Interference Ratio}
  \acro{SLNR}{Signal-to-Leakage-plus-Noise Ratio}
  \acro{SMA}{Simple Moving Average}
  \acro{SNR}{\LU{S}{s}ignal to \LU{N}{n}oise \LU{R}{r}atio}
  \acro{SORA}{Satisfaction Oriented Resource Allocation}
  \acro{SORA-NRT}{Satisfaction-Oriented Resource Allocation for Non-Real Time Services}
  \acro{SORA-RT}{Satisfaction-Oriented Resource Allocation for Real Time Services}
  \acro{SPF}{Single-Carrier Proportional Fair}
  \acro{SRA}{Sequential Removal Algorithm}
  \acro{SRS}{sounding reference signal}
  \acro{SU-MIMO}{Single-User Multiple Input Multiple Output}
  \acro{SU}{Single-User}
  \acro{SVD}{Singular Value Decomposition}
  \acro{SVM}{\LU{S}{s}upport \LU{V}{v}ector \LU{M}{m}achine}
  \acro{SWIPT}{simultaneous wireless information and power transfer}
  \acro{TCP}{Transmission Control Protocol}
  \acro{TDD}{time division duplex}
  \acro{TDMA}{\LU{T}{t}ime \LU{D}{d}ivision \LU{M}{m}ultiple \LU{A}{a}ccess}
  \acro{TNFD}{three node full duplex}
  \acro{TETRA}{Terrestrial Trunked Radio}
  \acro{TP}{Transmit Power}
  \acro{TPC}{Transmit Power Control}
  \acro{TTI}{transmission time interval}
  \acro{TTR}{Time-To-Rendezvous}
  \acro{TSM}{Throughput-based Satisfaction Maximization}
  \acro{TU}{Typical Urban}
  \acro{UAV}{Unmanned Aerial Vehicle}
  \acro{UE}{\LU{U}{u}ser \LU{E}{e}quipment}
  \acro{UEPS}{Urgency and Efficiency-based Packet Scheduling}
  \acro{UL}{uplink}
  \acro{UMTS}{Universal Mobile Telecommunications System}
  \acro{URI}{Uniform Resource Identifier}
  \acro{URM}{Unconstrained Rate Maximization}
  \acro{VR}{Virtual Resource}
  \acro{VoIP}{Voice over IP}
  \acro{WAN}{Wireless Access Network}
  \acro{WCDMA}{Wideband Code Division Multiple Access}
  \acro{WF}{Water-filling}
  \acro{WiMAX}{Worldwide Interoperability for Microwave Access}
  \acro{WINNER}{Wireless World Initiative New Radio}
  \acro{WLAN}{Wireless Local Area Network}
  \acro{WMMSE}{weighted minimum mean square error}
  \acro{WMPF}{Weighted Multicarrier Proportional Fair}
  \acro{WPF}{Weighted Proportional Fair}
  \acro{WSN}{Wireless Sensor Network}
  \acro{WWW}{World Wide Web}
  \acro{XIXO}{(Single or Multiple) Input (Single or Multiple) Output}
  \acro{ZF}{zero-forcing}
  \acro{ZMCSCG}{Zero Mean Circularly Symmetric Complex Gaussian}
\end{acronym}

%% file: sample-now.bib
@Article{Du2020,
  author   = {R. {Du} and S. {Magnusson} and C. {Fischione}},
  title    = {{The Internet of Things as a Deep Neural Network}},
  journal  = {IEEE Communications Magazine},
  year     = {2020},
  volume   = {58},
  number   = {9},
  pages    = {20-25},
  month    = sep,
  doi      = {10.1109/MCOM.001.2000015},
}

@Article{Chen2021,
	author   = {M. Chen and H. V. Poor and W. Saad and S. Cui},
	title    = {{Convergence Time Optimization for Federated Learning Over Wireless Networks}},
	journal  = {IEEE Transactions on Wireless Communications},
	year     = {2021},
	volume   = {20},
	number   = {4},
	pages    = {2457-2471},
	month    = apr,
	doi      = {10.1109/TWC.2020.3042530},
}

@Article{Xu2021b,
	author  = {Jie Xu and Heqiang Wang and Lixing Chen},
	title   = {{Bandwidth Allocation for Multiple Federated Learning Services in Wireless Edge Networks}},
  journal={{arXiv}},
	year    = {2021},
	month   = {1},
    volume={abs/2101.03627},
}

@Article{Xu2021,
	author   = {Jie Xu and Heqiang Wang},
	title    = {{Client Selection and Bandwidth Allocation in Wireless Federated Learning Networks: A Long-Term 
	Perspective}},
	journal  = {IEEE Transactions on Wireless Communications},
	year     = {2021},
	volume   = {20},
	number   = {2},
	pages    = {1188-1200},
	month    = feb,
	doi      = {10.1109/TWC.2020.3031503},
}

@Article{Xia2021,
	author   = {W. Xia and W. Wen and K. -K. Wong and T. Q. S. Quek and J. Zhang and H. Zhu},
	title    = {{Federated-Learning-Based Client Scheduling for Low-Latency Wireless Communications}},
	journal  = {IEEE Wireless Communications},
	year     = {2021},
	volume   = {28},
	number   = {2},
	pages    = {32-38},
	month    = apr,
	doi      = {10.1109/MWC.001.2000252},
}

@InProceedings{Yang2020,
	author    = {H. H. Yang and A. Arafa and T. Q. S. Quek and H. Vincent Poor},
	title     = {{Age-Based Scheduling Policy for Federated Learning in Mobile Edge Networks}},
	booktitle = {Proceedings of the IEEE International Conference on Acoustics, Speech and Signal Processing (ICASSP)},
	year      = {2020},
	month     = 5,
	doi       = {10.1109/ICASSP40776.2020.9053740},
}

@Article{Dinh2021,
	author   = {Canh T. Dinh and Nguyen H. Tran and Minh N. H. Nguyen and Choong Seon Hong and Wei Bao and Albert Y. 
	Zo5a and Vincent Gramoli},
	title    = {{Federated Learning Over Wireless Networks: Convergence Analysis and Resource Allocation}},
	journal  = {IEEE/ACM Transactions on Networking},
	year     = {2021},
	volume   = {29},
	number   = {1},
	pages    = {398-409},
	month    = feb,
	doi      = {10.1109/TNET.2020.3035770},
}

@Article{Salehi2021,
	author   = {Mohammad Salehi and Ekram Hossain},
	title    = {{Federated Learning in Unreliable and Resource-Constrained Cellular Wireless Networks}},
	journal  = {{IEEE Transactions on Communications}},
	year     = {2021},
	month     = {8},
	doi      = {10.1109/TCOMM.2021.3081746},
}

@Article{Zeng2021b,
	author        = {Tengchan Zeng and Omid Semiari and Mingzhe Chen and Walid Saad and Mehdi Bennis},
	title         = {{Federated Learning on the Road: Autonomous Controller Design for Connected and Autonomous 
	Vehicles}},
	year          = {2021},
	month         = {2},
    journal={{arXiv}},
    volume={abs/2102.03401},
}

@Article{Elbir2020,
	author        = {Ahmet M. Elbir and Burak Soner and Sinem Coleri},
	title         = {{Federated Learning in Vehicular Networks}},
	year          = {2020},
	month         = {6},
    journal={{arXiv}},
    volume={abs/2006.01412},
	url           = {http://arxiv.org/abs/2006.01412},
}

@Article{DuWu2020,
	author   = {Z. {Du} and C. {Wu} and T. {Yoshinaga} and K. -L. A. {Yau} and Y. {Ji} and J. {Li}},
	title    = {{Federated Learning for Vehicular Internet of Things: Recent Advances and Open Issues}},
	journal  = {IEEE Open Journal of the Computer Society},
	year     = {2020},
	month    = {5},
	pages    = {45-61},
}

@article{park2021large,
  title={{Large-Scale Water Quality Prediction Using Federated Sensing and Learning: A Case Study with Real-World Sensing Big-Data}},
  author={Park, Soohyun and Jung, Soyi and Lee, Haemin and Kim, Joongheon and Kim, Jae-Hyun},
  journal={Sensors},
  volume={21},
  number={4},
  pages={1462},
  year={2021},
  month={2},
  publisher={Multidisciplinary Digital Publishing Institute}
}

@article{hard2018federated,
  title={{Federated Learning for Mobile Keyboard Prediction}},
  author={Hard, Andrew and Rao, Kanishka and Mathews, Rajiv and Ramaswamy, Swaroop and Beaufays, Fran{\c{c}}oise and Augenstein, Sean and Eichner, Hubert and Kiddon, Chlo{\'e} and Ramage, Daniel},
  journal={{arXiv}},
  year={2018},
  month={11},
  volume={abs/1811.03604},
}

@article{nguyen2021federated,
  title={{Federated Learning for Industrial Internet of Things in Future Industries}},
  author={Nguyen, Dinh C and Ding, Ming and Pathirana, Pubudu N and Seneviratne, Aruna and Li, Jun and Niyato, Dusit and Poor, H Vincent},
  journal={{arXiv}},
  year={2021},
  month={5},
  volume={abs/2105.14659},
}

@Article{Samarakoon2020,
	author   = {S. {Samarakoon} and M. {Bennis} and W. {Saad} and M. {Debbah}},
	title    = {{Distributed Federated Learning for Ultra-Reliable Low-Latency Vehicular Communications}},
	journal  = {IEEE Transactions on Communications},
	year     = {2019},
	month    = {11},
	volume   = {68},
	number   = {2},
	pages    = {1146--1159},
	issn     = {15580857},
}

@Article{Clerckx2021,
	author        = {Bruno Clerckx and Kaibin Huang and Lav R. Varshney and Sennur Ulukus and Mohamed-Slim Alouini},
	title         = {{Wireless Power Transfer for Future Networks: Signal Processing, Machine Learning, Computing, and 
	Sensing}},
    journal={{arXiv}},
	year          = {2021},
	month         = {1},
    volume={abs/2101.04810},
	primaryclass  = {cs.IT},
}

@Article{Silva2021,
	author        = {José Mairton B. da Silva Jr. and Konstantinos Ntougias and Ioannis Krikidis and Gabor Fodor and 
	Carlo Fischione},
	title         = {{Simultaneous Wireless Information and Power Transfer for Federated Learning}},
    journal={{arXiv}},
	year          = {2021},
	month         = apr,
    volume={abs/2104.12749},
}

@Article{Zeng2021,
	author        = {Qunsong Zeng and Yuqing Du and Kaibin Huang},
	title         = {{Wirelessly Powered Federated Edge Learning: Optimal Tradeoffs Between Convergence and Power 
	Transfer}},
    journal={{arXiv}},
	year          = {2021},
	month         = {2},
    volume={abs/2102.12357},
}

@InCollection{Li2020b,
	author    = {Tian Li and Anit Kumar Sahu and Manzil Zaheer and Maziar Sanjabi and Ameet Talwalkar and Virginia 
	Smith},
	title     = {{Federated Optimization in Heterogeneous Networks}},
	booktitle = {Proceedings of Machine Learning and Systems},
	year      = {2020},
	month     = {3},
	pages     = {429--450},
}

@BOOK{yang2019book,
	author={Q. {Yang} and Y. {Liu} and Y. {Cheng} and Y. {Kang} and T. {Chen} and H. {Yu}},
	booktitle={Federated Learning}, 
	title={Federated Learning}, 
	year={2019},
	publisher={Morgan \& Claypool Publishers},
	series={{Synthesis Lectures on Artificial Intelligence and Machine Learning}},
	month=dec,
}

@article{kairouz2019survey,
	title={{Advances and Open Problems in Federated Learning}},
	author={Peter Kairouz and H. Brendan McMahan and Brendan Avent and Aur{\'e}lien Bellet and Mehdi Bennis and Arjun 
	Nitin Bhagoji and Keith Bonawitz and Zachary Charles and Graham Cormode and Rachel Cummings and Rafael G. L. 
	D'Oliveira and Salim El Rouayheb and David Evans and Josh Gardner and Zachary A. Garrett and Adri{\`a} Gasc{\'o}n 
	and Badih Ghazi and Phillip B. Gibbons and Marco Gruteser and Za{\"i}d Harchaoui and Chaoyang He and Lie He and 
	Zhouyuan Huo and Ben Hutchinson and Justin Hsu and Martin Jaggi and Tara Javidi and Gauri Joshi and Mikhail Khodak 
	and Jakub Kone\v{c}n{\'y} and Aleksandra Korolova and Farinaz Koushanfar and Oluwasanmi Koyejo and Tancr{\`e}de 
	Lepoint and Yang Liu and Prateek Mittal and Mehryar Mohri and Richard Nock and Ayfer {\"O}zg{\"u}r and Rasmus Pagh 
	and Mariana Raykova and Hang Qi and Daniel Ramage and Ramesh Raskar and Dawn Xiaodong Song and Weikang Song and 
	Sebastian U. Stich and Ziteng Sun and Ananda Theertha Suresh and Florian Tram{\`e}r and Praneeth Vepakomma and 
	Jianyu Wang and Li Xiong and Zheng Xu and Qiang Yang and Felix X. Yu and Han Yu and Sen Zhao},
	journal={ArXiv},
	year={2019},
	month={12},
	volume={abs/1912.04977}
}

@article{makonecny2017,
	author = {Chenxin Ma and Jakub Kone\v{c}n{\'y} and Martin Jaggi and Virginia Smith and Michael I. Jordan and Peter 
	Richt{\'a}rik and Martin Tak{\'a}\v{c}},
	title = {Distributed optimization with arbitrary local solvers},
	journal = {Optimization Methods and Software},
	volume = {32},
	number = {4},
	pages = {813-848},
	year  = {2017},
	month={2},
	publisher = {Taylor & Francis},
}

@ARTICLE{tianli2020,  
	author={T. {Li} and A. K. {Sahu} and A. {Talwalkar} and V. {Smith}},  
	journal={IEEE Signal Processing Magazine},   
	title={{Federated Learning: Challenges, Methods, and Future Directions}},
	year={2020},  
	month={5},
	volume={37},  
	number={3},  
	pages={50-60},
}

@book{goodfellow2016book,
	title={Deep Learning},
	author={Ian Goodfellow and Yoshua Bengio and Aaron Courville},
	publisher={MIT Press},
	note={\url{http://www.deeplearningbook.org}},
	year={2016}
}

@ARTICLE{nedic2018survey,  
	author={A. {Nedi\'{c}} and A. {Olshevsky} and M. G. {Rabbat}},  
	journal={Proceedings of the IEEE},   
	title={{Network Topology and Communication-Computation Tradeoffs in Decentralized Optimization}},   
	year={2018},  
	month={4},
	volume={106},  
	number={5},  
	pages={953-976},
}

@book{hastie2009,
  title={{The Elements of Statistical Learning: Data Mining, Inference, and Prediction}},
  author={T. Hastie and R. Tibshirani and J.H. Friedman},
  series={Springer series in statistics},
  year={2009},
  publisher={Springer}
}

@article{bottou2018,
	author = {L\'{e}on Bottou and Frank E. Curtis and Jorge Nocedal},
	title = {{Optimization Methods for Large-Scale Machine Learning}},
	journal = {SIAM Review},
	volume = {60},
	number = {2},
	pages = {223-311},
	year = {2018},	
	month = {5}
}

@article{bennun2018survey,
	author = {Tal Ben-Nun and Torsten Hoefler},
	title = {{Demystifying Parallel and Distributed Deep Learning: An In-Depth Concurrency Analysis}},
	year = {2019},
	issue_date = {9 2019},
	publisher = {Association for Computing Machinery},
	address = {New York, NY, USA},
	volume = {52},
	number = {4},
	journal = {ACM Computing Surveys},
	month = aug,
	articleno = {65},
	numpages = {43}
}

@article{park2019wireless,
  title={{Wireless Network Intelligence at the Edge}},
  author={Park, Jihong and Samarakoon, Sumudu and Bennis, Mehdi and Debbah, M{\'e}rouane},
  journal={Proceedings of the IEEE},
  volume={107},
  number={11},
  pages={2204--2239},
  year={2019},
  month={10},
  publisher={IEEE}
}

@article{wang2020convergence,
  title={{Convergence of Edge Computing and Deep Learning: A Comprehensive Survey}},
  author={Wang, Xiaofei and Han, Yiwen and Leung, Victor CM and Niyato, Dusit and Yan, Xueqiang and Chen, Xu},
  journal={IEEE Communications Surveys \& Tutorials},
  volume={22},
  number={2},
  pages={869--904},
  year={2020},
  month={1},
  publisher={IEEE}
}

@article{zhu2018towards,
  title={{Towards an Intelligent Edge: Wireless Communication meets Machine Learning}},
  author={Zhu, Guangxu and Liu, Dongzhu and Du, Yuqing and You, Changsheng and Zhang, Jun and Huang, Kaibin},
  journal={{arXiv}},
  year={2018},
  month={9},
  volume={abs/1809.00343},
}

@misc{ericssonmobility,
    title={{Ericsson Mobility Report}},
    howpublished={available at \url{https://www.ericsson.com/en/press-releases/2019/6/ericsson-mobility-report-5g-uptake-even-faster-than-expected}, accessed: 2022-01-19, published: Jun. 2019}
}

@inproceedings{mcmahan2017communication,
  title={{Communication-efficient Learning of Deep Networks from Decentralized Data}},
  author={McMahan, Brendan and Moore, Eider and Ramage, Daniel and Hampson, Seth and y Arcas, Blaise Aguera},
  booktitle={{Artificial intelligence and Statistics}},
  pages={1273--1282},
  year={2017},
  month={4},
  organization={PMLR}
}

@article{zhu2019broadband,
  title={{Broadband Analog Aggregation for Low-Latency Federated Edge Learning}},
  author={Zhu, Guangxu and Wang, Yong and Huang, Kaibin},
  journal={IEEE Transactions on Wireless Communications},
  VOLUME = {19},
  NUMBER = {1},
  PAGES = {491-506},
  YEAR = {2020},
  MONTH = jan,
}

@inproceedings{huang2018communication,
  title={Communication, Computing, and Learning on the Edge},
  author={Huang, Kaibin and Zhu, Guangxu and You, Changsheng and Zhang, Jun and Du, Yuqing and Liu, Dongzhu},
  booktitle={2018 IEEE International Conference on Communication Systems (ICCS)},
  pages={268--273},
  year={2018},
  organization={IEEE}
}

@article{amiri2020machine,
  title={{Machine Learning at the Wireless Edge: Distributed Stochastic Gradient Descent Over-the-Air}},
  author={Amiri, Mohammad Mohammadi and Gunduz, Deniz},
  journal={IEEE Transactions on Signal Processing},
  VOLUME = {68},
  NUMBER = {},
  PAGES = {2155-2169},
  YEAR = {2020},
  MONTH = {3},
}

@article{zhu2021over,
  title={{Over-the-Air Computing for Wireless Data Aggregation in Massive IoT}},
  author={Zhu, Guangxu and Xu, Jie and Huang, Kaibin and Cui, Shuguang},
  journal={IEEE Wireless Communications},
  volume={28},
  number={4},
  pages={57--65},
  year={2021},
  month={9},
  publisher={IEEE}
}

@inproceedings{ahn2019wireless,
  title={{Wireless Federated Distillation for Distributed Edge Learning with Heterogeneous Data}},
  author={Ahn, Jin-Hyun and Simeone, Osvaldo and Kang, Joonhyuk},
  booktitle={Proceedings of the 2019 IEEE 30th Annual International Symposium on Personal, Indoor and Mobile Radio Communications (PIMRC)},
  pages={1--6},
  year={2019},
  month={7},
  organization={IEEE}
}

@inproceedings{amiri2019collaborative,
  title={{Collaborative Machine Learning at the Wireless Edge with Blind Transmitters}},
  author={{M. M. Amiri} and T. M. Duman and D. G\"und\"uz},
  booktitle={Proc. IEEE Global Conference on Signal and Information Processing (GlobalSIP)},
   MONTH = nov,
   PAGES = {1-5},
   YEAR = {2019},
   ADDRESS = {Ottawa, ON, Canada}
}

@article{YoSebMohammadMIMOFL,
   TITLE = {{A Compressive Sensing Approach for Federated Learning over Massive {MIMO} Communication Systems}},
   AUTHOR = {Y.-S. Jeon and M. M. Amiri and J. Li and H. V. Poor},
   JOURNAL = {IEEE Transactions on Wireless Communications},
   VOLUME = {20},
   NUMBER = {3},
   PAGES = {1990-2004},
   year = {2021},
   MONTH = mar,
}

@article{YoSebMohammadCommEffiMIMOFL,
   TITLE = {{Communication-Efficient Federated Learning over {MIMO} Multiple Access Channels}},
   AUTHOR = {Y.-S. Jeon and M. M. Amiri and N. Lee},
   JOURNAL = {},
   VOLUME = {},
   NUMBER = {},
   PAGES = {},
   YEAR = {2021},
   MONTH = {},
   note = {under review}
}

@inproceedings{carlini2019secret,
  title={{The Secret Sharer: Evaluating and Testing Unintended Memorization in Neural Networks}},
  author={Carlini, Nicholas and Liu, Chang and Erlingsson, {\'U}lfar and Kos, Jernej and Song, Dawn},
  booktitle={Proceedings of the 28th $\{$USENIX$\}$ Security Symposium ($\{$USENIX$\}$ Security 19)},
  pages={267--284},
  year={2019},
  month={8}
}

@inproceedings{MelisUnintendedLeakage2019,
  title={{Exploiting Unintended Feature Leakage in Collaborative Learning}},
  author={L. Melis and C. Song and E. D. Cristofaro and V. Shmatikov},
  booktitle={Proceedings of the IEEE Symposium on Security and Privacy},
   MONTH = 5,
   PAGES = {691-706},
   YEAR = {2019},
   ADDRESS = {San Francisco, CA, USA}
}

@inproceedings{MohammadUpdateAwareISIT2020,
  title={{Update Aware Device Scheduling for Federated Learning at the Wireless Edge}},
  author={M. M. Amiri and D. G\"und\"uz and S. R. Kulkarni and H. V. Poor},
  booktitle={Proceedings of the IEEE International Symposium on Information Theory},
   MONTH = jun,
   PAGES = {2598-2603},
   YEAR = {2020},
   ADDRESS = {Los Angeles, CA, USA}
}

@Article{Timotheou2014,
  author   = {S. {Timotheou} and I. {Krikidis} and G. {Zheng} and B. {Ottersten}},
  title    = {{Beamforming for MISO Interference Channels with QoS and RF Energy Transfer}},
  journal  = {IEEE Transactions on Wireless Communications},
  year     = {2014},
  volume   = {13},
  number   = {5},
  pages    = {2646-2658},
  month    = 5,
  issn     = {1558-2248},
  doi      = {10.1109/TWC.2014.032514.131199},
}

@inproceedings{SeifTandonDPFLISIT2020,
  title={{Wireless Federated Learning with Local Differential Privacy}},
  author={M. Seif and R. Tandon and M. Li},
  booktitle={Proceedings of the IEEE International Symposium on Information Theory},
   MONTH = {8},
   PAGES = {2604–2609},
   YEAR = {2020}
}

@article{amiri2020federated,
  title={{Federated Learning over Wireless Fading Channels}},
  author={Amiri, Mohammad Mohammadi and G{\"u}nd{\"u}z, Deniz},
  journal={IEEE Transactions on Wireless Communications},
  volume={19},
  number={5},
  pages={3546--3557},
  year={2020},
  month={2},
  publisher={IEEE}
}

@article{ren2020accelerating,
  title={{Accelerating DNN Training in Wireless Federated Edge Learning Systems}},
  author={Ren, Jinke and Yu, Guanding and Ding, Guangyao},
  journal={IEEE Journal on Selected Areas in Communications},
  volume={39},
  number={1},
  pages={219--232},
  year={2020},
  month={11},
  publisher={IEEE}
}

@article{goldsmith1998adaptive,
  title={{Adaptive Coded Modulation for Fading Channels}},
  author={Goldsmith, Andrea J and Chua, S-G},
  journal={IEEE Transactions on communications},
  volume={46},
  number={5},
  pages={595--602},
  year={1998},
  month={5},
  publisher={IEEE}
}

@inproceedings{amiri2019over,
  title={{Over-the-Air Machine Learning at the Wireless Edge}},
  author={Amiri, Mohammad Mohammadi and G{\"u}nd{\"u}z, Deniz},
  booktitle={Proceedings of the IEEE 20th International Workshop on Signal Processing Advances in Wireless Communications (SPAWC)},
  pages={1--5},
  year={2019},
  month={8},
  organization={IEEE}
}

@article{sery2020analog,
  title={{On Analog Gradient Descent Learning over Multiple Access Fading Channels}},
  author={Sery, Tomer and Cohen, Kobi},
  journal={IEEE Transactions on Signal Processing},
  volume={68},
  pages={2897--2911},
  year={2020},
  month={4},
  publisher={IEEE}
}

@inproceedings{liu2019wireless,
  title={{Wireless Data Acquisition for Edge Learning: Importance-Aware Retransmission}},
  author={Liu, Dongzhu and Zhu, Guangxu and Zhang, Jun and Huang, Kaibin},
  booktitle={2019 IEEE 20th International Workshop on Signal Processing Advances in Wireless Communications (SPAWC)},
  pages={1--5},
  year={2019},
  month={7},
  organization={IEEE}
}

@article{liu2020data,
  title={{Data-Importance Aware User Scheduling for Communication-efficient Edge Machine Learning}},
  author={Liu, Dongzhu and Zhu, Guangxu and Zhang, Jun and Huang, Kaibin},
  journal={IEEE Transactions on Cognitive Communications and Networking},
  volume={7},
  number={1},
  pages={265--278},
  year={2020},
  month={6},
  publisher={IEEE}
}

@article{MaddahAliCentralized,
   TITLE = {Fundamental limits of caching},
   AUTHOR = {M. A. Maddah-Ali and U. Niesen},
   JOURNAL = {IEEE Transactions on Information Theory},
   VOLUME = {60},
   NUMBER = {5},
   PAGES = {2856-2867},
   YEAR = {2014},
   MONTH = 5,
}

@article{VuNgoCellFreeFLMIMOTWC2020,
  title={{Cell-Free Massive {MIMO} for Wireless Federated Learning}},
  author={T. T. Vu and D. T. Ngo and N. H. Tran and H. Q. Ngo and M. N. Dao and R. H. Middleton},
  journal={IEEE Transactions on Wireless Communications},
  volume={19},
  number={10},
  pages={6377–6392},
  year={2020},
  month=oct
}

@article{ExpandingRedClientResFL,
   TITLE = {{Expanding the Reach of Federated Learning by Reducing Client Resource Requirements}},
   AUTHOR = {S. Caldas and J. Konecny and H. B. McMahan and A. Talwalkar},
  journal={{arXiv}},
   NUMBER = {},
   PAGES = {},
   YEAR = {2019},
   MONTH = jan,
   volume={abs/1812.07210},
}

@inproceedings{DoubleSqueezeTangConf,
   TITLE = {{DoubleSqueeze}: {Parallel} Stochastic Gradient Descent with Double-pass Error-Compensated Compression},
   AUTHOR = {H. Tang and X. Lian and C. Yu and T. Zhang and J. Liu},
   BOOKTITLE = {Proceedings of the International Conference on Machine Learning},
   MONTH = {6},
   PAGES = {},
   YEAR = 2019,
   ADDRESS = {Long Beach, CA}
}

@article{MohammadConvUpdateAwareTWC2021,
  title={Convergence of Update Aware Device Scheduling for Federated Learning at the Wireless Edge},
  author={M. M. Amiri and D. G\"und\"uz and S. R. Kulkarni and H. V. Poor},
  journal={IEEE Transactions on Wireless Communications},
  volume={20},
  number={6},
  pages={3643 - 3658},
  year={2021},
  month=jun
}

@article{MDSV_NIPS2020,
   TITLE = {{Federated Learning With Quantized Global Model Updates}},
   AUTHOR = {M. {M. Amiri} and D. G\"und\"uz and S. R. Kulkarni and H. V. Poor},
  journal={{arXiv}},
   NUMBER = {},
   PAGES = {},
   YEAR = {2020},
   MONTH = jun,
   volume={\\abs/2006.10672},
}

@article{MDSVNoisyDownlinkJournal2020,
  title={Convergence of federated learning over a noisy downlink},
  author={Amiri, Mohammad Mohammadi and G{\"u}nd{\"u}z, Deniz and Kulkarni, Sanjeev R and Poor, H Vincent},
  journal={IEEE Transactions on Wireless Communications},
  year={2021},
  publisher={IEEE}
}

@inproceedings{JinHyunAhnFLNoisyDownlink,
   TITLE = {{Cooperative Learning via Federated Distillation over Fading Channels}},
   AUTHOR = {Jin-Hyun Ahn and Osvaldo Simeone and Joonhyuk Kang},
   BOOKTITLE = {Proceedings of the IEEE International Conference on Acoustics, Speech and Signal Processing (ICASSP)},
   MONTH = 5,
   PAGES = {8856-8860},
   YEAR = 2020,
   ADDRESS = {Barcelona, Spain}
}

@article{YangJiangFLMIMOTWC2020,
  title={Federated learning via over-the-air computation},
  author={K. Yang and T. Jiang and Y. Shi and Z. Ding},
  journal={IEEE Transactions on Wireless Communications},
  volume={19},
  number={3},
  pages={2022 – 2035},
  year={2020},
  month=mar
}

@inproceedings{nishio2019client,
  title={{Client Selection for Federated Learning with Heterogeneous Resources in Mobile Edge}},
  author={Nishio, Takayuki and Yonetani, Ryo},
  booktitle={{Proceedings of the 2019 IEEE International Conference on Communications (ICC)}},
  pages={1--7},
  year={2019},
  month={7},
  organization={IEEE}
}

@article{mohammadTolgaBlindFLTWC2021,
  title={Blind federated edge learning},
  author={M. M. Amiri and T. M. Duman and D. G\"und\"uz and S. R. Kulkarni and H. V. Poor},
  journal={IEEE Transactions on Wireless Communications},
  year={2021},
  month={3},
  publisher={Early Access}
}

@article{chen2020joint,
  title={{A Joint Learning and Communications Framework for Federated Learning over Wireless Networks}},
  author={Chen, Mingzhe and Yang, Zhaohui and Saad, Walid and Yin, Changchuan and Poor, H Vincent and Cui, Shuguang},
  journal={IEEE Transactions on Wireless Communications},
  volume={20},
  number={1},
  pages={269--283},
  year={2020},
  month={10},
  publisher={IEEE}
}

@inproceedings{zeng2020energy,
  title={{Energy-Efficient Radio Resource Allocation for Federated Edge Learning}},
  author={Zeng, Qunsong and Du, Yuqing and Huang, Kaibin and Leung, Kin K},
  booktitle={Proceedings of the 2020 IEEE International Conference on Communications Workshops (ICC Workshops)},
  pages={1--6},
  year={2020},
  month={7},
  organization={IEEE}
}

@article{settles2012active,
  title={{Active Learning}},
  author={Settles, Burr},
  journal={Synthesis Lectures on Artificial Intelligence and Machine Learning},
  volume={6},
  number={1},
  pages={1--114},
  year={2012},
  publisher={Morgan \& Claypool Publishers}
}

@inproceedings{sun2020energy,
  title={{Energy-Aware Analog Aggregation for Federated Learning with Redundant Data}},
  author={Sun, Yuxuan and Zhou, Sheng and G{\"u}nd{\"u}z, Deniz},
  booktitle={Proceedings of the 2020 IEEE International Conference on Communications (ICC)},
  pages={1--7},
  year={2020},
  month={7},
  organization={IEEE}
}

@article{LiuOsvaldoDPFLJSAC2021,
  title={{Privacy for Free: {Wireless} Federated Learning via Uncoded Transmission With Adaptive Power Control}},
  author={D. Liu and O. Simeone},
  journal={IEEE Journal on Selected Areas in Communications},
  volume={39},
  number={1},
  pages={170--185},
  year={2021},
  month=jan
}

@inproceedings{shinkuma2019data,
  title={{Data Assessment and Prioritization in Mobile Networks for Real-Time Prediction of Spatial Information with Machine Learning}},
  author={Shinkuma, Ryoichi and Nishio, Takayuki},
  booktitle={Proceedings of the IEEE First International Workshop on Network Meets Intelligent Computations (NMIC)},
  pages={1--6},
  year={2019},
  month={8},
  organization={IEEE}
}

@inproceedings{KodaDPAircompGlobecom2020,
  title={{Differentially Private Aircomp Federated Learning with Power Adaptation Harnessing Receiver Noise}},
  author={Y. Koda and K. Yamamoto and T. Nishio and M. Morikura},
  booktitle={Proceedings of the IEEE Global Communications Conference},
  pages={1--6},
  year={2021},
  month={1},
  address={Taipei, Taiwan}
}

@article{inagaki2019prioritization,
  title={{Prioritization of Mobile IoT Data Transmission Based on Data Importance Extracted From Machine Learning Model}},
  author={Inagaki, Yuichi and Shinkuma, Ryoichi and Sato, Takehiro and Oki, Eiji},
  journal={IEEE Access},
  volume={7},
  pages={93611--93620},
  year={2019},
  month={7},
  publisher={IEEE}
}

@inproceedings{huang2013active,
  title={{Active Query Driven by Uncertainty and Diversity for Incremental Multi-Label Learning}},
  author={Huang, Sheng-Jun and Zhou, Zhi-Hua},
  booktitle={2013 IEEE 13th International Conference on Data Mining},
  pages={1079--1084},
  year={2014},
  month={2},
  organization={IEEE}
}

@inproceedings{shi2020device,
  title={{Device Scheduling with Fast Convergence for Wireless Federated Learning}},
  author={Shi, Wenqi and Zhou, Sheng and Niu, Zhisheng},
  booktitle={Proceedings of the 2020 IEEE International Conference on Communications (ICC)},
  pages={1--6},
  year={2020},
  month={7},
  organization={IEEE}
}

@article{goetz2019active,
  title={{Active Federated Learning}},
  author={Goetz, Jack and Malik, Kshitiz and Bui, Duc and Moon, Seungwhan and Liu, Honglei and Kumar, Anuj},
  journal={{arXiv}},
  year={2019},
  month={9},
  volume={abs/1909.12641},
}

@article{wang2020machine,
  title={{Machine Intelligence at the Edge With Learning Centric Power Allocation}},
  author={Wang, Shuai and Wu, Yik-Chung and Xia, Minghua and Wang, Rui and Poor, H Vincent},
  journal={IEEE Transactions on Wireless Communications},
  volume={19},
  number={11},
  pages={7293--7308},
  year={2020},
  month={7},
  publisher={IEEE}
}

@inproceedings{zhang2006hot,
  title={{Hot Topic: Physical-Layer Network Coding}},
  author={Zhang, Shengli and Liew, Soung Chang and Lam, Patrick P},
  booktitle={{Proceedings of the 12th Annual International Conference on Mobile Computing and Networking}},
  pages={358--365},
  year={2006},
  month={9},
  organization={ACM}
}

@inproceedings{MSV_DownUserSel_Conf2021,
  title={{Federated Learning With Downlink Device Selection}},
  author={M. {M. Amiri} and S. R. Kulkarni and H. V. Poor},
  booktitle={Proceedings of the IEEE International Workshop on Signal Processing Advances in Wireless Communications},
  MONTH = sep,
  PAGES = {},
  YEAR = 2021,
  ADDRESS = {Lucca, Italy}
}

@article{goldenbaum2013robust,
  title={{Robust Analog Function Computation via Wireless Multiple-Access Channels}},
  author={Goldenbaum, Mario and Stanczak, Slawomir},
  journal={IEEE Transactions on Communications},
  volume={61},
  number={9},
  pages={3863--3877},
  year={2013},
  month={8},
  publisher={IEEE}
}

@article{abari2016over,
  title={{Over-the-Air Function Computation in Sensor Networks}},
  author={Abari, Omid and Rahul, Hariharan and Katabi, Dina},
  journal={{arXiv}},
  year={2016},
  month={12},
  volume={abs/1612.02307},
}

@inproceedings{abari2015airshare,
  title={{Airshare: Distributed Coherent Transmission Made Seamless}},
  author={Abari, Omid and Rahul, Hariharan and Katabi, Dina and Pant, Mondira},
  booktitle={Proceedings of the 2015 IEEE Conference on Computer Communications (INFOCOM)},
  pages={1742--1750},
  year={2015},
  month={8},
  organization={IEEE}
}

@misc{ltetimingadvance,
  title = {{Timing Advance (TA) in LTE}},
  howpublished = {\url{http://http://4g5gworld.com/blog/timing-advance-ta-lte}},
  note = {Accessed: 2019-12-09}
}

@article{dong2020blind,
  title={Blind over-the-air computation and data fusion via provable wirtinger flow},
  author={Dong, Jialin and Shi, Yuanming and Ding, Zhi},
  journal={IEEE Transactions on Signal Processing},
  volume={68},
  pages={1136--1151},
  year={2020},
  publisher={IEEE}
}

@article{nazer2007computation,
  title={{Computation over Multiple-Access Channels}},
  author={Nazer, Bobak and Gastpar, Michael},
  journal={IEEE Transactions on Information Theory},
  volume={53},
  number={10},
  pages={3498--3516},
  year={2007},
  month={9},
  publisher={IEEE}
}

@inproceedings{sery2019sequential,
  title={{A Sequential Gradient-Based Multiple Access for Distributed Learning over Fading Channels}},
  author={Sery, Tomer and Cohen, Kobi},
  booktitle={Proceedings of the 2019 57th Annual Allerton Conference on Communication, Control, and Computing (Allerton)},
  pages={303--307},
  year={2019},
  month={12},
  organization={IEEE}
}

@article{lin2017deep,
  title={{Deep Gradient Compression: Reducing the Communication Bandwidth for Distributed Training}},
  author={Lin, Yujun and Han, Song and Mao, Huizi and Wang, Yu and Dally, William J},
  journal={{arXiv}},
  year={2017},
  month={12},
  volume={abs/1712.01887},
}

@article{simonyan2014very,
  title={{Very Deep Convolutional Networks for Large-Scale Image Recognition}},
  author={Simonyan, Karen and Zisserman, Andrew},
  journal={{arXiv}},
  year={2014},
  month={9},
  volume={abs/1409.1556},
}

@article{jeong2018communication,
  title={{\\Communication-Efficient On-Device Machine Learning: Federated Distillation and Augmentation Under non-IID Private Data}},
  author={Jeong, Eunjeong and Oh, Seungeun and Kim, Hyesung and Park, Jihong and Bennis, Mehdi and Kim, Seong-Lyun},
  journal={{arXiv}},
  year={2018},
  month={11},
  volume={\\abs/1811.11479},
}

@inproceedings{bertin2011million,
  title={The million song dataset},
  author={Bertin-Mahieux, Thierry and Ellis, Daniel PW and Whitman, Brian and Lamere, Paul},
  booktitle={Proceedings of the 11th International Conference on Music Information Retrieval (ISMIR)},
  year={2011},
  month={10}
}

@article{lecun1998gradient,
  title={{Gradient-Based Learning Applied to Document Recognition}},
  author={LeCun, Yann and Bottou, L{\'e}on and Bengio, Yoshua and Haffner, Patrick and others},
  journal={Proceedings of the IEEE},
  volume={86},
  number={11},
  pages={2278--2324},
  year={1998},
  month={11}
}

@article{zhao2018federated,
  title={{Federated Learning With non-IID Data}},
  author={Zhao, Yue and Li, Meng and Lai, Liangzhen and Suda, Naveen and Civin, Damon and Chandra, Vikas},
  journal={{arXiv}},
  year={2018},
  month={6},
  volume={abs/1806.00582},
}

@article{goldenbaum2014channel,
  title={{On the Channel Estimation Effort for Analog Computation over Wireless Multiple-Access Channels}},
  author={Goldenbaum, Mario and Stanczak, Slawomir},
  journal={IEEE Wireless Communications Letters},
  volume={3},
  number={3},
  pages={261--264},
  year={2014},
  month={3},
  publisher={IEEE}
}

@article{ngo2017cell,
  title={{Cell-free massive {MIMO} versus small cells}},
  author={Ngo, Hien Quoc and Ashikhmin, Alexei and Yang, Hong and Larsson, Erik G and Marzetta, Thomas L},
  journal={IEEE Transactions on Wireless Communications},
  volume={16},
  number={3},
  pages={1834--1850},
  year={2017},
  month={1},
  publisher={IEEE}
}

@article{liu2020wireless,
  title={{Wireless Data Acquisition for Edge Learning: Data-Importance Aware Retransmission}},
  author={Liu, Dongzhu and Zhu, Guangxu and Zeng, Qunsong and Zhang, Jun and Huang, Kaibin},
  journal={Proceedings of the IEEE Transactions on Wireless Communications},
  volume={20},
  number={1},
  pages={406--420},
  year={2020},
  month={9},
  publisher={IEEE}
}

@techreport{settles2009active,
  title={{Active Learning Literature Survey}},
  author={Settles, Burr},
  year={2009},
  institution={University of Wisconsin-Madison Department of Computer Sciences}
}

@inproceedings{ha2019coded,
  title={{Coded Federated Computing in Wireless Networks With Straggling Devices and Imperfect {CSI}}},
  author={Ha, Sukjong and Zhang, Jingjing and Simeone, Osvaldo and Kang, Joonhyuk},
  booktitle={Proceedings of the IEEE International Symposium on Information Theory (ISIT)},
  pages={2649--2653},
  year={2019},
  month={9},
  organization={IEEE}
}

@inproceedings{yu2001constant,
  title={{On Constant Power Water-Filling}},
  author={Yu, Wei and Cioffi, John M},
  booktitle={IEEE International Conference on Communications. Conference Record (ICC)},
  volume={6},
  pages={1665--1669},
  year={2002},
  month={8},
  organization={IEEE}
}

@inproceedings{xue2009max,
  title={{Max-Min Fairness Based Radio Resource Management in Fourth Generation Heterogeneous Networks}},
  author={Xue, Peng and Gong, Peng and Park, Jae Hyun and Park, Daeyoung and Kim, Duk Kyung},
  booktitle={Proceedings of the 9th International Symposium on Communications and Information Technology},
  pages={208--213},
  year={2009},
  month={9},
  organization={IEEE}
}

@article{zhu2020toward,
  title={{Toward an Intelligent Edge: Wireless Communication Meets Machine Learning}},
  author={Zhu, Guangxu and Liu, Dongzhu and Du, Yuqing and You, Changsheng and Zhang, Jun and Huang, Kaibin},
  journal={IEEE Communications Magazine},
  volume={58},
  number={1},
  pages={19--25},
  year={2020},
  month={1},
  publisher={IEEE}
}

@article{boyd2011distributed,
  title={{Distributed Optimization and Statistical Learning via the Alternating Direction Method of Multipliers}},
  author={Boyd, Stephen and Parikh, Neal and Chu, Eric and Peleato, Borja and Eckstein, Jonathan and others},
  journal={{Foundations and Trends{\textregistered} in Machine Learning}},
  volume={3},
  number={1},
  pages={1--122},
  year={2011},
  publisher={Now Publishers, Inc.}
}

@inproceedings{zinkevich2010parallelized,
  title={{Parallelized Stochastic Gradient Descent}},
  author={Zinkevich, Martin and Weimer, Markus and Li, Lihong and Smola, Alex J},
  booktitle={Advances in Neural Information Processing systems},
  pages={2595--2603},
  year={2010},
  month={12}
}

@inproceedings{jaggi2014communication,
  title={{Communication-efficient Distributed Dual Coordinate Ascent}},
  author={Jaggi, Martin and Smith, Virginia and Tak{\'a}c, Martin and Terhorst, Jonathan and Krishnan, Sanjay and Hofmann, Thomas and Jordan, Michael I},
  booktitle={Advances in Neural Information Processing Systems},
  pages={3068--3076},
  year={2014},
  month={12}
}

@inproceedings{he2016deep,
  title={{Deep Residual Learning for Image Recognition}},
  author={He, Kaiming and Zhang, Xiangyu and Ren, Shaoqing and Sun, Jian},
  booktitle={Proceedings of the IEEE Conference on Computer Vision and Pattern Recognition},
  pages={770--778},
  year={2016},
  month={6}
}

@article{collobert2011natural,
  title={{Natural Language Processing (almost) from Scratch}},
  author={Collobert, Ronan and Weston, Jason and Bottou, L{\'e}on and Karlen, Michael and Kavukcuoglu, Koray and Kuksa, Pavel},
  journal={Journal of Machine Learning Research},
  volume={12},
  pages={2493--2537},
  year={2011},
  month={10}
}

@article{zhou2019edge,
  title={{Edge Intelligence: Paving the Last Mile of Artificial Intelligence with Edge Computing}},
  author={Zhou, Zhi and Chen, Xu and Li, En and Zeng, Liekang and Luo, Ke and Zhang, Junshan},
  journal={Proceedings of the IEEE},
  volume={107},
  number={8},
  pages={1738--1762},
  year={2019},
  month={6},
  publisher={IEEE}
}

@article{shi2020communication,
  title={{Communication-efficient Edge AI: Algorithms and Systems}},
  author={Shi, Yuanming and Yang, Kai and Jiang, Tao and Zhang, Jun and Letaief, Khaled B},
  journal={{IEEE Communications Surveys \& Tutorials}},
  volume={22},
  number={4},
  pages={2167--2191},
  year={2020},
  month={7},
  publisher={IEEE}
}

@inproceedings{seide20141,
  title={{1-bit Stochastic Gradient Descent and its Application to Data-Parallel Distributed Training of Speech DNNs}},
  author={Seide, Frank and Fu, Hao and Droppo, Jasha and Li, Gang and Yu, Dong},
  booktitle={Proceedings of the 15th Annual Conference of the International Speech Communication Association},
  year={2014},
  month={9}
}

@article{saad2019vision,
  title={{A Vision of 6{G} Wireless Systems: Applications, Trends, Technologies, and Open Research Problems}},
  author={Saad, Walid and Bennis, Mehdi and Chen, Mingzhe},
  journal={IEEE Network},
  volume={34},
  number={3},
  pages={134--142},
  year={2019},
  month={10},
  publisher={IEEE}
}

@article{zhang20196g,
  title={{6{G} Wireless Networks: Vision, Requirements, Architecture, and Key Technologies}},
  author={Zhang, Zhengquan and Xiao, Yue and Ma, Zheng and Xiao, Ming and Ding, Zhiguo and Lei, Xianfu and Karagiannidis, George K and Fan, Pingzhi},
  journal={IEEE Vehicular Technology Magazine},
  volume={14},
  number={3},
  pages={28--41},
  year={2019},
  month={7},
  publisher={IEEE}
}

@article{rost2017network,
  title={{Network Slicing to Enable Scalability and Flexibility in 5G Mobile Networks}},
  author={Rost, Peter and Mannweiler, Christian and Michalopoulos, Diomidis S and Sartori, Cinzia and Sciancalepore, Vincenzo and Sastry, Nishanth and Holland, Oliver and Tayade, Shreya and Han, Bin and Bega, Dario and others},
  journal={IEEE Communications Magazine},
  volume={55},
  number={5},
  pages={72--79},
  year={2017},
  month={5},
  publisher={IEEE}
}

@article{bennis2018ultrareliable,
  title={Ultrareliable and Low-Latency Wireless Communication: Tail, Risk, and Scale},
  author={Bennis, Mehdi and Debbah, M{\'e}rouane and Poor, H Vincent},
  journal={Proceedings of the IEEE},
  volume={106},
  number={10},
  pages={1834--1853},
  year={2018},
  month={9},
  publisher={IEEE}
}

@article{hinton2015distilling,
  title={{Distilling the Knowledge in a Neural Network}},
  author={Hinton, Geoffrey and Vinyals, Oriol and Dean, Jeff},
  journal={{arXiv}},
  year={2015},
  month={3},
  volume={abs/1503.02531},
}

@article{niknam2020federated,
  title={{Federated Learning for Wireless Communications: Motivation, Opportunities, and Challenges}},
  author={Niknam, Solmaz and Dhillon, Harpreet S and Reed, Jeffrey H},
  journal={IEEE Communications Magazine},
  volume={58},
  number={6},
  pages={46--51},
  year={2020},
  month={7},
  publisher={IEEE}
}

@article{deng2020edge,
  title={{Edge Intelligence: the Confluence of Edge Computing and Artificial Intelligence}},
  author={Deng, Shuiguang and Zhao, Hailiang and Fang, Weijia and Yin, Jianwei and Dustdar, Schahram and Zo5a, Albert Y},
  journal={IEEE Internet of Things Journal},
  year={2020},
  month={4},
  publisher={IEEE}
}

@inproceedings{bhagoji2019analyzing,
  title={{Analyzing Federated Learning Through an Adversarial Lens}},
  author={Bhagoji, Arjun Nitin and Chakraborty, Supriyo and Mittal, Prateek and Calo, Seraphin},
  booktitle={Proceedings of the International Conference on Machine Learning},
  pages={634--643},
  year={2019},
  month={6},
  organization={PMLR}
}

@article{fung2018mitigating,
  title={{Mitigating Sybils in Federated Learning Poisoning}},
  author={Fung, Clement and Yoon, Chris JM and Beschastnikh, Ivan},
  journal={{arXiv}},
  year={2018},
  month={8},
  volume={abs/1808.04866},
}

@inproceedings{bagdasaryan2020backdoor,
  title={{How to Backdoor Federated Learning}},
  author={Bagdasaryan, Eugene and Veit, Andreas and Hua, Yiqing and Estrin, Deborah and Shmatikov, Vitaly},
  booktitle={Proceedings of the International Conference on Artificial Intelligence and Statistics},
  pages={2938--2948},
  year={2020},
  month={6},
  organization={PMLR}
}

@article{zhu2018broadband,
  title={{Broadband Analog Aggregation for Low-Latency Federated Edge Learning (extended version)}},
  author={Zhu, Guangxu and Wang, Yong and Huang, Kaibin},
  journal={{arXiv}},
  year={2018},
  month={12},
  volume={abs/1812.11494},
}

@article{hussain2020machine,
  title={{Machine Learning for Resource Management in Cellular and IoT Networks: Potentials, Current Solutions, and Open Challenges}},
  author={Hussain, Fatima and Hassan, Syed Ali and Hussain, Rasheed and Hossain, Ekram},
  journal={IEEE Communications Surveys \& Tutorials},
  volume={22},
  number={2},
  pages={1251--1275},
  year={2020},
  month={1},
  publisher={IEEE}
}

@article{branco2016survey,
  title={{A Survey of Predictive Modeling on Imbalanced Domains}},
  author={Branco, Paula and Torgo, Lu{\'\i}s and Ribeiro, Rita P},
  journal={ACM Computing Surveys (CSUR)},
  volume={49},
  number={2},
  pages={1--50},
  year={2016},
  month={11},
  publisher={ACM New York, NY, USA}
}

@inproceedings{saito2013non,
  title={{Non-Orthogonal Multiple Access (NOMA) for Cellular Future Radio Access}},
  author={Saito, Yuya and Kishiyama, Yoshihisa and Benjebbour, Anass and Nakamura, Takehiro and Li, Anxin and Higuchi, Kenichi},
  booktitle={Proceedings of the 2013 IEEE 77th Vehicular Technology Conference (VTC Spring)},
  pages={1--5},
  year={2013},
  month={6},
  organization={IEEE}
}

@article{islam2019nonorthogonal,
  title={{Nonorthogonal Multiple Access (NOMA): How It Meets 5G and Beyond}},
  author={Islam, SM Riazul and Zeng, Ming and Dobre, Octavia A and Kwak, Kyung-Sup},
  journal={Wiley 5G Ref: The Essential 5G Reference Online},
  pages={1--28},
  year={2019},
  month={12},
  publisher={Wiley Online Library}
}

@article{zhu2018over,
  title={{Over-the-Air Computation in MIMO Multi-Access Channels: Beamforming and Channel Feedback}},
  author={Zhu, Guangxu and Chen, Li and Huang, Kaibin},
  journal={CoRR, vol. abs/1803.11129},
  year={2018},
  month={3}
}

@article{wen2019overview,
  title={{An Overview of Data-Importance Aware Radio Resource Management for Edge Machine Learning}},
  author={Wen, Dingzhu and Li, Xiaoyang and Zeng, Qunsong and Ren, Jinke and Huang, Kaibin},
  journal={Journal of Communications and Information Networks},
  volume={4},
  number={4},
  pages={1--14},
  year={2019},
  month={12},
  publisher={PTP}
}

@article{cheng2017survey,
  title={{A Survey of Model Compression and Acceleration for Deep Neural Networks}},
  author={Cheng, Yu and Wang, Duo and Zhou, Pan and Zhang, Tao},
  journal={{arXiv}},
  year={2017},
  month={10},
  volume={abs/1710.09282},
}

@article{zhu2017prune,
  title={{To Prune, or not to Prune: Exploring the Efficacy of Pruning for Model Compression}},
  author={Zhu, Michael and Gupta, Suyog},
  journal={{arXiv}},
  year={2017},
  month={10},
  volume={abs/1710.01878},
}

@article{zhu2020one,
  title={{One-Bit Over-the-Air Aggregation for Communication-Efficient Federated Edge Learning: Design and Convergence Analysis}},
  author={Zhu, Guangxu and Du, Yuqing and G{\"u}nd{\"u}z, Deniz and Huang, Kaibin},
  journal={IEEE Transactions on Wireless Communications},
  volume={20},
  number={3},
  pages={2120--2135},
  year={2020},
  month={11},
  publisher={IEEE}
}

@inproceedings{bernstein2018signsgd,
  title={{signSGD: Compressed Optimisation for Non-Convex Problems}},
  author={Bernstein, Jeremy and Wang, Yu-Xiang and Azizzadenesheli, Kamyar and Anandkumar, Animashree},
  booktitle={Proceedings of the International Conference on Machine Learning},
  pages={560--569},
  year={2018},
  month={7},
  organization={PMLR}
}

@article{ren2020scheduling,
  title={{Scheduling for Cellular Federated Edge Learning With Importance and Channel Awareness}},
  author={Ren, Jinke and He, Yinghui and Wen, Dingzhu and Yu, Guanding and Huang, Kaibin and Guo, Dongning},
  journal={IEEE Transactions on Wireless Communications},
  volume={19},
  number={11},
  pages={7690--7703},
  year={2020},
  month={8},
  publisher={IEEE}
}

@article{elgabli2021harnessing,
  title={{Harnessing Wireless Channels for Scalable and Privacy-Preserving Federated Learning}},
  author={Elgabli, Anis and Park, Jihong and Issaid, Chaouki Ben and Bennis, Mehdi},
  journal={IEEE Transactions on Communications},
  year={2021},
  month={5},
  publisher={IEEE}
}

@article{mao2018deep,
  title={{Deep Learning for Intelligent Wireless Networks: A Comprehensive Survey}},
  author={Mao, Qian and Hu, Fei and Hao, Qi},
  journal={IEEE Communications Surveys \& Tutorials},
  volume={20},
  number={4},
  pages={2595--2621},
  year={2018},
  month={6},
  publisher={IEEE}
}

@inproceedings{wadu2020federated,
  title={{Federated Learning Under Channel Uncertainty: Joint Client Scheduling and Resource Allocation}},
  author={Wadu, Madhusanka Manimel and Samarakoon, Sumudu and Bennis, Mehdi},
  booktitle={Proceedings of the 2020 IEEE Wireless Communications and Networking Conference (WCNC)},
  pages={1--6},
  year={2020},
  month={6},
  organization={IEEE}
}

@inproceedings{fredrikson2015model,
  title={{Model Inversion Attacks That Exploit Confidence Information and Basic Countermeasures}},
  author={Fredrikson, Matt and Jha, Somesh and Ristenpart, Thomas},
  booktitle={{Proceedings of the 22nd ACM SIGSAC Conference on Computer and Communications Security}},
  pages={1322--1333},
  year={2015},
  month = {10}
}

@article{zhang2021gradient,
  title={{Gradient Statistics Aware Power Control for Over-the-Air Federated Learning}},
  author={Zhang, Naifu and Tao, Meixia},
  journal={IEEE Transactions on Wireless Communications},
  year={2021},
  month={3},
  publisher={IEEE}
}

@inproceedings{cao2021optimizedfl,
  title={{Optimized Power Control for Over-the-Air Federated Edge Learning}},
  author={Cao, Xiaowen and Zhu, Guangxu and Xu, Jie and Cui, Shuguang},
  booktitle={ICC 2021-IEEE International Conference on Communications},
  pages={1--6},
  year={2021},
  organization={IEEE}
}

@article{cao2020optimized,
  title={{Optimized Power Control for Over-the-Air Computation in Fading Channels}},
  author={Cao, Xiaowen and Zhu, Guangxu and Xu, Jie and Huang, Kaibin},
  journal={IEEE Transactions on Wireless Communications},
  volume={19},
  number={11},
  pages={7498--7513},
  year={2020},
  month={8},
  publisher={IEEE}
}

@article{zang2020over,
  title={{Over-the-Air Computation Systems: Optimal Design with Sum-Power Constraint}},
  author={Zang, Xin and Liu, Wanchun and Li, Yonghui and Vucetic, Branka},
  journal={IEEE Wireless Communications Letters},
  volume={9},
  number={9},
  pages={1524--1528},
  year={2020},
  month={5},
  publisher={IEEE}
}

@article{liu2020over,
  title={{Over-the-Air Computation Systems: Optimization, Analysis and Scaling Laws}},
  author={Liu, Wanchun and Zang, Xin and Li, Yonghui and Vucetic, Branka},
  journal={IEEE Transactions on Wireless Communications},
  volume={19},
  number={8},
  pages={5488--5502},
  year={2020},
  month={5},
  publisher={IEEE}
}

@inproceedings{hellstrom2021retransmit,
  title={{Over-the-Air Federated Learning with Retransmissions}},
  author={Hellström, Henrik and Fodor, Viktoria and Fischione, Carlo},
  booktitle={Proceedings of the IEEE 22nd International Workshop on Signal Processing Advances in Wireless Communications (SPAWC)},
  pages={1--5},
  year={2021},
  month={9},
  organization={IEEE}
}

@article{hellstrom2021retransmit2,
  title={{Over-the-Air Federated Learning with Retransmissions (Extended Version)}},
  author={Hellström, Henrik and Fodor, Viktoria and Fischione, Carlo},
  journal={{arXiv}},
  year={2021},
  month={11},
  volume={\\abs/2111.10267},
}

@article{fan2021temporal,
  title={{Temporal-Structure-Assisted Gradient Aggregation for Over-the-Air Federated Edge Learning}},
  author={Fan, Dian and Yuan, Xiaojun and Zhang, Ying-Jun Angela},
  journal={{arXiv}},
  year={2021},
  month={3},
  volume={abs/2103.02270},
}

@article{wang2021federated,
  title={{Federated Learning via Intelligent Reflecting Surface}},
  author={Wang, Zhibin and Qiu, Jiahang and Zhou, Yong and Shi, Yuanming and Fu, Liqun and Chen, Wei and Letaief, Khaled B},
  journal={IEEE Transactions on Wireless Communications},
  year={2021},
  month={7},
  publisher={IEEE}
}

@inproceedings{jiang2019over,
  title={{Over-the-Air Computation via Intelligent Reflecting Surfaces}},
  author={Jiang, Tao and Shi, Yuanming},
  booktitle={Proceedings of the 2019 IEEE Global Communications Conference (GLOBECOM)},
  pages={1--6},
  year={2019},
  month={12},
  organization={IEEE}
}

@article{liu2021csit,
  title={{CSIT-Free Federated Edge Learning via Reconfigurable Intelligent Surface}},
  author={Liu, Hang and Yuan, Xiaojun and Zhang, Ying-Jun Angela},
  journal={{arXiv}},
  month={2},
  year={2021},
  volume={\\abs/1905.10497},
}

@article{liu2021reconfigurable,
  title={{Reconfigurable Intelligent Surface Enabled Federated Learning: A Unified Communication-Learning Design Approach}},
  author={Liu, Hang and Yuan, Xiaojun and Zhang, Ying-Jun Angela},
  journal={IEEE Transactions on Wireless Communications},
  year={2021},
  month={6},
  publisher={IEEE}
}

@article{ning2021survey,
  title={{A Survey on Metaverse: the State-of-the-art, Technologies, Applications, and Challenges}},
  author={Ning, Huansheng and Wang, Hang and Lin, Yujia and Wang, Wenxi and Dhelim, Sahraoui and Farha, Fadi and Ding, Jianguo and Daneshmand, Mahmoud},
  journal={{arXiv}},
  year={2021},
  month={11},
  volume={abs/2111.09673}
}

@article{yagol2018new,
  title={{New Trends in Using Augmented Reality Apps for Smart City Contexts}},
  author={Yagol, Pravesh and Ramos, Francisco and Trilles, Sergio and Torres-Sospedra, Joaqu{\'\i}n and Perales, Francisco J},
  journal={ISPRS International Journal of Geo-Information},
  volume={7},
  number={12},
  pages={478},
  year={2018},
  month={10},
  publisher={Multidisciplinary Digital Publishing Institute}
}

@inproceedings{jiang2020cluster,
  title={{Cluster-Based Cooperative Digital Over-the-Air Aggregation for Wireless Federated Edge Learning}},
  author={Jiang, Ruichen and Zhou, Sheng},
  booktitle={2020 IEEE/CIC International Conference on Communications in China (ICCC)},
  pages={887--892},
  year={2020},
  month={11},
  organization={IEEE}
}

@article{qin2021federated,
  title={{Federated Learning and Wireless Communications}},
  author={Qin, Zhijin and Li, Geoffrey Ye and Ye, Hao},
  journal={IEEE Wireless Communications},
  year={2021},
  moneth = {9},
  publisher={IEEE}
}

@INPROCEEDINGS{9414785,  author={Hu, Y. and Chen, M. and Chen, M. and Yang, Z. and Shikh-Bahaei, M. and Poor, H. V. and Cui, S.},  booktitle={Proceedings of the IEEE International Conference on Acoustics, Speech and Signal Processing (ICASSP)},   title={{Energy Minimization for Federated Learning with {IRS}-Assisted Over-the-Air Computation}},   year={2021},  address={Toronto, ON, Canada}, month={6},}

@article{zhao2019survey,
  title={{A Survey of Intelligent Reflecting Surfaces (IRSs): Towards 6G Wireless Communication Networks}},
  author={Zhao, Jun},
  journal={{arXiv}},
  year={2019},
  month={7},
  volume={abs/1907.04789},
}

@article{hayat2016survey,
  title={{Survey on Unmanned Aerial Vehicle Networks for Civil Applications: A Communications Viewpoint}},
  author={Hayat, Samira and Yanmaz, Ev{\c{s}}en and Muzaffar, Raheeb},
  journal={IEEE Communications Surveys \& Tutorials},
  volume={18},
  number={4},
  pages={2624--2661},
  year={2016},
  month={4},
  publisher={IEEE}
}

@article{polka2017use,
  title={{The Use of UAV's for Search and Rescue Operations}},
  author={P{\'o}{\l}ka, Marzena and Ptak, Szymon and Kuziora, {\L}ukasz},
  journal={Procedia Engineering},
  volume={192},
  pages={748--752},
  year={2017},
  publisher={Elsevier}
}

@article{brik2020federated,
  title={{Federated Learning for UAVs-enabled Wireless Networks: Use Cases, Challenges, and Open Problems}},
  author={Brik, Bouziane and Ksentini, Adlen and Bouaziz, Maha},
  journal={IEEE Access},
  volume={8},
  pages={53841--53849},
  year={2020},
  month={3},
  publisher={IEEE}
}

@ARTICLE{8851408,  author={Chen, M. and Semiari, O. and Saad, W. and Liu, X. and Yin, C.},  journal={IEEE Transactions on Wireless Communications},   title={{Federated Echo State Learning for Minimizing Breaks in Presence in Wireless Virtual Reality Networks}},   year={2020},  volume={19},  number={1},  pages={177-191}, month={1},}

@article{imteaj2021survey,
  title={{A Survey on Federated Learning for Resource-Constrained IoT Devices}},
  author={Imteaj, Ahmed and Thakker, Urmish and Wang, Shiqiang and Li, Jian and Amini, M Hadi},
  journal={IEEE Internet of Things Journal},
  year={2021},
  month={7},
  publisher={IEEE}
}

@article{gafni2021federated,
  title={{Federated Learning: A Signal Processing Perspective}},
  author={Gafni, Tomer and Shlezinger, Nir and Cohen, Kobi and Eldar, Yonina C and Poor, H Vincent},
  journal={{arXiv}},
  year={2021},
  month={3},
  volume={\\abs/2103.17150},
}

@article{konecny2016,
	title={{Federated Optimization: Distributed Machine Learning for On-Device Intelligenc}e},
	author={Jakub Kone\v{c}n{\'y} and H. Brendan McMahan and Daniel Ramage and Peter Richt{\'a}rik},
	journal={ArXiv},
	year={2016},
	month={10},
	volume={abs/1610.02527},
}

@ARTICLE{9264742,  author={Yang, Z. and Chen, M. and Saad, W. and Hong, C. S. and Shikh-Bahaei, M.},  journal={IEEE Transactions on Wireless Communications},   title={{Energy Efficient Federated Learning Over Wireless Communication Networks}},   year={2021},  volume={20},  number={3},  pages={1935-1949}, month={3},}

@article{strinati20196g,
  title={{6G: The Next Frontier: From Holographic Messaging to Artificial Intelligence Using
Subterahertz and Visible Light Communication}},
  author={Strinati, Emilio Calvanese and Barbarossa, Sergio and Gonzalez-Jimenez, Jose Luis and Ktenas, Dimitri and Cassiau, Nicolas and Maret, Luc and Dehos, Cedric},
  journal={IEEE Vehicular Technology Magazine},
  volume={14},
  number={3},
  pages={42--50},
  year={2019},
  month={8},
  publisher={IEEE}
}

@inproceedings{reisizadeh2020fedpaq,
  title={{FedPAQ: A Communication-Efficient Federated Learning Method with Periodic Averaging and Quantization}},
  author={Reisizadeh, Amirhossein and Mokhtari, Aryan and Hassani, Hamed and Jadbabaie, Ali and Pedarsani, Ramtin},
  booktitle={Proceedings of the International Conference on Artificial Intelligence and Statistics},
  pages={2021--2031},
  year={2020},
  month={6},
  organization={PMLR}
}

@article{wei2020federated,
  title={{Federated Learning with Differential Privacy: Algorithms and Performance Analysis}},
  author={Wei, Kang and Li, Jun and Ding, Ming and Ma, Chuan and Yang, Howard H and Farokhi, Farhad and Jin, Shi and Quek, Tony QS and Poor, H Vincent},
  journal={IEEE Transactions on Information Forensics and Security},
  volume={15},
  pages={3454--3469},
  year={2020},
  month={4},
  publisher={IEEE}
}

@article{li2019fair,
  title={Fair Resource Allocation in Federated Learning},
  author={Li, Tian and Sanjabi, Maziar and Beirami, Ahmad and Smith, Virginia},
  journal={{arXiv}},
  year={2019},
  month={5},
  volume={abs/1905.10497},
}

@inproceedings{tran2019federated,
  title={{Federated Learning over Wireless Networks: Optimization Model Design and Analysis}},
  author={Tran, Nguyen H and Bao, Wei and Zo5a, Albert and Nguyen, Minh NH and Hong, Choong Seon},
  booktitle={Proceedings of the IEEE INFOCOM 2019 Conference on Computer Communications},
  pages={1387--1395},
  year={2019},
  month={6},
  organization={IEEE}
}

@misc{krizhevsky2009learning,
  title = {{Learning Multiple Layers of Features from Tiny Images}},
  howpublished = {\href{https://www.cs.toronto.edu/~kriz/learning-features-2009-TR.pdf}{[Online] University of Toronto}},
  note = {Available at \url{https://www.cs.toronto.edu/~kriz/learning-features-2009-TR.pdf}, accessed: 2021-11-11}
}

@article{so2021codedprivateml,
  title={{CodedPrivateML: A Fast and Privacy-Preserving Framework for Distributed Machine Learning}},
  author={So, Jinhyun and G{\"u}ler, Ba{\c{s}}ak and Avestimehr, A Salman},
  journal={IEEE Journal on Selected Areas in Information Theory},
  volume={2},
  number={1},
  pages={441--451},
  year={2021},
  month={1},
  publisher={IEEE}
}

@article{wang2020thirty,
  title={{Thirty Years of Machine Learning: The Road to Pareto-Optimal Wireless Networks}},
  author={Wang, Jingjing and Jiang, Chunxiao and Zhang, Haijun and Ren, Yong and Chen, Kwang-Cheng and Hanzo, Lajos},
  journal={IEEE Communications Surveys \& Tutorials},
  volume={22},
  number={3},
  pages={1472--1514},
  year={2020},
  month={1},
  publisher={IEEE}
}

@article{xia2021survey,
  title={{A Survey of Federated Learning for Edge Computing: Research Problems and Solutions}},
  author={Xia, Qi and Ye, Winson and Tao, Zeyi and Wu, Jindi and Li, Qun},
  journal={High-Confidence Computing},
  pages={100008},
  year={2021},
  publisher={Elsevier}
}

@article{abdulrahman2020survey,
  title={{A Survey on Federated Learning: The Journey from Centralized to Distributed On-Site Learning and Beyond}},
  author={Abdulrahman, Sawsan and Tout, Hanine and Ould-Slimane, Hakima and Mourad, Azzam and Talhi, Chamseddine and Guizani, Mohsen},
  journal={IEEE Internet of Things Journal},
  volume={8},
  number={7},
  pages={5476--5497},
  year={2020},
  month={10},
  publisher={IEEE}
}

@article{guan2021customized,
  title={{Customized Slicing for 6G: Enforcing Artificial Intelligence on Resource Management}},
  author={Guan, Wanqing and Zhang, Haijun and Leung, Victor CM},
  journal={IEEE Network},
  year={2021},
  month={3},
  publisher={IEEE}
}

@inproceedings{sun2017revisiting,
  title={{Revisiting Unreasonable Effectiveness of Data in Deep Learning Era}},
  author={Sun, Chen and Shrivastava, Abhinav and Singh, Saurabh and Gupta, Abhinav},
  booktitle={Proceedings of the IEEE international conference on computer vision},
  pages={843--852},
  month={10},
  year={2017}
}

@article{jin2022communication,
  title={{Communication Efficient Federated Learning with Energy Awareness over Wireless Networks}},
  author={Jin, Richeng and He, Xiaofan and Dai, Huaiyu},
  journal={IEEE Transactions on Wireless Communications},
  year={2022},
  month={1},
  publisher={IEEE}
}

@article{cao2022transmission,
  title={{Transmission Power Control for Over-the-Air Federated Averaging at Network Edge}},
  author={Cao, Xiaowen and Zhu, Guangxu and Xu, Jie and Cui, Shuguang},
  journal={IEEE Journal on Selected Areas in Communications},
  year={2022},
  month={1},
  publisher={IEEE}
}

@article{leng2022client,
  title={{Client Scheduling in Wireless Federated Learning Based on Channel and Learning Qualities}},
  author={Leng, Jichao and Lin, Zihuai and Ding, Ming and Wang, Peng and Smith, David and Vucetic, Branka},
  journal={IEEE Wireless Communications Letters},
  year={2022},
  month={1},
  publisher={IEEE}
}

@article{ding2019gradient,
  title={{Gradient Information for Representation and Modeling}},
  author={Ding, Jie and Calderbank, Robert and Tarokh, Vahid},
  journal={Advances in Neural Information Processing Systems},
  volume={32},
  pages={2396--2405},
  year={2019},
  month={12}
}

@inproceedings{elkhalil2021fisher,
  title={{Fisher Auto-Encoders}},
  author={Elkhalil, Khalil and Hasan, Ali and Ding, Jie and Farsiu, Sina and Tarokh, Vahid},
  booktitle={International Conference on Artificial Intelligence and Statistics},
  pages={352--360},
  year={2021},
  month={4},
  organization={PMLR}
}

@ARTICLE{9121290,
  author={Oh, Seungeun and Park, Jihong and Jeong, Eunjeong and Kim, Hyesung and Bennis, Mehdi and Kim, Seong-Lyun},
  journal={IEEE Communications Letters}, 
  title={{Mix2FLD: Downlink Federated Learning After Uplink Federated Distillation With Two-Way Mixup}}, 
  year={2020},
  volume={24},
  number={10},
  month={6},
  pages={2211-2215},
  doi={10.1109/LCOMM.2020.3003693}}

@article{park2019distilling,
  title={{Distilling on-Device Intelligence at the Network Edge}},
  author={Park, Jihong and Wang, Shiqiang and Elgabli, Anis and Oh, Seungeun and Jeong, Eunjeong and Cha, Han and Kim, Hyesung and Kim, Seong-Lyun and Bennis, Mehdi},
  journal={arXiv preprint arXiv:1908.05895},
  month={8},
  year={2019}
}

@article{zhang2017mixup,
  title={{mixup: Beyond Empirical Risk Minimization}},
  author={Zhang, Hongyi and Cisse, Moustapha and Dauphin, Yann N and Lopez-Paz, David},
  journal={arXiv preprint arXiv:1710.09412},
  year={2017},
  month={10}
}

@article{sifaou2021robust,
   TITLE = {{Robust Federated Learning via Over-The-Air Computation}},
   AUTHOR = {Sifaou, Houssem and Li, Geoffrey Ye},
  journal={{arXiv}},
   NUMBER = {},
   PAGES = {},
   YEAR = {2021},
   MONTH = 11,
   volume={\\abs/2111.01221},
}

@article{vempaty2013distributed,
  title={{Distributed Inference with Byzantine Data: State-of-the-Art Review on Data Falsification Attacks}},
  author={Vempaty, Aditya and Tong, Lang and Varshney, Pramod K},
  journal={IEEE Signal Processing Magazine},
  volume={30},
  number={5},
  pages={65--75},
  year={2013},
  month={8},
  publisher={IEEE}
}

@article{blanchard2017machine,
  title={{Machine Learning with Adversaries: Byzantine Tolerant Gradient Descent}},
  author={Blanchard, Peva and El Mhamdi, El Mahdi and Guerraoui, Rachid and Stainer, Julien},
  journal={Advances in Neural Information Processing Systems},
  volume={30},
  year={2017},
  month={12}
}

@article{prakash2020byzantine,
   TITLE = {{Byzantine-Resilient Federated Learning with Heterogeneous Data Distribution}},
   AUTHOR = {Prakash, Saurav and Hashemi, Hanieh and Wang, Yongqin and Annavaram, Murali and Avestimehr, Salman},
  journal={{arXiv}},
   NUMBER = {},
   PAGES = {},
   YEAR = {2020},
   MONTH = 10,
   volume={\\abs/2010.07541},
}

@inproceedings{shamir2014communication,
  title={{Communication-Efficient Distributed Optimization Using an Approximate Newton-Type Method}},
  author={Shamir, Ohad and Srebro, Nati and Zhang, Tong},
  booktitle={International conference on machine learning},
  pages={1000--1008},
  year={2014},
  month={6},
  organization={PMLR}
}

@inproceedings{xing2020decentralized,
  title={{Decentralized Federated Learning via SGD over Wireless D2D Networks}},
  author={Xing, Hong and Simeone, Osvaldo and Bi, Suzhi},
  booktitle={2020 IEEE 21st International Workshop on Signal Processing Advances in Wireless Communications (SPAWC)},
  pages={1--5},
  year={2020},
  organization={IEEE}
}

@article{sundhar2010distributed,
  title={{Distributed Stochastic Subgradient Projection Algorithms for Convex Optimization}},
  author={Sundhar Ram, S and Nedi{\'c}, Angelia and Veeravalli, Venugopal V},
  journal={Journal of optimization theory and applications},
  volume={147},
  number={3},
  pages={516--545},
  year={2010},
  month={7},
  publisher={Springer}
}

@inproceedings{shi2021over,
  title={{Over-the-Air Decentralized Federated Learning}},
  author={Shi, Yandong and Zhou, Yong and Shi, Yuanming},
  booktitle={2021 IEEE International Symposium on Information Theory (ISIT)},
  pages={455--460},
  year={2021},
  month={7},
  organization={IEEE}
}

@article{pu2021distributed,
  title={{Distributed Stochastic Gradient Tracking Methods}},
  author={Pu, Shi and Nedi{\'c}, Angelia},
  journal={Mathematical Programming},
  volume={187},
  number={1},
  pages={409--457},
  year={2020},
  month={3},
  publisher={Springer}
}

@article{xin2020decentralized,
  title={{Decentralized Stochastic Optimization and Machine Learning: A Unified Variance-Reduction Framework for Robust Performance and Fast Convergence}},
  author={Xin, Ran and Kar, Soummya and Khan, Usman A},
  journal={IEEE Signal Processing Magazine},
  volume={37},
  number={3},
  pages={102--113},
  year={2020},
  month={5},
  publisher={IEEE}
}

@ARTICLE{9681911,
  author={Yue, Sheng and Ren, Ju and Xin, Jiang and Zhang, Deyu and Zhang, Yaoxue and Zhuang, Weihua},
  journal={IEEE Journal on Selected Areas in Communications}, 
  title={{Efficient Federated Meta-Learning over Multi-Access Wireless Networks}}, 
  year={2022},
  month={1},
  volume={Early Access},
  number={},
  pages={1-1},
  doi={10.1109/JSAC.2022.3143259}}

@ARTICLE{9562546,
  author={Liu, Dongzhu and Simeone, Osvaldo},
  journal={IEEE Journal on Selected Areas in Communications}, 
  title={Channel-Driven Monte Carlo Sampling for Bayesian Distributed Learning in Wireless Data Centers}, 
  year={2022},
  volume={40},
  number={2},
  pages={562-577},
  month={2},
  doi={10.1109/JSAC.2021.3118406}}

@article{bory2019deep,
  title={{Deep New: The Shifting Narratives of Artificial intelligence from Deep Blue to AlphaGo}},
  author={Bory, Paolo},
  journal={Convergence: The International Journal of Research into New Media Technologies},
  volume={25},
  number={4},
  pages={627--642},
  year={2019},
  month={2},
  publisher={SAGE Publications Sage UK: London, England}
}

@article{lee2016human,
  title={{Human vs. Computer Go: Review and Prospect [Discussion Forum]}},
  author={Lee, Chang-Shing and Wang, Mei-Hui and Yen, Shi-Jim and Wei, Ting-Han and Wu, I-Chen and Chou, Ping-Chiang and Chou, Chun-Hsun and Wang, Ming-Wan and Yan, Tai-Hsiung},
  journal={IEEE Computational intelligence magazine},
  volume={11},
  number={3},
  pages={67--72},
  year={2016},
  month={7},
  publisher={IEEE}
}

@inproceedings{finn2017model,
  title={{Model-Agnostic Meta-Learning for Fast Adaptation of Deep Networks}},
  author={Finn, Chelsea and Abbeel, Pieter and Levine, Sergey},
  booktitle={International conference on machine learning},
  pages={1126--1135},
  year={2017},
  month={8},
  organization={PMLR}
}

@article{chen2018federated,
  title={{Federated Meta-Learning with Fast Convergence and Efficient Communication}},
  author={Chen, Fei and Luo, Mi and Dong, Zhenhua and Li, Zhenguo and He, Xiuqiang},
  journal={arXiv preprint arXiv:1802.07876},
  year={2018}
}

@article{lakshminarayanan2017simple,
  title={{Simple and Scalable Predictive Uncertainty Estimation Using Deep Ensembles}},
  author={Lakshminarayanan, Balaji and Pritzel, Alexander and Blundell, Charles},
  journal={Advances in neural information processing systems},
  volume={30},
  year={2017}
}

@article{wilson2020bayesian,
  title={{Bayesian Deep Learning and a Probabilistic Perspective of Generalization}},
  author={Wilson, Andrew G and Izmailov, Pavel},
  journal={Advances in neural information processing systems},
  volume={33},
  pages={4697--4708},
  year={2020}
}

@article{wu2017autonomous,
  title={{An Autonomous Wireless Body Area Network Implementation Towards IoT Connected Healthcare Applications}},
  author={Wu, Taiyang and Wu, Fan and Redoute, Jean-Michel and Yuce, Mehmet Rasit},
  journal={IEEE access},
  volume={5},
  pages={11413--11422},
  year={2017},
  month={6},
  publisher={IEEE}
}

@article{broadley2018methods,
  title={{Methods for the Real-World Evaluation of Fall Detection Technology: A Scoping Review}},
  author={Broadley, Robert W and Klenk, Jochen and Thies, Sibylle B and Kenney, Laurence PJ and Granat, Malcolm H},
  journal={Sensors},
  volume={18},
  number={7},
  pages={2060},
  year={2018},
  month={6},
  publisher={Multidisciplinary Digital Publishing Institute}
}
